\documentclass[aps,pra,onecolumn,superscriptaddress,nofootinbib,letterpaper]{revtex4-1}

\usepackage{amsmath,amsfonts,amssymb,array}
\usepackage{hyperref}
\usepackage{color}
\usepackage{graphicx}
\usepackage{cleveref}
\usepackage[normalem]{ulem}

\graphicspath{{figures/}}

\def \be {\begin{enumerate}}
\def \ee {\end{enumerate}}
\def \beq {\begin{equation}}
\def \eeq {\end{equation}}
\def \ba {\begin{eqnarray}}
\def \ea {\end{eqnarray}}

\def \ban {\begin{eqnarray*}}
\def \ean {\end{eqnarray*}}
\def \bfl {\begin{flalign*}}
\def \efl {\end{flalign}}
\def \bsp {\begin{split}}

\def \intd{\int \mbox{d}}
\def \l {\left}
\def \r {\right}

\newcommand{\ketbrad}[1]{|#1\rangle\!\langle #1|}

\newcommand{\braket}[2]{\langle#1|#2\rangle}

\newcommand{\Mean}[1]{\left\langle#1\right\rangle}
\def\ket#1{\left| #1\right>}
\def\bra#1{\left< #1\right|}

\newcommand{\pdp}[2]{\frac{\partial #1}{\partial #2}}

\newcommand{\tr}[1] {\mbox{tr} \left\{ #1 \right \}}

\begin{document}
\title{Tunable inductive coupling of superconducting qubits in the strongly nonlinear regime}

\author{Dvir Kafri}
\affiliation{Google Inc., Venice, CA 90291, USA}
\author{Chris Quintana}
\affiliation{Department of Physics, University of California, Santa Barbara, CA 93106, USA}
\author{Yu Chen}
\affiliation{Google Inc., Santa Barbara, CA 93117, USA}
\author{Alireza Shabani}
\affiliation{Google Inc., Venice, CA 90291, USA}
\author{John M. Martinis}
\affiliation{Department of Physics, University of California, Santa Barbara, CA 93106, USA}
\affiliation{Google Inc., Santa Barbara, CA 93117, USA}
\date{\today}
\author{Hartmut Neven}
\affiliation{Google Inc., Venice, CA 90291, USA}

\begin{abstract}
  For a variety of superconducting qubits, tunable interactions are achieved through mutual inductive coupling to a coupler circuit containing a nonlinear Josephson element. In this paper we derive the general interaction mediated by such a circuit under the Born-Oppenheimer Approximation. This interaction naturally decomposes into a classical part, with origin in the classical circuit equations, and a quantum part, associated with the coupler's zero-point energy. Our result is non-perturbative in the qubit-coupler coupling strengths and in the coupler nonlinearity. This can lead to significant departures from previous, linear theories for the inter-qubit coupling,  including non-stoquastic and many-body interactions. Our analysis provides explicit and efficiently computable series for any term in the interaction Hamiltonian and can be applied to any superconducting qubit type. We conclude with a numerical investigation of our theory using a case study of two coupled flux qubits, and in particular study the regime of validity of the Born-Oppenheimer Approximation.

\end{abstract}
\maketitle

\section{Introduction}
Nonlinearity is essential to superconducting circuit implementations of quantum information. It allows for an individually addressable qubit subspace and tunable interactions between qubit circuits. Qubit-qubit interactions in a variety of platforms are mediated by coupler circuits inductively coupled to the qubits, with tunability  provided by nonlinear Josephson elements~\cite{Hime2006, vanderPloag2007, Harris2007, Allman2010, Bialczak2011, Chen2014}. Several theoretical treatments of such circuits have been performed, including detailed analyses for tunably coupled flux qubits~\cite{Brink2005,Grajcar2006}, phase qubits~\cite{Pinto2010}, lumped-element resonators~\cite{Tian2008}, and transmon-type (gmon) qubits~\cite{Geller2015}. However, both previous classical and quantum analyses have either been linear or have treated the qubit-coupler coupling strengths perturbatively~\cite{Hutter2006}, and they are therefore expected to break down in the regime of strong coupling or large nonlinearities. In particular, the commonly used classical linear analysis can create the misconception that arbitrary inter-qubit coupling strengths can be achieved with a sufficiently nonlinear coupler circuit, an artifact of extending the linear equations beyond their applicable domain. One platform for which a non-perturbative treatment would be of immediate use is quantum annealing, where strong yet accurate two-qubit interactions are necessary and $k$-qubit or non-stoquastic~\cite{Bravyi2008} interactions are desirable, and where the ability to controllably operate in the strongly nonlinear regime could therefore be highly beneficial.

In this work we present a non-perturbative analysis of two or more superconducting qubits inductively coupled through a Josephson coupler circuit. Our treatment is generic in that, as long as the coupling takes the form depicted in Fig.~\ref{diagram}, it is independent of the individual qubit Hamiltonians. In fact, it applies within the infinite dimensional Hilbert space of the underlying circuits implementing the qubits (which can be highly nonlinear with any form for their individual potential energies) and only reduces to the qubit subspace to compute coupling matrix elements.  We numerically investigate the accuracy of our theory in various regimes, with focus on the interesting limit of large coupler nonlinearities $\beta_c \approx 2 \pi L_c I_c^{(c)}/\Phi_0 \lesssim 1$ within the monostable regime of the coupler and for large dimensionless coupling strengths $\alpha_j \equiv M_j/L_j$. Here, $L_c$ and $I_c^{(c)}$ are the coupler's inductor and junction (or DC-SQUID) parameters, and $M_j$ and $L_j$ are the mutual and self inductance of the $j$'th qubit, respectively.

To perform the analysis, we eliminate the coupler circuit using the Born-Oppenheimer Approximation. In this approximation, the coupler circuit's ground state energy dictates the qubit-qubit interaction potential. This potential naturally decomposes into a classical part, whose origin lies in the classical equations of motion, and a small but non-negligible quantum part originating from the coupler circuit's zero-point fluctuations. We derive an exact expression for the classical part and an approximate expression for the quantum part valid in the experimentally relevant limit of small coupler impedance. Using this interaction potential, we derive explicit and efficiently computable Fourier series for all terms in the effective inter-qubit interaction Hamiltonian, including non-stoquastic terms and $k$-body terms with $k>2$ (although these are found to be small for the investigated parameter regimes). Unlike previous results, the interaction is defined explicitly and not in terms of quantum mechanical averages of the coupler system. As a case study, we apply our results to two coupled flux qubits, using parameters from our recent flux qubit design, the fluxmon \cite{Quintana2016}. We find that our results agree with previous treatments in the appropriate limits, but significantly differ in the highly nonlinear regime. We quantify the accuracy of our results by comparing them to an exact numerical diagonalization of the full system, allowing us to study when the Born-Oppenheimer Approximation breaks down.

\section{Interaction mediated by nonlinear circuit}
\subsection{Qubit-coupler Hamiltonian}
\begin{figure}
\includegraphics{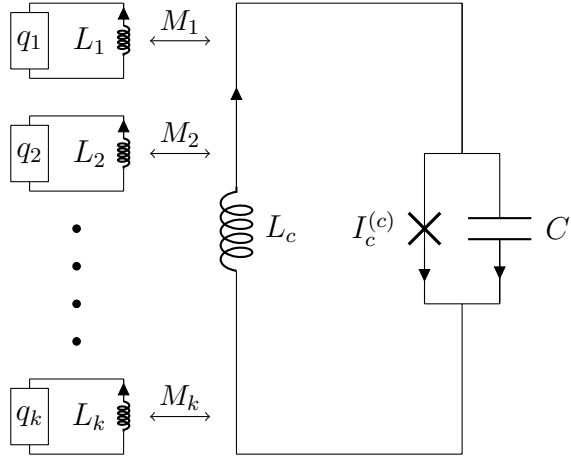}
\caption{A generic circuit for inductive coupling of two or more superconducting circuits (left column). Each smaller circuit $\{q_i,L_i\}$ represents a single qubit. The strength and type of coupling can be tuned via an external magnetic flux $\Phi_{cx}$ applied through the main coupler loop (right-hand side). The coupler's junction may alternatively be a DC SQUID forming an effective Josephson junction with tunable $I_c^{(c)}$ via a separate flux bias.}
\label{diagram}
\end{figure}

We wish to derive the interaction between $k$ circuits (the qubits) inductively coupled through an intermediate circuit (the coupler) as depicted in Fig.~\ref{diagram}. We begin by deriving the full Hamiltonian describing both qubits and coupler. While the coupler circuit is elementary (it contains just an inductor, capacitor, and Josephson junction in parallel), our only assumption about the qubit circuits is that they interact with the coupler through a geometric mutual inductance, $M_j$. Accordingly, we write the current equations defining their dynamics as~\cite{Devoret1995,Burkard2004},
\begin{align}
  \label{currentEqs}
  \begin{split}
    C \ddot \Phi_c + I_c^{(c)} \sin(2 \pi \Phi_c/\Phi_0) - I_{L,c} & = 0\\
    I_{j} - I^{*}_{j} & = 0\,, \quad (1\leq j \leq k)\,.
  \end{split}
\end{align}
For the first equation, $\Phi_c$ denotes the flux across the coupler's Josephson junction (and capacitor), $I_{L,c}$ denotes the current through the coupler's inductor, and $\Phi_0 = h/(2 e)$ is the flux quantum. The second equation simply states that the current $I_{j}$ through qubit $j$'s inductor is equal to the current $I^{*}_{j}$ flowing through the rest of the qubit circuit (represented by box `$q_j$' in the figure). The basic inductive and flux quantization relationships are then
\begin{align}
  \label{fluxQuant}
  \begin{split}
    L_c I_{L,c} + \sum_{j=1}^k M_j I_{j} & = \Phi_{L,c}\\
    L_{j} I_{j} + M_j I_{L,c} & = \Phi_{j}\\
    \Phi_{L,c} & = \Phi_{cx} - \Phi_c\,,
  \end{split}
\end{align}
where $\Phi_{cx}$ is the external flux bias applied to the coupler loop and $\Phi_j$ is the flux across qubit $j$'s inductor. Using these equations and some algebra one can rewrite the current equations in terms of the flux variables,
\begin{align}
  \label{currentEqs2}
  \begin{split}
    C \ddot \Phi_c + I_c^{(c)} \sin(2 \pi \Phi_c/\Phi_0) + \frac{1}{\tilde L_c }\l( \Phi_{c}  -\Phi_{cx}+ \sum_{j=1}^k \alpha_j \Phi_{j}   \r) & = 0\\
     \frac{\Phi_{j}}{L_j} + \alpha_j \frac{1}{\tilde L_c}\l( \Phi_{c}-\Phi_{cx}+ \sum_{j'=1}^k \alpha_{j'} \Phi_{j'}  \r)  - I^{*}_{j} & = 0\,,
  \end{split}
\end{align}
where
\begin{align*}
  \alpha_j& \equiv \frac{M_j}{L_j}\\
  \tilde L_c & \equiv L_c - \sum_{j=1}^k \alpha_j M_j\,.
\end{align*}
The rescaled coupler inductance, $\tilde L_c$, represents the shift in the coupler's inline inductance due to its interaction with the qubits. Although we could similarly rescale the qubit inductances in the second equation~\eqref{currentEqs2}, we instead keep separate all terms that depend on the mutual inductance, $\alpha_j$.

To complete the derivation of the Hamiltonian, we note that equations~\eqref{currentEqs2} are just the Euler-Lagrange equations for the qubits and coupler. Since the $\Phi$-dependent terms correspond to derivatives of the potential energy ($\pdp{U}{\Phi_c}$ and $\pdp{U}{\Phi_{j}}$), we quickly arrive at the corresponding Hamiltonian for the coupled systems
\begin{equation}
  \label{HExact}
  \hat H =   \frac{\hat Q_c^2}{2 C} - E_{J_c} \cos(2 \pi \hat \Phi_c/\Phi_0) + \frac{\l(  \hat \Phi_{c}-\Phi_{cx}+ \sum_{j=1}^k \alpha_j \hat \Phi_{j} \r)^2}{2 \tilde L_c} + \sum_{j=1}^k \hat H_{j} \,.
\end{equation}
Here $\hat H_{j}$ (obtained from $ \frac{\Phi_{j}}{L_j}-I^{*}_{j}$) denotes the Hamiltonian for qubit $j$ in the absence of the coupler (i.e., in the limit $\alpha_j\rightarrow 0$), $\hat Q_c$ is the canonical conjugate to $\hat \Phi_c$ satisfying $[\hat \Phi_c,\hat Q_c] = i \hbar$, and the coupler's Josephson energy is $E_{J_c} = \Phi_0 I_{c}^{(c)}/2 \pi$.

\subsection{Born-Oppenheimer Approximation}

To obtain the effective interaction between the qubits, we now eliminate the coupler's degree of freedom. In other words, we apply the Born-Oppenheimer Approximation~\cite{Born1927} by fixing the (slow) qubit degrees of freedom and assuming that the (fast) coupler is always in its ground state. This is analogous to the Born-Oppenheimer Approximation in quantum chemistry, in which the nuclei (qubits) evolve adiabatically with respect to the electrons (coupler). The coupler's ground state energy (a function of the slow qubit variables, $\Phi_j$) then determines the interaction potential between the qubits. This approximation is valid as long as the coupler's intrinsic frequency is much larger than other energy scales in the system, namely the qubits' characteristic frequencies and qubit-coupler coupling strength.

We begin by considering the coupler-dependent part of the Hamiltonian, ${\hat H_c = \hat H - \sum_j \hat H_{j}}$. We re-express this operator in terms of standard dimensionless parameters,
\begin{align}
  \label{Hc}
  \begin{split}
    \hat H_c  &=  E_{\tilde L_c}\l( 4 \zeta_c^2 \, \frac{\hat q_c^2}{2} + U(\hat \varphi_c; \varphi_x) \r)\\
    U( \varphi_c; \varphi_x ) & = \frac{\l( \varphi_c -  \varphi_x\r)^2}{2 } + \beta_c\cos( \varphi_c) \,,
  \end{split}
\end{align}
where
\begin{align}
  \label{paramDefs}
  \begin{split}
    E_{\tilde L_c} & = \frac{(\Phi_0/2 \pi)^2}{\tilde L_c}\\
    \zeta_c & =  \frac{2 \pi e}{\Phi_0}\sqrt{\frac{\tilde L_c}{C}} = 4 \pi \tilde Z_c /R_K\\
    \beta_c & = 2 \pi \tilde L_c I_{c}^{(c)}/\Phi_0 = E_{J_c}/E_{\tilde L_c}  \\
    \hat q_c & = \frac{\hat Q_c}{2 e}\\
    \hat \varphi_c & = \frac{2 \pi}{\Phi_0}\hat \Phi_c + \pi\\
    \varphi_{cx} & = \frac{2 \pi}{\Phi_0} \Phi_{cx} + \pi\\
    \hat \varphi_j & =  \frac{2 \pi}{\Phi_0} \hat \Phi_{j}\\
    \varphi_x & = \varphi_{cx} -  \l(\sum_{j=1}^k \alpha_j \varphi_{j}   \r) \\
         [\hat \varphi_c,\hat q_c] & = i\,.
  \end{split}
\end{align}
Note that we have defined $\hat \varphi_c$ and $\varphi_{cx}$ with an explicit $\pi$ phase shift, which flipped the sign in front of $\beta_c \cos(\varphi_c)$. Typical coupler inductive energies are on the order of $E_{\tilde L_c}/h \sim 0.5-2$ THz~\cite{Harris2007,Allman2010,Allman2014}. For reasons that will become clear shortly, we assume ${\beta_c \lesssim 1}$ (monostable coupler regime) and low impedance (${\zeta_c\ll 1}$), consistent with typical qubit-coupler implementations\footnote{ For example, $\zeta_c$ is estimated to be $0.013$ in Ref.~\cite{Allman2010}, $0.04$ in the most recent gmon device \cite{Neill2016}, and $0.05$ in our initial fluxmon coupler design~\cite{Quintana2017}.}. Importantly, we are momentarily treating the external flux $\varphi_x$ as a scalar parameter of the Hamiltonian. This is analogous to the Born-Oppenheimer Approximation in quantum chemistry, where the nuclear degrees of freedom are treated as scalar parameters modifying the electron Hamiltonian. Since $\varphi_x$ is a function of the qubit fluxes $\varphi_j$, the coupler's ground state energy $E_g(\varphi_x)$ acts as an effective potential between the qubit circuits. The full effective qubit Hamiltonian under Born-Oppenheimer is then $\hat H_{\textrm{BO}} = \sum_j  \hat H_j + E_g(\hat \varphi_x)$, where the variable $\varphi_x$ is promoted back to an operator. (See Appendix Sections~\ref{BOValidity} and~\ref{BONonadiabatic} for a detailed discussion of this approximation.)

In order to derive an analytic expression for the ground state energy, $E_g(\varphi_x)$, we must first decompose it into classical and quantum parts. This natural decomposition allows for a very precise approximation to the ground state energy, because the classical part (corresponding to the classical minimum value of $H_c$) is the dominant contribution to the energy and can be derived {\it exactly}. The quantum part (corresponding to the zero-point energy) is the only approximate contribution, though it is relatively small for typical circuit parameters.

To begin our analysis, we write the potential energy $U(\hat \varphi_c; \varphi_x )$ in a more suggestive form,
\begin{equation}
  U(\hat \varphi_c; \varphi_x) = U_{\textrm{min}}(\varphi_x) + U_{\textrm{ZP}}(\hat \varphi_c;  \varphi_x)\,.
\end{equation}
Here the scalar
$$U_{\textrm{min}}(\varphi_x) = \min_{\varphi_c} U(\varphi_c ; \varphi_x ) =  \frac{(\varphi_c^{(*)}-\varphi_x)^2}{2} + \beta_c \cos(\varphi_c^{(*)})$$
is the value of the coupler potential at its minimum point $\varphi_c^{(*)}$, i.e. its `height' (overall offset) above zero. Setting $E_g(\varphi_x)$ equal to only $E_{\tilde L_c} U_{\textrm{min}}(\varphi_x)$ corresponds to a completely classical analysis of the coupler dynamics (originating from equation~\eqref{currentEqs2}, prior to quantizing the Hamiltonian; see Appendix Section~\ref{classical}). Unlike $U_{\textrm{min}}(\varphi_x)$, the operator
$$U_{\textrm{ZP}}(\hat \varphi_c; \varphi_x) = U(\hat \varphi_c; \varphi_x) - U_{\textrm{min}}(\varphi_x)$$
does not have a classical analogue -- it corresponds to extra energy due to the finite width of the coupler's ground state wave-function. Combining this operator with the charging energy defines the coupler's zero-point energy
\begin{equation}
  \label{ZPE}
  U_{\textrm{ZPE}}(\varphi_x) = \min_{\langle \psi \ket{\psi} = 1} \bra{\psi} \l( 4 \zeta_c^2 \, \frac{\hat q_c^2}{2} + U_{\textrm{ZP}}(\hat \varphi_c; \varphi_x) \r) \ket{\psi}\,.
\end{equation}
(This minimization picks out the ground state.) The coupler's ground state energy is then the sum of the classical and zero-point energy terms,
\begin{equation}
  E_g/E_{\tilde L_c} = U_{\textrm{min}}(\varphi_x) + U_{\textrm{ZPE}}(\varphi_x)\,.
\end{equation}
Both contributions to the energy are parameterized by the qubit-dependent flux $\varphi_x$, which is what allows us to treat $E_g$ as an effective qubit-qubit interaction potential. In the following two sections we compute an exact expression for $U_{\textrm{min}}(\varphi_x)$ and an approximate expression for $U_{\textrm{ZPE}}(\varphi_x)$ as Fourier series in $\varphi_x$. These are combined in Section~\ref{interactionHamiltonian} to produce an expression for the full qubit-qubit interaction Hamiltonian \eqref{Hint}, the key result of our work.

\subsection{Classical contribution to the interaction potential}
We first discuss the classical component of the coupler's ground state energy. From equation~\eqref{Hc}, the minimum value $U_{\textrm{min}}(\varphi_x)$ can be expressed in terms of the minimum point $\varphi_c^{(*)}$ as
\begin{equation}
  \label{EcInter}
  U_{\textrm{min}}(\varphi_x) = U(\varphi_c^{(*)}; \varphi_x) = \frac{\l(\beta_c \sin(\varphi_c^{(*)} ) \r)^2}{2} + \beta_c \cos(\varphi_c^{(*)})\,,
\end{equation}
where we have used the fact that $\varphi_c^{(*)}$ is a critical point,
\begin{equation}
  \label{trans}
  \l. \partial_{\varphi_c} U(\varphi_{c}; \varphi_x) \r|_{\varphi_c = \varphi_c^{(*)}} = \varphi_c^{(*)} - \varphi_x - \beta_c \sin(\varphi_c^{(*)}) = 0\,.
\end{equation}
Importantly, the parameter $\varphi_c^{(*)}$ is a function of $\varphi_x$ and is defined implicitly as the solution to equation~\eqref{trans}. This equation is identical to the classical current equation~\eqref{currentEqs2} in the large coupler plasma frequency limit $\tilde L_c C \rightarrow 0$ (Appendix Section~\ref{classical}).

Although equation \eqref{EcInter} is exact, it is not useful unless we can express $\varphi_c^{(*)}$ as an explicit function of the qubit degrees of freedom (i.e., the variable $\varphi_x$). To motivate how to do this, we observe that the transcendental equation~\eqref{trans} is unchanged under the transformation $\varphi_{c}^{(*)}\rightarrow \varphi_{c}^{(*)} + 2 \pi$, $\varphi_{x}\rightarrow \varphi_x + 2 \pi$, and similarly $U_{\textrm{min}}(\varphi_x)$ is a periodic function of $\varphi_{c}^{(*)}$ (equation~\eqref{EcInter}). This suggests that we can express $U_{\textrm{min}}(\varphi_x)$ as a Fourier series in $\varphi_x$. Indeed, as shown in Appendix Section~\ref{expMuSeries}, for every integer $\mu$,
\begin{equation}
  \label{transInvert}
  e^{ i \mu \varphi_c^{(*)}} =  \sum_{\nu} e^{i \nu \varphi_x} A_{\nu}^{(\mu)}\,,
\end{equation}
where
\begin{equation}
  \label{fourierCoefs}
  A_{\nu}^{(\mu)} = \l\{\begin{array}{cc}
  \delta_{\mu,0} - \frac{\beta_c}{2}(\delta_{\mu,1} + \delta_{\mu,-1})& \nu = 0\\
  \frac{\mu J_{\nu-\mu}(\beta_c \nu)}{\nu} & \nu \neq 0
  \end{array}\r.\,,
\end{equation}
and $J_\nu(x)$ denotes the Bessel function of the first kind. (Unless otherwise specified, summation indices in this text go over all integers.) Using this equation with $\sin(\varphi_c^{(*)}) = \frac{1}{2 i}\l(e^{i \varphi_c^{(*)}} - e^{-i \varphi_c^{(*)}}\r)$, we define
\begin{align}
  \label{sinBeta}
  \begin{split}
    \sin_{\beta_c}(\varphi_x) & \equiv \sin(\varphi_c^{(*)}) \\
    &= \sum_{\nu} e^{i \nu \varphi_x} \frac{1}{2 i}\l(A_\nu^{(1)} - A_\nu^{(-1)}  \r)\\
    & = \sum_{\nu >0}  \frac{2 J_{\nu}({\beta_c} \nu)}{  {\beta_c} \nu} \sin(\nu \varphi_x)\,.
  \end{split}
\end{align}
The function $\sin_{\beta_c}(\varphi_x)$ is the explicit solution to $\sin(\varphi_c^{(*)})$ satisfying equation~\eqref{trans}, and therefore satisfies the identity
\begin{equation}
  \label{sinBetaCharacteristicEquation}
  \sin_{\beta_c}(\varphi_x) = \sin(\varphi_x + \beta_c \sin_{\beta_c}(\varphi_x))\,.
\end{equation}
In the context of Josephson junctions, $\sin_{\beta_c}(\varphi_x)$ represents the current through the junction as a function of the external flux bias\footnote{For a loop containing only a linear inductor and a Josephson junction, the current through the junction as a function of external bias satisfies $I_J/I_{c}^{(c)} = \sin(\varphi_{cx} + \beta_c I_J/I_{c}^{(c)})$. This is exactly the defining relation of the $\sin_{\beta_c}$ function, equation~\eqref{sinBetaCharacteristicEquation}.}.
Since $\sin_{\beta_c}(\varphi_x) = \sin(\varphi_c^{(*)})$ we can also explicitly write $\varphi_c^{(*)}$ as
\begin{equation}
  \label{varphicstar}
  \varphi_c^{(*)} = \varphi_x + \beta_c \sin_{\beta_c}(\varphi_x)\,.
\end{equation}
Substituting these results into equation~\eqref{EcInter}, we get an explicit expression for the minimum value $U_{\textrm{min}}(\varphi_x)$,
\begin{equation}
  \label{EcInter2}
  U_{\textrm{min}}(\varphi_x) = \frac{(\beta_c \sin_{\beta_c}( \varphi_x))^2}{2} + \beta_c \cos(\varphi_x + \beta_c \sin_{\beta_c}(\varphi_x))\,.
\end{equation}

We now derive the Fourier series for $U_{\textrm{min}}(\varphi_x)$ as a function of $\varphi_x$. Taking the derivative of Equation~\eqref{EcInter2} with respect to $\varphi_x$, one may verify that
\begin{equation}
  \label{EcDeriv}
  \partial_{\varphi_x} U_{\textrm{min}}(\varphi_x) = - \beta_c \sin_{\beta_c}(\varphi_x)\,.
\end{equation}
Here we have used the identity,
\begin{equation}
  \label{sinBetaDeriv}
  \partial_{\varphi_x} \sin_{\beta_c}(\varphi_x) = \frac{\cos(\varphi_x + \beta_c \sin_{\beta_c}(\varphi_x))}{1 - \beta_c \cos(\varphi_x + \beta_c \sin_{\beta_c}(\varphi_x))}\,,
\end{equation}
which can be derived directly from equation~\eqref{sinBetaCharacteristicEquation}. Equation~\eqref{EcDeriv} is analogous to
$$\partial_{\varphi_x}\cos(\varphi_x) = -\sin(\varphi_x)\,,$$ which suggests that we define $U_{\textrm{min}}(\varphi_x)$ as
\begin{equation}
  \label{EcFinal}
  U_{\textrm{min}}(\varphi_x) = \beta_c \cos_{\beta_c}(\varphi_x)\,.
\end{equation}
In analogy with the sine and cosine functions, we define the $\cos_{\beta}(\varphi)$ function as the formal integral of $\sin_{\beta}(\varphi)$,
\begin{align}
  \label{cosBetaSeries}
  \begin{split}
    \cos_{\beta}(\varphi_x) & \equiv  1 - \int_{0}^{\varphi_x} \sin_{\beta}(\theta) \,\mbox{d}\theta \\
    & = \frac{\beta}{2}\l( \sin_{\beta}(\varphi_x)\r)^2 + \cos(\varphi_x + \beta \sin_{\beta}(\varphi_x))\\
    & = 1 + \sum_{\nu >0}  \frac{2 J_{\nu}({\beta} \nu)}{  {\beta} \nu^2} \l(\cos(\nu \varphi_x) - 1\r) \\
    & =  -\frac{\beta}{4} + \sum_{\nu \neq 0}  \frac{J_{\nu}({\beta} \nu)}{  {\beta} \nu^2} e^{i \nu \varphi_x} \,.
  \end{split}
\end{align}
We prove the equality of each of these expressions in Appendix~\ref{cosBetaIdentity}. Equations \eqref{EcFinal} and ~\eqref{cosBetaSeries} exactly characterize the classical part of the coupler's ground state energy, $E_g$. As shown in Fig.~\ref{EgPlots}(a), $U_{\textrm{min}}(\varphi_x)$ is the dominant contribution to $E_g$ in the small impedance limit $\zeta_c \ll 1$. Substituting the definition $\varphi_x = \varphi_{cx} - \sum_j \alpha_j \varphi_j$ into equation~\eqref{EcFinal}, we can interpret $U_{\textrm{min}}(\varphi_x) = \beta_c\cos_{\beta_c}\l(\varphi_{cx} - \sum_j \alpha_j \varphi_j \r)$ as a {\it scalar potential} mediating an interaction between the qubit circuits\footnote{This potential emerges from the conservative vector field, ${\bar S(\varphi_1,\varphi_2,\,...\,,\varphi_k)} ={\beta_c \sin_{\beta_c}(\varphi_x) \sum_j \alpha_j \bar e_j} ={ -\beta_c \nabla  \cos_{\beta_c}\l(\varphi_{cx} - \sum_j \alpha_j \varphi_j\r)}$, where $\bar e_j$ denotes the unit vector associated with the degree of freedom $\varphi_j$.}.

\subsection{Quantum contribution to the interaction potential}
\label{ZPEDerivation}

We now discuss the quantum part of the coupler ground state energy. This is given by the ground state energy of $\hat H_c - E_{\tilde L_c}U_{\textrm{min}}(\varphi_x)$ (equation~\eqref{ZPE}), which represents the coupler's zero-point energy. To approximate this energy we expand the zero-point potential, $U_{\textrm{ZP}} = U(\hat \varphi_c;\varphi_x) - U_{\textrm{min}}(\varphi_x)$, about the classical minimum point $\varphi_c^{(*)}$. Since $U_{\textrm{ZP}}(\varphi_c; \varphi_x)$ and its derivative vanish at the minimum point $\varphi_c^{(*)}$, the Taylor series of  $U_{\textrm{ZP}}$ is of the form
\begin{equation}
  \label{linHam}
  \hat H_c/E_{\tilde L_c}   = U_{\textrm{min}}(\varphi_x) + \l( 4 \zeta_c^2\, \frac{\hat q_c^2}{2}   + \frac{U_{\textrm{ZP}}''(\varphi_c^{(*)}; \varphi_x)}{2}(\hat \varphi_c - \varphi_c^{(*)})^2   \r)  + O\l((\hat \varphi_c - \varphi_c^{(*)})^3\r)\,,
\end{equation}
where
\begin{equation}
  U_{\textrm{ZP}}''(\varphi_c; \varphi_x) =  \partial_{\varphi_c}^2 U(\varphi_c; \varphi_x)  = 1 - \beta_c \cos(\varphi_c)\,.
\end{equation}
If we neglect the terms of order $O((\hat \varphi_c - \varphi_c^{(*)})^3)$, the zero-point energy of $\hat H_c$ is the same as for a harmonic oscillator,
\begin{align}
  \label{UZPEfirst}
  \begin{split}
   U_{\textrm{ZPE}} &\simeq \frac{1}{2}\sqrt{4 \zeta_c^2 U''_{\textrm{ZP}}(\varphi_c^{(*)}; \varphi_x)} \\
    & =\zeta_c \sqrt{1 - \beta_c \cos(\varphi_c^{(*)})}\,.
  \end{split}
\end{align}
The harmonic approximation is the second approximation we use to derive the qubit-qubit interaction potential. (The zero-point energy $E_{\tilde L_c} U_{\textrm{ZPE}} \rightarrow E_{\tilde L_c} \zeta_c = \frac{\hbar}{2\sqrt{\tilde L_c C}}$ in the limit $\beta_c\rightarrow 0$, as expected for the linear coupler limit.)

As we did for the classical component $U_{\textrm{min}}(\varphi_x)$, we wish to compute the Fourier series of $U_{\textrm{ZPE}}$ in the qubit-dependent flux parameter $\varphi_x$. To do so, we first write $U_{\textrm{ZPE}}$ as a Fourier series in $\varphi_c^{(*)}$,
\begin{equation}
  \sqrt{1 - \beta \cos(\varphi_c^{(*)})} = \sum_{\mu} G_\mu(\beta)e^{i \mu \varphi_c^{(*)}}\,,
\end{equation}
where the functions $G_{\mu}(\beta)$ satisfy\footnote{ The generalized binomial $\binom{z}{k}= \frac{1}{k!}(z)(z-1)(z-2)\,...\,(z-k+1)$ for integer $k\geq0$ and is zero for negative integers $k$.  }
\begin{align}
  \begin{split}
  G_\mu(\beta) & = \sum_{l\geq 0} \binom{1/2}{\mu+2l}\binom{\mu+2l}{l}\l(-\frac{\beta}{2} \r)^{\mu+2l}\\
  & = \l(-\frac{\beta}{2}\r)^\mu \binom{1/2}{\mu} {_2F_1}\l(\frac{\mu}{2}-\frac{1}{4},\frac{\mu}{2}+\frac{1}{4};1+\mu; \beta^2 \r)\,,
  \end{split}
\end{align}
and ${_2F_1(a,b;c;z)}$ is the confluent hypergeometric function. Combining this with equation~\eqref{transInvert} in the previous section, we obtain the desired series,
\begin{align}
  \label{UZPE}
  \begin{split}
    U_{\textrm{ZPE}}(\varphi_x) & = \zeta_c\l( G_{0}(\beta_c)- \beta_c G_{1}(\beta_c)+ \sum_{\nu\neq 0}e^{i \nu \varphi_x} \l(\frac{1 }{\nu} \sum_\mu \mu\, G_{\mu}(\beta_c) J_{\nu-\mu}(\beta_c \nu)  \r) \r)\,.
  \end{split}
\end{align}
We derive the above identities in Appendix~\ref{sqrtCosFS}. The functions $G_{\mu}(\beta_c)$ decay exponentially in $\mu$, so numerical evaluation of the inner sum typically requires only a few terms (see Appendix~\ref{truncationError}). In Fig~\ref{EgPlots}(b). we compare our approximate value for $U_{\textrm{ZPE}}$ (equations~\eqref{UZPEfirst} and \eqref{UZPE}) to the numerically exact zero-point energy (equation \eqref{ZPE}).

\subsection{Total interaction Hamiltonian}
\label{interactionHamiltonian}

Having computed both classical and quantum parts of the coupler ground state energy $E_g$, we now set this quantity equal to the qubit-qubit interaction potential.  In the language of physical chemistry, $E_g(\varphi_x)$ is the potential energy surface that varies with the qubit flux variables, $\varphi_j$.  We can immediately read off this value from equations~\eqref{EcFinal} and~\eqref{UZPE},
\begin{align}
  \label{EgFinal}
  \begin{split}
     E_{g}(\varphi_x)/E_{\tilde L_c} & = \beta_c \cos_{\beta_c}(\varphi_x) + U_{\textrm{ZPE}}(\varphi_x) \\
     & = \sum_{\nu} e^{i\nu \varphi_x} B_{\nu}\,,
  \end{split}
\end{align}
where
\begin{equation}
  \label{interactionSeries}
  B_{\nu}  = \l\{\begin{array}{cc}
  -\frac{\beta_c^2}{4}+ \zeta_c\l( G_{0}(\beta_c)- \beta_c G_{1}(\beta_c)\r)  & \nu = 0\\
  \frac{ J_{\nu}(\beta_c \nu)}{\nu^2} + \zeta_c \l(\sum_{\mu} \frac{\mu}{\nu} G_{\mu}(\beta_c) J_{\nu-\mu}(\beta_c \nu) \r)& \nu \neq 0
  \end{array}\r.\,.
\end{equation}
With this result we can complete the Born-Oppenheimer Approximation: substituting for $\varphi_x = \varphi_{cx} - \sum_j \alpha_j \varphi_{j}$, the interaction potential mediated by the coupler is thus
\begin{equation}
  \label{Hint}
  \hat H_{\textrm{int}} = E_{g}\l(\varphi_{cx} - \sum_{j} \alpha_j \hat \varphi_j\r)    = E_{\tilde L_c} \sum_{\nu}B_{\nu} e^{i \nu \varphi_{cx}} e^{-i \nu \l(\sum_{j}\alpha_j \hat \varphi_j\r)} \,.
\end{equation}
We note that~\cite{Abramowitz1964}, since $J_{-\nu}(x) = J_{\nu}(-x) = (-1)^\nu J(x)$ and $G_{-\mu}(\beta_c) = G_{\mu}(\beta_c)$, the Fourier coefficients are symmetric, $B_\nu = B_{-\nu}$. Thus $\hat H_{\textrm{int}}$ is an Hermitian operator (as expected) and can be expressed as a Fourier cosine series.

\begin{figure}[h]
\includegraphics[width = \textwidth,trim={2cm 0 2cm 0},clip]{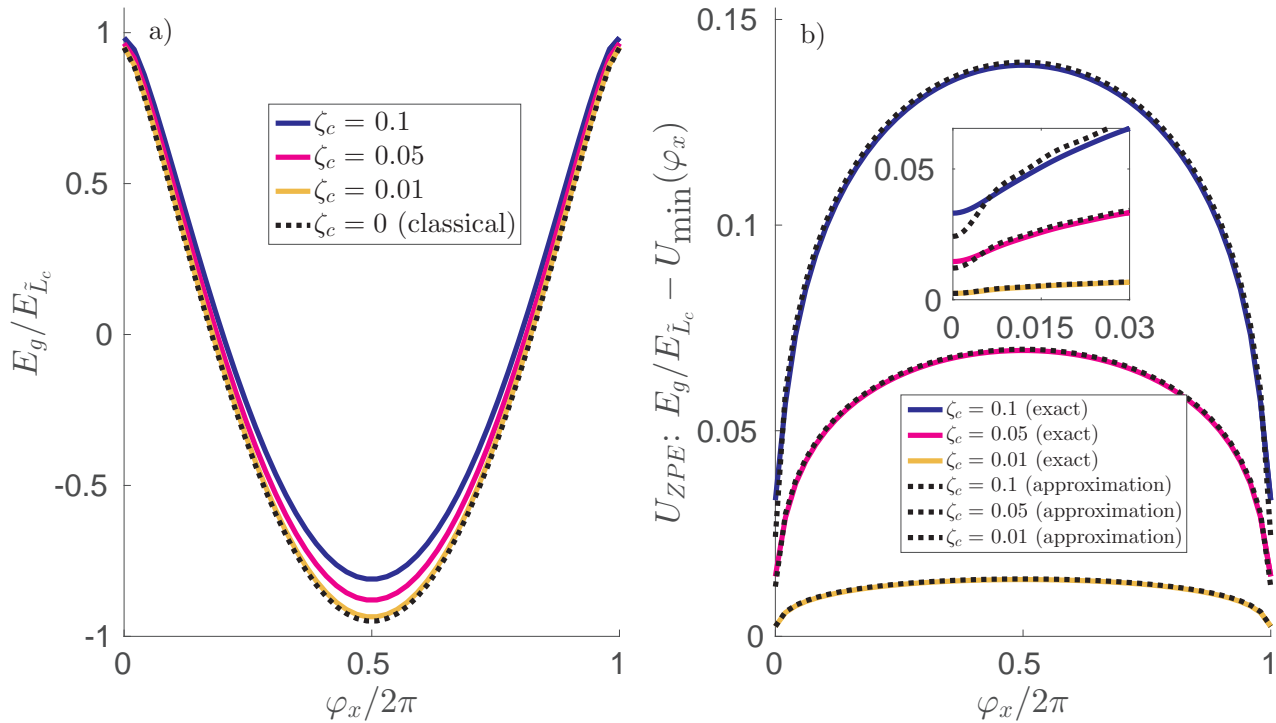}
\caption{a) Coupler ground state energy as a function of external flux bias, $\varphi_x$. Solid lines: exact ground state energy of $\hat H_c$ (equation~\eqref{Hc}) computed by diagonalizing in the first $50$ harmonic oscillator basis states. The coupler parameters correspond to $\beta_c = 0.95$ and $\zeta_c = 0.1$ (dark blue), $0.05$ (magenta), $0.01$ (light orange), respectively. Dashed, black line: classical component of the coupler ground state energy, computed using the scalar function $U_{\textrm{min}}(\varphi_x) = \beta_c \cos_{\beta_c}(\varphi_x)$. b) Coupler zero-point energy as a function of external flux bias, $\varphi_x$. Solid lines: difference between the exact ground state energy $E_g/E_{\tilde L_c}$ (computed numerically as above) and the classical energy contribution, $U_{\textrm{min}}(\varphi_x)$. Overlayed dashed lines: linearized approximation to the coupler zero-point energy, computed using equation~\eqref{UZPE} and truncating the series at $|\nu|\leq \nu_{\textrm{max}} = 100$. Inset are the same curves, restricted to the bias range $\varphi_x \in [0,0.03] \times 2 \pi$.}
\label{EgPlots}
\end{figure}

We stress two important points related to equation \eqref{Hint}, which is the central result of our work. First, our result leads to quantitatively different predictions from previous treatments~\cite{Brink2005,Tian2008,Geller2015}. These expand the coupler ground state energy to second order in the flux variables $\hat \varphi_j$ to derive an `effective mutual inductance' between the qubits. As we shall see, the discrepancy between these results is most pronounced when the qubit-coupler interaction $\alpha_j$ is large or when the coupler nonlinearity $\beta_c$ approaches 1.\footnote{Note that as $\alpha_j$ or $\beta_c$ increase, one must also ensure that the Born-Oppenheimer Approximation remains valid.} Second, since the Fourier coefficients $B_\nu$ decay quickly to zero~\cite[equation 9.1.63]{Abramowitz1964}, the interaction $\hat H_{\textrm{int}}$ is a smooth, bounded function of the qubit flux operators. This remains true even in the regime of large coupler nonlinearity ($\beta_c \approx 1$), and it reinforces the physical intuition that the qubit-qubit coupling strength cannot diverge as $\beta_c\rightarrow 1$.\footnote{Indeed, since $\exp(-i \nu \sum_j \alpha_j \hat \varphi_j)$ is a unitary operator, every matrix element of $\hat H_{\textrm{int}}/E_{\tilde L_c}$ is bounded by $\sum_j |B_{\nu}|\leq \beta_c(1 + \beta_c/4) -\zeta_c\l(\sqrt{1-\beta_c} -G_0(\beta_c) + \beta_c G_1(\beta_c) \r)$. See Appendix Section~\ref{truncationError}.}

We conclude this section by discussing the approximations used to reach equation~\eqref{Hint}. First, the Born-Oppenheimer Approximation is used to replace the coupler Hamiltonian with its ground state energy. This is equivalent to assuming the full system wavefunction (in the flux basis) is of the form
  \begin{equation}
    \label{BOansatz0}
  \Psi(\varphi_c,\bar \varphi_q,t) = \psi_{g}(\varphi_c; \bar \varphi_q) \, \chi(\bar \varphi_q,t)\,.
\end{equation}
  Here the function $ \psi_{g}(\varphi_c; \bar \varphi_q)$ denotes the ground state of the coupler Hamiltonian $\hat H_c$, equation~\eqref{Hc}. Like $\hat H_c$, we view this wavefunction as parameterized by the qubit flux variables, $\bar \varphi_q = (\varphi_1,\varphi_2,\,...\,\varphi_k)$. Inserting this ansatz into the full Hamiltonian's ($\hat H_c + \sum_j \hat H_j$) Schroedinger equation, in Appendix Section~\ref{BOValidity} we integrate out the coupler degree of freedom and obtain a reduced equation of motion for just the qubit wavefunction, $\chi(\bar \varphi_q)$. Up to a small correction (discussed below), the resulting dynamics corresponds to an effective qubit Hamiltonian, $\hat H_{\textrm{BO}} = E_g(\hat \varphi_x) + \sum_j \hat H_j$. Although intuitively similar, the ansatz wavefunction used above is distinct from standard adiabatic elimination\cite{Brion2007}, since that approximation accounts for virtual transitions into higher energy excited states.

  Born-Oppenheimer is a valid approximation when transitions out of the coupler ground state (the ansatz~\eqref{BOansatz0}) are suppressed. Heuristically, this holds when the characteristic qubit energy scale $\hbar \omega_q$ is much less than the gap between coupler's ground and first excited state energies. For $\beta_c<1$ not too close to one, a good bound for this condition is
\begin{equation}
  \hbar \omega_q  \ll  \frac{\hbar}{\sqrt{\tilde L_c C}}\sqrt{1 -\beta_c}\,,
\end{equation}
where on the right hand side we have approximated the coupler's energy gap by twice its (linearized) minimum zero point energy\footnote{The right hand side is only an approximate lower bound for the coupler's energy gap, which in fact does not vanish as $\beta_c \rightarrow 1$.}. More concretely, there are two corrections to Born-Oppenheimer that determine when it breaks down. First, the Born-Oppenheimer Diagonal Correction~\cite{Tully1976,Valeev2003} is a direct modification to the coupler mediated potential, $E_g(\hat \varphi_x)$, which requires no change to the ansatz wavefunction~\eqref{BOansatz0}. We analyze this correction in Appendix Sections~\ref{BOValidity} and find that it is negligible for typical circuit parameters. More important are non-adiabatic corrections to Born-Oppenheimer, which are associated with transitions from the ansatz wavefunction $\psi_{g}(\varphi_c; \bar \varphi_q) \, \chi(\bar \varphi_q,t)$ to excited states of the coupler. We derive formal expressions for these corrections in Appendix Section~\ref{BONonadiabatic}, though due to their complexity we do not have concise analytical expressions bounding their size. Instead we have carried out a numerical study (Section~\ref{numStudy}) to validate our approximation for typical flux qubit circuit parameters.

The second approximation used to derive equation~\eqref{Hint} is the harmonic approximation to the coupler's zero-point energy (equation~\eqref{linHam}). This is mainly a concern when the coupler bias is close to peak coupling, $\varphi_{cx} \approx 0$ (mod $2 \pi$), and the coupler nonlinearity $\beta_c$ approaches $1$ (cf. inset of Fig.~\ref{EgPlots}b); in that limit the harmonic approximation to the zero-point energy ($U_{ZPE} = \zeta_c \sqrt{1 - \beta_c \cos(\varphi_c^{(*)})}$) vanishes and the quartic correction to $\hat H_c$ becomes relevant. As we shall see below, the zero-point energy component of $E_g$ does have a non-negligible effect on the qubit dynamics, but for typical coupler impedances and non-zero bias $\varphi_{cx}$ the inaccuracy in the harmonic approximation is small (see also Fig~\ref{ZPEComparison} in the Appendix).

\section{Projection into the qubit basis}

\label{reductionQubit}
We now describe an efficient method for computing the qubit dynamics mediated by the coupler. It applies to any number of qubits interacting through a single coupler and arises from the generic qubit Hamiltonian derived in the previous section,
\begin{equation}
  \label{HFull}
   \hat H =  \hat H_{\textrm{int}} +\sum_{j =1}^k \hat H_{j}\,.
\end{equation}
Here $\hat H_{j}$ is the local Hamiltonian of qubit $j$ in the absence of the coupler and $\hat H_{\textrm{int}}$ is the general interaction Hamiltonian of equation \eqref{Hint}. Our method is based on the Fourier decomposition of $\hat H_{\textrm{int}}$, a sum of operators of the form ${\exp(i \nu \sum_j \alpha_j \hat \varphi_j) = \prod_j \exp(-i \nu \alpha_j \hat \varphi_j)}$. This product form means we need only compute matrix elements of single qubit operators (cf. equation \eqref{cjDef}). Accordingly, the cost of this method scales only linearly in the number of distinct qubits. The effect of the local Hamiltonians $\hat H_j$ on the qubit dynamics is implementation dependent.

To compute the dynamics induced by the coupler, we restrict our analysis to the `qubit subspace' of each qubit Hamiltonian. (Typically these are spanned by the ground and first excited state of $\hat H_{j}$.) Accordingly, we let $\ket{0}_j$ and $\ket{1}_j$ denote a basis for the local qubit subspace of $\hat H_{j}$. The projection operator into this space is then
\begin{equation}
  \hat P_{j} = \ketbrad{0}_j + \ketbrad{1}_j \,.
\end{equation}
Within this convention we define the Pauli operators $(I,\sigma_x,\sigma_y,\sigma_z)$ in the usual way. We now consider the projection of the exponential operators used in the Fourier series description of $\hat H_{\textrm{int}}$ (equation~\eqref{Hint}). Written within the qubit subspace, we have:
\begin{align}
  \label{expPhi}
  \begin{split}
   \hat P_{j}\, e^{-i s \hat \varphi_j}\, \hat P_{j} &=  \sum_\eta c^{(j)}_\eta(s) \sigma_\eta^{(j)} \,,
  \end{split}
\end{align}
where $\eta \in \{ I,x,y,z\}$ indexes the identity operator and three Pauli operators acting on qubit $j$. Using the identity
 \begin{equation}
   \label{pauliId}
   \mbox{tr}[\sigma_{\alpha}\sigma_{\beta}  ]/2 = \delta_{\alpha \beta}\,,
 \end{equation}
we see that
\begin{equation}
  \label{cjDef}
  c_\eta^{(j)}(s) = \mbox{tr}\l[\frac{\sigma^{(j)}_\eta}{2}e^{-i s\hat \varphi_j}  \r]\,,
\end{equation}
or more explicitly (and dropping the qubit index $j$),
\begin{align}
  \label{cDef}
\begin{split}
  c_I(s) & = \frac{\bra{0}e^{-i s \hat \varphi}\ket{0} + \bra{1}e^{-i s \hat \varphi}\ket{1}}{2}\\
  c_x(s) & = \frac{\bra{0}e^{-i s \hat \varphi}\ket{1}+\bra{1}e^{-i s \hat \varphi}\ket{0}}{2} \\
  c_y(s) & = i \frac{\bra{0}e^{-i s \hat \varphi}\ket{1}-\bra{1}e^{-i s \hat \varphi}\ket{0}}{2} \\
  c_z(s) & = \frac{\bra{0}e^{-i s \hat \varphi}\ket{0} - \bra{1}e^{-i s \hat \varphi}\ket{1}}{2} \,.
\end{split}
\end{align}
We note that in general these coefficients are complex valued and differ between each qubit.

To finish our analysis we also project $\hat H_{\textrm{int}}$ into the qubit subspace. We again write this projection as a sum of Pauli operators,
\begin{equation}
  \label{HintProjection}
  \hat P_q \hat H_{\textrm{int}} \hat P_q  = \sum_{\bar \eta} g_{\bar \eta} \, \sigma_{\bar \eta} \,,
\end{equation}
where $\hat P_q = \hat P_{1}\otimes\hat P_{2}\otimes\,...\,\otimes \hat P_{k}$ and the vector ${\bar \eta} = (\eta_1,\eta_2,\,...\,\eta_{k})$ denotes the corresponding product of Pauli operators,
\begin{equation}
  \label{sigmaAlpha}
  \sigma_{\bar \eta} = \sigma_{\eta_1}^{(1)}\otimes  \sigma_{\eta_2}^{(2)}\otimes\,...\,\otimes  \sigma_{\eta_k}^{(k)} \,.
\end{equation}
With this decomposition we directly compute
\begin{align}
  \label{gEta}
  \begin{split}
    g_{\bar \eta} & = \mbox{tr}\l[\frac{\sigma_{\bar \eta}}{2^k} \hat H_{\textrm{int}} \r]\\
    & = E_{\tilde L_c} \sum_{\nu } \mbox{tr}\l[   \frac{\sigma_{\bar \eta}}{2^k} B_{\nu} e^{i \nu \varphi_{cx}} e^{-i \nu \l(\sum_{j}\alpha_j \hat \varphi_j\r)}  \r]\\
& =  E_{\tilde L_c} \sum_{\nu } B_{\nu} e^{i \nu \varphi_{cx}} \prod_{j = 1}^k \mbox{tr}\l[   \frac{\sigma_{\eta_j}^{(j)}}{2}   e^{-i \nu \alpha_j \hat \varphi_j } \r]  \\
& =   E_{\tilde L_c} \sum_{\nu } B_\nu e^{i \nu \varphi_{cx}} \prod_{j = 1}^k c_{\eta_j}^{(j)}(\nu \alpha_j) \,.
  \end{split}
\end{align}
Each line of the above calculation follows from \eqref{pauliId}, \eqref{Hint}, \eqref{sigmaAlpha}, and \eqref{cjDef}, respectively. (This equation also encompasses the individual qubit operators induced by the presence of the coupler, e.g., for $\bar \eta = (x,I,I,\,...\,,I)$.) Thus the calculation of $g_{\bar \eta}$ reduces to computing the single qubit coefficients $c_{\eta_j}^{(j)}(\nu \alpha_j)$ and evaluating the sum in \eqref{gEta}. For realistic calculations the sum~\eqref{gEta} must be truncated at some maximum value $\nu_{\textrm{max}}$, though for $\beta_c < 1$ the truncation error decays rapidly with $\nu_{\textrm{max}}$ (since the functions defining $B_{\nu}$ decay exponentially in $\nu$, see \cite[equation 9.1.63]{Abramowitz1964}). We give a technique for bounding this error in Appendix Section~\ref{truncationError}.

We remark that the reduction into the qubit subspace is actually an approximation of the qubit dynamics. This is because $\hat H_{\textrm{int}}$ generally has non-zero matrix elements between the qubit subspace $\mathcal{P}$ (represented by projector $\hat P_q$) and its complement, $\mathcal{Q}$. Hence the projection in equation~\eqref{HintProjection} is valid only in the limit that transitions into $\mathcal{Q}$ are suppressed. This occurs if there is a large energy gap between $\mathcal{P}$ and $\mathcal{Q}$, but unfortunately this is not always the case. For example, for three distinct qubits with low nonlinearity, it is possible to observe a resonance\footnote{The energy splittings $E_{mn}^{(j)} = E_m^{(j)} - E_n^{(j)}$ are defined with respect to the local qubit Hamiltonian, $\hat H_j$. } of the form $E_{20}^{(1)} = E_{10}^{(2)} + E_{10}^{(3)}$. The multi-qubit transition $\ket{g, e_1, e_1}\rightarrow \ket{e_2, g, g}$ (where $\ket{g}, \ket{e_m}$ denote the ground and $m$th excited state) thus conserves energy with respect to the local Hamiltonian $\sum_j \hat H_j$. Such accidental degeneracies can occur even in the highly nonlinear case where the qubit energies are far from evenly spaced.  As long as these resonant transitions correspond to non-negligible matrix elements of $\hat H_{\textrm{int}}$, over time the composite qubit system can be mapped outside of the qubit subspace $\mathcal{P}$. One must therefore take special care to account for degeneracies when using equation~\eqref{gEta}, especially when more than two qubits interact through the same coupler. A standard technique accounting for the higher energy states is the Schrieffer-Wolff transformation\cite{Bravyi2011}. This treatment is based on algebraic transformations acting on a Hilbert space with more than four states, so applying it to continuous variable circuits would likely preclude any analytical results as we have obtained for the Born-Oppenheimer Approximation\footnote{The Schrieffer-Wolff transformation is not equivalent to the standard Born-Oppenheimer Approximation applied in our text. Indeed, while the former explicitly depends on matrix elements involving higher energy excited states, the latter is only explicitly dependent on the (scalar) ground state energy of the coupler degree of freedom.}. A practical approach would be to use the Schrieffer-Wolff transformation to account for the higher energy qubit states after using the Born-Oppenheimer Approximation to account for the coupler. This has the advantage of first removing the coupler Hilbert space, which greatly reduces the numerical cost of applying Schrieffer-Wolff.

We note that result~\eqref{gEta} in principle allows for couplings absent in linear theories describing $\hat H_{\textrm{int}}$. For example, it predicts non-zero $k$-body ($k>2$) couplings between multiple qubits, which could be a powerful feature in a quantum annealer where `tall and narrow' potential barriers allow quantum tunneling to outperform classical counterparts~\cite{Boixo2016}. From a quantum information perspective it would also be interesting to engineer tunable non-commuting couplings, for example $\sigma_x\otimes\sigma_x$ and $\sigma_x\otimes\sigma_z +\sigma_z\otimes\sigma_x$. Interactions of this second type are non-stoquastic, i.e. they may have positive off-diagonal elements in any computational basis. These are believed necessary to observe exponential quantum speedups over classical algorithms~\cite{Bravyi2008,Biamonte2008}. The presented analytic derivation in this paper makes it possible to consider inductive couplings to implement such non-stoquastic terms. We consider these kinds of couplings in Section~\ref{kLocalNonStoquastic}.

\section{Two qubit case and linear approximations}

\begin{figure}
\includegraphics{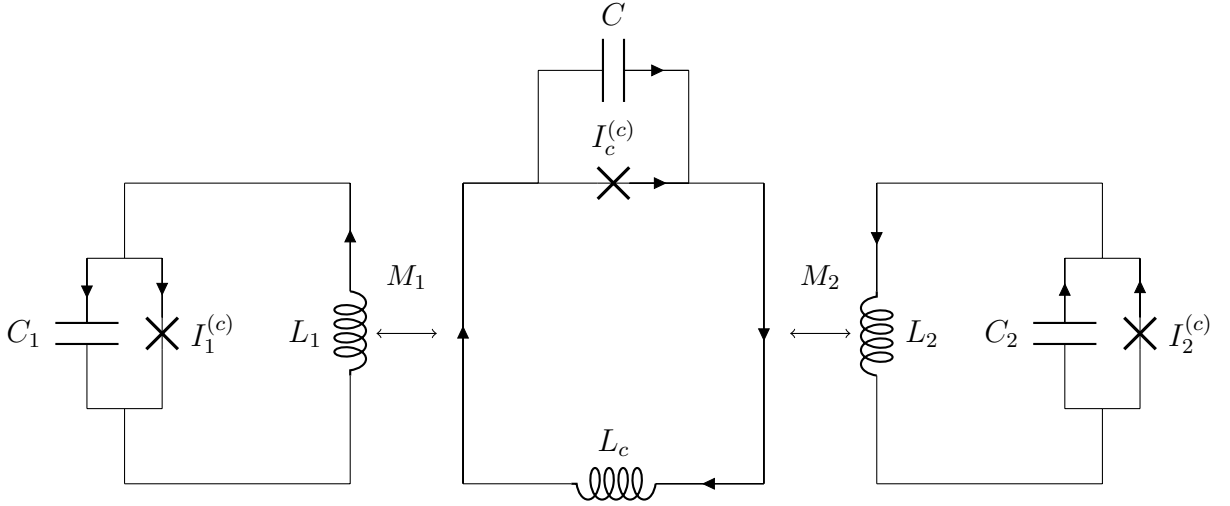}
\caption{Standard flux qubits with interaction mediated by an inductive coupler.}
\label{diagram2Qubits}
\end{figure}

In this section we limit our consideration to the case of two coupled flux qubits (Fig.~\ref{diagram2Qubits}). To compare our analysis to previous work, we linearize the coupler-mediated interaction potential $E_g(\varphi_{x})$ (equation~\eqref{EgFinal}) about the {\it qubit} degrees of freedom and show that it reproduces the standard picture of an effective mutual inductance mediated by the coupler~\cite{Brink2005,Tian2008,Geller2015}. This result is perturbative in the qubit-coupler interaction strength $\alpha_j = M_j/L_j$ and is therefore equivalent to the weak coupling limit. In the subsequent section we will compare the predictions of this linear theory our nonlinear result. We conclude this section with a different treatment of the qubit-qubit coupling, valid when the qubit basis states have a definite parity. Interestingly, where the linear theory treats the coupling in terms of the second derivative of $E_g$, this (more precise) theory expresses it as a second order {\it finite difference}\cite{Averin2003}. This distinction between continuous and discrete derivatives allows us to bound the error between the linear theory and nonlinear theory of the previous section.

\subsection{Flux qubit Hamiltonian}
\label{fluxQubit}

We begin by describing the flux qubit Hamiltonian. The circuit diagrams of these qubits are identical to those of the coupler, though their characteristic frequencies are necessarily smaller. Similarly to the coupler, they are characterized by three parameters\footnote{Other forms of flux qubit also exist~\cite{Mooij1999,Koch2007,Yan2016}. Our analysis can be similarly applied in these cases, with resulting numerical examples showing the same qualitative trends.}:
\begin{align}
  \begin{split}
    E_{L_j} & = \frac{ (\Phi_0/2\pi)^2}{L_j}\\
    \zeta_j & = \frac{2 \pi e}{\Phi_0}\sqrt{\frac{ L_j}{C_j}}= 4 \pi Z_j/R_K \\
    \beta_j & = \frac{2 \pi}{\Phi_0} L_j I_{j}^{(c)}= E_{J_j}/E_{L_j} \,,
  \end{split}
\end{align}
Here $E_{L_j}$ represents the characteristic energy of the qubit's linear inductor and the dimensionless parameter $\zeta_j$ represents its characteristic impedance. These parameters are related to the $LC$ plasma frequency through ${f_{LC,j} = \frac{1}{2 \pi\sqrt{L_j C_j}} = 2  \zeta_j E_{L_j}/h}$. For typical flux qubit implementations of this type \cite{Harris2010, Quintana2017} $E_{L_j}/h$ is on the order of hundreds of GHz while $\zeta_j$ is between $0.01$ and $0.1$, so that $f_{LC,j}$ ranges from a few to tens of GHz.  The parameter $\beta_j$ represents the nonlinearity in the qubit circuit due to the Josephson element. This parameter can vary between circuit designs, and unlike the coupler, within our analysis it is relevant to consider regimes where $\beta_j>1$ (corresponding to a multi-well potential). The qubit Hamiltonian has an identical form to the coupler Hamiltonian of equation~\eqref{Hc},
\begin{equation}
  \label{HQubit}
  \hat H_j = E_{L_j}\l(4 \zeta_j^2 \frac{\hat q_j^2}{2}  + \frac{(\hat \varphi_j - \varphi_{jx})^2}{2}+\beta_j \cos(\hat \varphi_j )  \r)\,,
\end{equation}
where the qubit charge and flux variables satisfy $[\hat \varphi_j,\hat q_j] = i$ and $\varphi_{jx}$ denotes an external flux bias. In the following sections, the basis we use for the qubit subspace is the ground and first excited state of $\hat H_j$.

\subsection{Linearization of the ground state energy}
\label{EgLinearized}
To linearize the qubit-qubit interaction potential we assume the weak coupling limit, $\alpha_j = M_j/L_j \ll 1$. This allows us to expand the coupler's ground state energy to second order in $\alpha_j$, leading to a quadratic interaction within the Born-Oppenheimer Approximation. To begin, we use equations ~\eqref{EgFinal} and \eqref{HQubit} to write the full Hamiltonian for the system,
\begin{equation}
  \label{2QubitHam}
  \sum_{j} \hat H_j + E_g(\hat \varphi_x)\,,
\end{equation}
where $\hat \varphi_x$ is defined as
\begin{equation}
  \label{varphiX2Qubit}
  \hat \varphi_x = \varphi_{cx} -\alpha_1 \hat \varphi_1 -  \alpha_2 \hat \varphi_2\,,
\end{equation}
 and
\begin{align}
  \label{U12}
    E_{g}(\varphi_x)/E_{\tilde L_c} = \beta_c \cos_{\beta_c}(\varphi_x)+  \zeta_c \sqrt{1 - \beta_c \cos(\varphi_x + \beta_c \sin_{\beta_c}(\varphi_x))} \,. 
\end{align}
Here $\varphi_{cx}$ denotes the external flux applied to the coupler's inductive loop. We have also used equation~\eqref{UZPEfirst} for the definition of the zero-point energy (it will not be necessary to compute its Fourier series) and substituted equation~\eqref{varphicstar} for $\varphi_c^{(*)}$.

We now expand the interaction potential $E_g(\varphi_x)$ to second order in the mutual inductance parameters $\alpha_j$ (i.e., about the point $\l. \varphi_x \r|_{\alpha_j = 0} = \varphi_{cx}$). Using the fact that $\pdp{\hat \varphi_x}{\alpha_j} = - \hat \varphi_j$ (cf. equation~\eqref{varphiX2Qubit}), from equation~\eqref{2QubitHam} we compute the effective Hamiltonian
\begin{align}
  \label{Heff}
  \begin{split}
    \hat H_{\textrm{eff}} & = \sum_j \hat H_j + \l( E_g'(\varphi_{cx})(\hat \varphi_x - \varphi_{cx}) + \frac{1}{2}E_g''(\varphi_{cx})(\hat \varphi_x - \varphi_{cx})^2 \r) + O(\alpha^3) \\
    &= \sum_j \l(\hat H_j - \alpha_j E_g'(\varphi_{cx})\hat \varphi_j\r) + \frac{1}{2}  E_g''(\varphi_{cx}) \sum_{j,k} \alpha_j \alpha_k \hat \varphi_j \hat \varphi_k + O(\alpha^3)\,.
  \end{split}
\end{align}
We use equations~\eqref{U12} and~\eqref{sinBetaDeriv} to compute the dependence of these terms on the coupler bias $\varphi_{cx}$,
\begin{align}
  \label{U12p}
   E_g'(\varphi_{cx})/E_{\tilde L_c} & = - \beta_c \sin_{\beta_c}(\varphi_{cx})\l(1 - \frac{\zeta_c}{2}\l(1 - \beta_c \cos(\varphi_{cx} + \beta_c \sin_{\beta_c}(\varphi_{cx})) \r)^{-3/2} \r)\\
  \begin{split}
    \label{U12pp}
    E_g''(\varphi_{cx})/E_{\tilde L_c}  = & - \frac{\beta_c \cos(\varphi_{cx} + \beta_c \sin_{\beta_c}(\varphi_{cx}))}{1 - \beta_c \cos(\varphi_{cx} + \beta_c \sin_{\beta_c}(\varphi_{cx}))}\\
    & + \zeta_c \beta_c \l(\frac{\cos(\varphi_{cx} + \beta_c \sin_{\beta_c}(\varphi_{cx})) - \beta_c - \beta_c\sin_{\beta_c}^2(\varphi_{cx})/2}{ 2\l(1 - \beta_c \cos(\varphi_{cx} + \beta_c \sin_{\beta_c}(\varphi_{cx}))\r)^{7/2}}\r)\,.
  \end{split}
\end{align}
The first order terms in equation~\eqref{Heff} (proportional to $E_g'$) correspond to local fields acting on individual qubits, while the second order terms are equivalent to an effective mutual inductance between the qubits. Note that we have neglected the constant term $E_g(\varphi_{cx})$ since it has a trivial effect on the qubit dynamics\footnote{On the other hand, it was not valid to ignore the potential minimum when we computed the ground state energy of the coupler. In that case the potential minimum $U_{\textrm{min}}(\varphi_x)$ varied with the qubit flux variables, whereas here it is completely independent of the qubits' state.}.

Let us compare the local field terms in equation~\eqref{Heff} to the quantum treatment in Ref.~\cite[Section 4]{Brink2005}. These terms ($\propto E_g'$) can be incorporated into each qubit Hamiltonian as a shift in its external flux bias,
\begin{align}
  \begin{split}
    \varphi_{jx} &\rightarrow \varphi_{j x} + \delta \varphi_{j x}\\
    \delta \varphi_{j x}  &= - \alpha_j \frac{E_g'(\varphi_{cx})}{E_{L_j}}  \\
    &= - \frac{M_j}{\tilde L_c}E_g'(\varphi_{cx})/E_{\tilde L_c}\\
    & = \frac{2 \pi}{\Phi_0} M_j I_c \,.
  \end{split}
\end{align}
In the last line we equated our result to equation~(44) of Ref.~\cite{Brink2005}, which identifies $\delta \varphi_{j x }$ with the current through the coupler's inductor. Indeed, rearranging terms and using $\beta_c = \frac{2 \pi}{\Phi_0} \tilde L_c I_c^{(c)}$ and equation~\eqref{U12p}, we get
\begin{equation}
  I_c = I_{c}^{(c)}\sin_{\beta_c}(\varphi_{cx})\l(1 + O(\zeta_c) \r)\,.
\end{equation}
As expected, the first ($\zeta_c$-independent) term is exactly the current flowing through the coupler's Josephson junction. On the other hand, the second term (proportional to $\zeta_c$) has an inherently quantum origin: the coupler's zero-point energy (equation~\eqref{EgFinal}).

The description of the coupling terms ($\propto E_g''$) in $\hat H_{\textrm{eff}}$ is analogous to that of the local fields. Writing the qubit `current operator' as $\hat I_j = \frac{\Phi_0}{2 \pi L_j} \hat \varphi_j$, the interaction in equation~\eqref{Heff} is described in terms of an effective mutual inductance~\cite{Brink2005},
\begin{equation}
  \label{MEffDef}
 E_g''(\varphi_{cx}) \alpha_1 \alpha_2 \hat \varphi_1 \hat \varphi_2  = \l(M_{1} M_2 \chi_c\r) \hat I_1 \hat I_2 \,,
\end{equation}
where the coupler's linear susceptibility is
\begin{equation}
  \chi_c = \frac{1}{\tilde L_c} E_g''(\varphi_{cx})/E_{\tilde L_c}\,.
\end{equation}
As it was for the coupler current $I_c$, the first term describing $\chi_c$ (cf. equation~\eqref{U12pp}) is in agreement with previous works~\cite{Brink2005,Tian2008} and corresponds to an essentially classical treatment. Again, the $\zeta_c$-dependent term is an added quantum contribution due to the coupler's zero-point energy. Finally, we note that equation~\eqref{Heff} also includes corrections proportional to $\chi_c \hat\varphi_{j}^2$. These are a source of `nonlinear cross talk' typical in flux qubit experiments and have the effect of shifting each qubit's linear inductance (and therefore energy gap)~\cite{Harris2010,Allman2010,Chen2014}.

To calculate the qubit dynamics within the linear theory, we project the coupler-dependent terms of $\hat H_{\textrm{eff}}$ (equation~\eqref{Heff}) into the qubit subspace. We define the basis for this subspace as the ground and first excited state of the qubit Hamiltonian, $\hat H_j$. The local and coupling terms then become
\begin{align}
  \label{gEtaLin}
  \begin{split}
    g_{\eta_1 \eta_2}^{\textrm{lin}} & =\tr{\frac{\sigma_{\eta_1}^{(1)}\otimes \sigma_{\eta_{2}}^{(2)}}{4} \l(E_g'(\varphi_{cx})\l(\alpha_1 \hat \varphi_1 + \alpha_2 \hat \varphi_2 \r) +\frac{1}{2}E_g''(\varphi_{cx})\l(\alpha_1 \hat \varphi_1 + \alpha_2 \hat \varphi_2 \r)^2 \r)}\,,
  \end{split}
\end{align}
where $E_g'$ and $E_g''$ are defined in equations~\eqref{U12p} and ~\eqref{U12pp}. For the interaction term $\sigma_x^{(1)}\otimes \sigma_x^{(2)}$, this expression simplifies to
\begin{align}
  \label{gxxlin}
  \begin{split}
    g_{x x}^{\textrm{lin}} &= E_g''(\varphi_{cx}) \alpha_1 \alpha_2 \bra{ 0 0}\hat \varphi_1 \hat \varphi_2 \ket{1 1} \\
    & =  \chi_c(\varphi_{cx}) M_1 M_2 I_{p}^{(1)} I_{p}^{(2)}\,,
  \end{split}
\end{align}
where we have used equation~\eqref{MEffDef} and defined the persistent current\footnote{In the absence of bias $\varphi_{j x}$, the $\hat H_j$ eigenstates have either even or odd parity wave-functions. This is in contrast to the `persistent current' basis commonly used in double-well flux qubits, which correspond to $\ket{\pm} = \frac{1}{\sqrt{2}}(\ket{0} \pm \ket{1})$. In that case, we would interchange $\sigma_x \leftrightarrow \sigma_z$ and redefine $I_p^{(j)}\rightarrow \frac{1}{2}\l((\hat I_j)_{00} - (\hat I_j)_{11}\r)$. },
\begin{equation}
  I_{p}^{(j)} = (\hat I_j)_{01} = \frac{\Phi_0}{2 \pi L_j} \bra{0} \hat \varphi_j \ket {1}\,.
\end{equation}
A similar calculation can be carried out for the local field terms.

We stress that equations~\eqref{U12p} and~\eqref{U12pp} are approximations. This is because, as with the nonlinear theory, the coupler's zero-point energy (the second term in equation~\eqref{U12}) is obtained by linearizing the coupler Hamiltonian about its classical minimum point. Indeed, the zero-point energy contributions ($\propto \zeta_c$) diverge even more rapidly as $\beta_c \rightarrow 1$ (for $\varphi_{cx} = 0$). As an alternative to this approximation, it is possible to compute $E_g'$ and $E_g''$ numerically using standard perturbation theory. Specifically, for any eigenstate $\ket{\psi_m}$ of $\hat H_c$ (parameterized by $\varphi_x$) with eigenvalue $E_m$, we observe that
\begin{align}
  \label{EigenDeriv}
  \begin{split}
    \partial_{\varphi_x} E_m /E_{\tilde L_c}& = \bra{\psi_m} \l( \partial_{\varphi_x} \hat H_c /E_{\tilde L_c} \r) \ket{\psi_m} \\
    &= \bra{\psi_m} \l( \varphi_x - \hat \varphi_c \r) \ket{\psi_m} \\
    \ket{\partial_{\varphi_x}\psi_m} & = - (E_m - \hat H_c)^{-1} \partial_{\varphi_x}\l((E_m - \hat H_c) \r) \ket{\psi_m}\\
    & = -\frac{E_{\tilde L_c}}{E_m - \hat H_c} \hat \varphi_c \ket{\psi_m}\,.
  \end{split}
\end{align}
(Here $(E_m - \hat H_c)^{-1}$ denotes the pseudo-inverse, which vanishes on $\ket{\psi_m}$.) Carrying out the second derivative for $m = g$ then gives
\begin{equation}
  \label{EigenDeriv2}
  \partial_{\varphi_x}^2 E_g(\varphi_x)/E_{\tilde L_c} = 1 + 2 \bra{\psi_g} \hat \varphi_c \frac{E_{\tilde L_c}}{E_g - \hat H_c} \hat \varphi_c \ket{\psi_g}\,.
\end{equation}
Thus the first and second derivatives of $E_g$ can be obtained diagonalizing $\hat H_c$ and performing the above matrix operations. While this calculation exactly accounts for the coupler's zero-point energy, it is computationally more expensive compared to the analytic theories.

\subsection{Coupling as a finite difference and errors in the linear theory}
\label{FDCoupling}
We now derive an approximate expression for the qubit-qubit coupling that is more refined than the linear approximation. What results is a nonlinear function of qubit flux variables' first and second moments. Whereas the linear theory coupling is proportional to the second derivative of the coupler energy ($E_g''$), this approximation expresses the coupling as a second order finite difference\cite{Averin2003}. It thus accounts for higher orders in the Taylor Series of $E_g$. This produces a more accurate approximation in the strong coupling limit that does not diverge as $\beta_c \rightarrow 1$. This analysis will also allow us to bound the error in the (analytic) linear theory.

We start by defining the `qubit subspace' of the qubit Hamiltonians. We set the basis as the ground and first excited state of each qubit's Hamiltonian. For simplicity, we assume identical qubits and also that the qubits' local potential energy functions are symmetric (e.g., zero external bias in equation~\eqref{HQubit}). This is reflected in the symmetry of the ground and excited state wave-functions. The wave-functions can then be written in terms of a reference wave-function,
\begin{equation}
  \label{qubitWF}
  \bra{\varphi} j \rangle = \frac{\psi_r(\varphi - \varphi_p) + (-1)^j \psi_r(-\varphi - \varphi_p)}{\sqrt{2}} \,.
\end{equation}
where $j = 0,1$ denotes the eigenstate index -- as well as the parity --  of each wave-function. The (normalized) reference wave-function $\psi_r(\varphi - \varphi_p) = \frac{1}{\sqrt{2}}(\bra{\varphi} 0 \rangle + \bra{\varphi} 1 \rangle)$ is defined with respect to an offset $\varphi_p$ so that it is approximately centered at the origin,
\begin{align}
  \begin{split}
    \int\mbox{d} \varphi \, \psi_r^2(\varphi) & = 1 \\
    \int\mbox{d} \varphi \, \psi_r^2(\varphi) &\varphi  = 0 \,.
  \end{split}
\end{align}
The flux offset $\varphi_p$ in equation~\eqref{qubitWF} is typically associated with the persistent current of the flux qubit,
\begin{equation}
  \varphi_p = \bra{0} \hat \varphi \ket{1} =   \frac{2 \pi}{\Phi_0} L_j I_p  \,.
\end{equation}
In the case of a two-well qubit potential, we can intuitively think of $\psi_r(\varphi- \varphi_p)$ as a having a single peak approximately centered at one of the local minima (near the point $\varphi = \varphi_p$). It will also prove useful to consider the second moment of $\hat \varphi$,
\begin{equation}
  2 \zeta_{\textrm{eff}} \equiv \int\mbox{d} \varphi \, \psi_r^2(\varphi) \varphi^2 = \frac{ \bra{0} (\hat \varphi - \varphi_p)^2 \ket{0} + \bra{1} (\hat \varphi - \varphi_p)^2 \ket{1} }{2}  \,.
\end{equation}
The effective impedance $\zeta_{\textrm{eff}}$ thus determines the characteristic width of $\psi_r$.\footnote{In the harmonic limit $\beta_j = 0$ (cf. Equation~\eqref{HQubit}), this definition of the effective impedance coincides with the qubit impedance, $\zeta_{j} = \zeta_{eff}$.} 

We now express the $xx$ coupling predicted by our nonlinear theory in terms of the reference wave-function. Since the eigenstate wave-functions are real valued, this coupling is equal to the matrix element $\bra{00} \hat H_{\textrm{int}} \ket{11}$. Using $\hat H_{\textrm{int}} = E_g(\varphi_{cx} - \alpha(\hat \varphi_1 + \hat \varphi_2))$, we substitute equation~\eqref{qubitWF} and integrate over the flux variables to get
\begin{align}
  \begin{split}
    g_{xx}  = & \int \mbox{d} \varphi_1 \, \mbox{d}\varphi_2 \, \braket{0}{\varphi_1}\braket{\varphi_1}{1}\braket{0}{\varphi_2}\braket{\varphi_2}{1}  E_g(\varphi_{cx} - \alpha( \varphi_1 +  \varphi_2)) \\
    = & \frac{1}{4}\int \mbox{d} \varphi_1 \, \mbox{d}\varphi_2 \, \l( \psi_r^2(\varphi_1 - \varphi_p) - \psi_r^2(\varphi_1 + \varphi_p) \r) \l( \psi_r^2(\varphi_2 - \varphi_p) - \psi_r^2(\varphi_2 + \varphi_p) \r)\\
    & \quad \times E_g(\varphi_{cx} - \alpha( \varphi_1 +  \varphi_2)) \\
    = & \frac{1}{4}\int \mbox{d} \varphi_1 \, \mbox{d}\varphi_2 \,  \psi_r^2(\varphi_1 ) \psi_r^2(\varphi_2 ) \,  E_g^{FD}(\varphi_{x})\,.
  \end{split}
\end{align}
In the last line we have shifted the flux variables $\varphi_1,\varphi_2$ by $\pm \varphi_p$ and introduced the second order finite difference of $E_g$,
\begin{equation}
  \label{EgFD}
  E_g^{FD}(\varphi_{x} ) = E_g(\varphi_{x} + 2 \alpha \varphi_p) + E_g(\varphi_{x} - 2 \alpha \varphi_p) - 2 E_g(\varphi_{x} )\,,
\end{equation}
where again we have written the total external coupler flux as
$$\varphi_x = \varphi_{cx} - \alpha(\varphi_1 + \varphi_2)\,.$$
Introducing the notation $\Mean{f(\hat \varphi_1, \hat \varphi_2)}_{r,r} = \int \mbox{d} \varphi_1 \, \mbox{d}\varphi_2 \,  \psi_r^2(\varphi_1 ) \psi_r^2(\varphi_2 ) f(\varphi_1,\varphi_2)$, we see that the coupling $g_{x x}$ can be written as the average of the finite difference of $E_g$ with respect to the reference wave-function $\psi_r$,
\begin{equation}
  \label{gxxFD}
  g_{xx} = \frac{1}{4}\Mean{E_g^{FD}(\hat \varphi_x )}_{r,r}\,.
\end{equation}
This definition for $g_{xx}$ is equivalent to the nonlinear theory result, equation~\eqref{gEta}.

We can approximate the coupling by assuming the reference wave-function $\psi_r(\varphi)$ is a Gaussian. Since its first two moments satisfy $\Mean{\hat \varphi}_r = 0$ and $\Mean{\hat \varphi^2}_r = 2 \zeta_{eff}$, we have
\begin{equation}
  \psi_{r}^{\textnormal{Gauss}}(\varphi) = (2 \pi \zeta_{eff})^{-1/4} \exp\l(-\frac{\varphi^2}{4 \zeta_{eff}}\r)\,.
\end{equation}
Substituting the explicit Fourier series~\eqref{EgFinal} into equation~\eqref{gxxFD} then gives a sum of Gaussian integrals,
\begin{align}
  \label{gxxGauss}
  \begin{split}
    g_{xx}^{\textrm{Gauss}} & = \frac{E_{\tilde L_c}}{4}\sum_{\nu} B_{\nu} e^{i \nu \varphi_{cx}  } \l( e^{i \nu  2 \alpha \varphi_p } + e^{-i \nu 2 \alpha \varphi_p} - 2 \r) \Mean{ e^{- i \nu \alpha(\hat \varphi_1 + \hat \varphi_2)}}_{r,r}\\
    & = -E_{\tilde L_c}\sum_{\nu} B_{\nu} e^{i \nu \varphi_{cx}} \sin^2(\nu \alpha \varphi_p) e^{-\alpha^2 \nu^2 \zeta_{eff}}\,.
  \end{split}
\end{align}
This approximation allows us to still incorporate higher order corrections in $\alpha_j$ while avoiding the need for computing any matrix elements beyond those in $\varphi_p$ and $\zeta_{eff}$.

We can recover the linear theory result of the previous section by making two approximations on equation~\eqref{gxxFD}. First, we notice that $E_g^{FD}(\varphi_{x} )/ (2 \alpha \varphi_p)^2$ is the finite difference approximation to the second derivative,
\begin{equation}
  \label{EgFDApprox}
  E_g^{FD}(\varphi_x) = E_g''(\varphi_{x})(2 \alpha \varphi_p)^2  + R_{1}\,,
\end{equation}
where the remainder term $R_{1}$ is bounded by\footnote{This bound can be derived by Taylor expanding $E_g(\varphi_x \pm 2 \alpha \varphi_p)$ to third order and using the Lagrange form for the (fourth order) remainder. Substituting into equation~\eqref{EgFD} causes the zeroth, first, and third order terms to cancel.}
\begin{align}
  \label{R1Bound}
  \begin{split}
    |R_{1}| & \leq  2 \frac{(2 \alpha \varphi_p)^4}{4!} \max_{ |\delta \varphi_x| \leq 2 \alpha \varphi_p} |E_g^{(4)}(\varphi_x + \delta \varphi_x)|\\
    & \leq  2 \frac{(2 \alpha \varphi_p)^4}{4!} \max_{ \varphi_x } |E_g^{(4)}(\varphi_x)|\,.
  \end{split}
\end{align}
Next, we expand $E_g''(\varphi_x)$ to first order about the point $\varphi_x = \varphi_{cx}$,
\begin{equation}
  \label{DDEgApprox}
  E_g''(\varphi_{x}) = E_g''(\varphi_{cx}) - \alpha (\varphi_1 + \varphi_2) E_g^{(3)}(\varphi_{cx}) + R_{2}\,,
\end{equation}
where the second remainder term is similarly bounded by
\begin{align}
  \label{R2Bound}
  \begin{split}
    |R_{2}| & \leq   \frac{ \alpha^2 (\varphi_1+\varphi_2)^2}{2} \max_{ |\delta \varphi_x| \leq |\alpha(\varphi_1 + \varphi_2) | } |E_g^{(4)}(\varphi_{cx} + \delta \varphi_x)|\\
    & \leq   \frac{ \alpha^2 (\varphi_1+\varphi_2)^2}{2} \max_{  \varphi_x } |E_g^{(4)}(\varphi_{x}) |\,.
  \end{split}
\end{align}
Finally, we substitute equations \eqref{EgFDApprox} and \eqref{DDEgApprox} into \eqref{gxxFD} to get\footnote{The third derivative term vanishes since the reference function is centered at zero, $\Mean{\hat \varphi_1 + \hat \varphi_2}_{r,r} = 0$.}
\begin{equation}
  g_{xx} = \frac{1}{4} \l((2 \alpha \varphi_p)^2 E_g''(\varphi_{cx}) + \Mean{(2 \alpha \varphi_p)^2 \hat R_2 + \hat R_1}_{r,r} \r)\,.
\end{equation}
The first term on the right hand side is exactly the linear theory result $g_{xx}^{\textrm{lin}}$, equation~\eqref{gxxlin}. Using equations~\eqref{R1Bound} and~\eqref{R2Bound} we can also bound the error in the linear theory,
\begin{equation}
  |g_{xx} - g_{xx}^{\textrm{lin}}| \leq \alpha^4 \varphi_p^2 \l(2 \zeta_{eff} + \frac{1}{3} \varphi_p^2\r) \max_{  \varphi_x } |E_g^{(4)}(\varphi_{x}) |\,.
\end{equation}
Further, if we only consider the classical part of $E_g(\varphi_x)$ (i.e., set $\zeta_c \rightarrow 0$), it is straightforward but tedious\footnote{Take two derivatives of~\eqref{U12pp} using~\eqref{sinBetaDeriv}.} to compute the maximum of $E_g^{(4)}(\varphi_{x})$,
\begin{equation}
   \max_{  \varphi_x }|E_g^{(4)}(\varphi_{x})| \stackrel{\zeta_c = 0}= |E_g^{(4)}(0)| = \frac{E_{\tilde L_c} \beta_c}{(1- \beta_c)^4}\,.
\end{equation}
Hence, assuming the quantum correction to $E_g$ is small, $g_{xx}^{\textrm{lin}}$ approximates $g_{xx}$ well in the limits
\begin{equation}
  \label{linApproxErrorBound}
   E_{\tilde L_c }\beta_c \l(\frac{\alpha}{1 - \beta_c}\r)^4 \varphi_p^2 \l(2 \zeta_{eff} + \frac{1}{3} \varphi_p^2\r)  \ll |g_{xx}^{\textnormal{lin}}|\,.
\end{equation}
This affirms physical intuition regarding the validity of the linear, analytic approximation: it is comparable to the nonlinear theory in the limits of weak qubit-coupler interaction ($\alpha = M_j/L_j \ll 1$), small qubit persistent current ($I_p \propto \varphi_p \ll 1$), and/or coupler nonlinearity $\beta_c$ not too close to one.

\section{ Numerical study}
\label{numStudy}
We have carried out a numerical study to evaluate the different approximations described in the text. Our first goal is to validate the Born-Oppenheimer Approximation. We numerically test the breakdown of this approximation in Section~\ref{BOBreakdown}.  The following section focuses on the different theories used to approximate the coupler ground state energy.  The main result of our work is the exact, analytic expression for the classical part of $E_g$ (i.e., the classical minimum of $H_c$) combined with the harmonic approximation to the coupler zero-point energy (equation~\eqref{linHam}). We refer to this treatment as {\it nonlinear, analytic} (NA) since it expresses $E_g$ as a Fourier Series in $\varphi_x$. As a simplification, we may Taylor expand our approximate expression to second order about the point $\varphi_{x} = \varphi_{cx}$ (i.e., $\alpha_j =0$) to get an {\it linear, analytic} (LA) form for $E_g$. Alternatively, instead of using the analytic expression for the first and second derivatives of $E_g$, we may numerically compute them about $\varphi_{x} = \varphi_{cx}$ using perturbation theory (see equation~\eqref{EigenDeriv}). We call this approximation to $E_g$ the {\it linear, numerical} (LN) theory. Our numerics will focus on distinguishing these theories. Specifically, we investigate the parameter regimes where each theory is valid and compare their effective qubit dynamics. Finally, we calculate the size of some non-stoquastic and $3$-local interactions predicted by the nonlinear theory.



\subsection{ Breakdown of Born-Oppenheimer Approximation}
\label{BOBreakdown}

We first numerically probe the limits of the Born-Oppenheimer Approximation\footnote{Most of the circuit parameters affect this approximation, so we can only note some qualitative trends. Detailed, quantitative discussions of corrections to Born-Oppenheimer are in Appendix Sections~\ref{BOValidity} and~\ref{BONonadiabatic}.}. To do so we have calculated the exact, low energy spectrum of two flux qubits interacting with a coupler circuit (treated as an independent degree of freedom). This is done by representing the full Hamiltonian in the harmonic oscillator eigenstate basis (see Appendix Section~\ref{numericsMethods} for details). We then compare the spectrum to the one predicted under the Born-Oppenheimer Approximation. That is, we consider the Hamiltonian $\hat H_{\textrm{BO}} = \hat H_1 + \hat H_2 + \hat H_{\textrm{int}}$, where $\hat H_j$ is the local Hamiltonian for qubit $j$ and $\hat H_{\textrm{int}} = E_g(\hat \varphi_x)$ is the qubit-dependent ground state energy of the coupler. As a reference, we consider a parameter regime where all of our approximations work well: $\zeta_j = \zeta_c = 0.05, \alpha_j = 0.05, \beta_c =0.75, E_{\tilde L_c}/E_{L_j} = 3,$ and $\beta_j \geq 0.5$. This can be seen in Fig.~\ref{BOBreakdownVaryBetaJReference}, which shows the different spectrum calculations at the maximum coupling bias point, $\varphi_{cx} = \varphi_{jx} = 0$. Tuning the coupler parameters far beyond this regime causes the Born-Oppenheimer Approximation to fail.

We modify the coupler circuit parameters away from the reference point to observe their effect on the Born-Oppenheimer Approximation. Generally, we find that Born-Oppenheimer is valid when the coupler Hamiltonian's ground state energy gap is much larger than the qubit energy gaps.  Since the coupler energy gap scales approximately linearly with $\zeta_c$ (for fixed $E_{\tilde L_c}$), we can test this intuition by decreasing the coupler impedance\footnote{ At the reference parameters and $\varphi_{x} =0$, the ground state energy gap of $\hat H_c$ is $\sim 5.32 \times 10^{-2} E_{\tilde L_c} = 1.60 \times 10^{-1} E_{L_j}$. Decreasing $\zeta_c$ to $0.02$ decreases the gap to $\sim 2.06 \times 10^{-2} E_{\tilde L_c} = 6.18 \times 10^{-2} E_{L_j}$, which is comparable to the observed qubit spectra.}. Comparing Fig.~\ref{BOBreakdownVaryBetaJSmallZetaC} to the reference regime (Fig.~\ref{BOBreakdownVaryBetaJReference}), we see that decreasing $\zeta_c$ from $0.05$ to $0.02$ causes all of the Born-Oppenheimer theories to break down. The theory also breaks down when the coupling strength $\alpha_j = M_j/L_j$ is too large, because a sufficiently strong qubit-coupler interaction allows the coupler to populate excited states beyond its ground state (cf. Section~\ref{BONonadiabatic}). This is seen in Fig.~\ref{BOBreakdownVaryBetaJBigAlpha}, where we increase the value of $\alpha_j$ from $0.05$ to $0.1$\footnote{An alternative reason for the mismatch in  Fig.~\ref{BOBreakdownVaryBetaJBigAlpha} is that our approximation to $E_g$ is inaccurate for large $\alpha_j$. But if that were the case, the nonlinear, analytic (NA) theory should still work since it describes $E_g$ to all orders in $\alpha_j$.}. We also consider the effect of coupler nonlinearity, $\beta_c$. In the limit of zero flux bias ($\varphi_{cx} = 0$ mod $2\pi$) corresponding to maximum coupling, the coupler gap closes exponentially quickly with increasing $\beta_c$, and therefore the Born-Oppenheimer Approximation breaks down\footnote{How quickly the gap closes depends on the coupler impedance. A larger impedance means exponential decay in the gap starts at larger values of $\beta_c$.}. In Fig.~\ref{BOBreakdownVaryBetaJBigBetaC} we see that increasing $\beta_c$ from $0.75$ to $0.95$ causes all of our theories to incorrectly predict the spectrum. However, in this case the mismatch in the spectrum could also be due to errors in the approximate representation of $E_g$, discussed below. Despite the observed spectrum mismatch, Born-Oppenheimer can still hold at large nonlinearity if the bias $\varphi_{cx}$ is finite: as seen in Fig.~\ref{BOBreakdownVaryPhiCX}, for $\varphi_{cx} \geq 0.02 \times 2 \pi$ there is good agreement between the exact spectrum and the one predicted by the NA theory. For sufficiently large $\varphi_{cx}$, the spectra of all theories for $E_g$ agree with the exact spectrum (cf. Fig.~\ref{BOBreakdownVaryPhiCXLargerRange}). Finally, the inductive energy $E_{\tilde L_c}$ sets the overall energy scale of the coupler, so it scales linearly with the coupler gap and increasing this parameter should improve the Born-Oppenheimer Approximation. Although $E_{\tilde L_c}$ also sets the energy scale of the coupling, we mention that for $k$ coupled qubits the coupling strength $\alpha_j \propto M_j$ is bounded by $\frac{1}{k} \sqrt{E_{L_j}/E_{\tilde L_c} }$, and for typical circuit implementations it should scale as $\propto E_{\tilde L_c}^{-1}$.  A qualitative summary of the observed trends can be found in Fig.~\ref{appTrends}. 
 
 \begin{figure}[t!]
   \begin{tabular*}{0.535 \textwidth}{|c | c | c | c | c| c|}
     \hline
     Increase:  & $E_{\tilde L_c}/E_{L_j}$ & $\alpha_j$ & $\zeta_c$ & $\beta_c$ & $|\varphi_{cx}|$\\
     \hline
     Born-Oppenheimer & better$^*$ & worse$^*$ & better & worse  & better \\
     \hline
     linear analytic (LA) $E_g$ & N/A & worse & worse & worse & better \\
     \hline
     linear numerical (LN) $E_g$ & N/A & worse & N/A & worse & better \\
     \hline
     nonlinear analytic (NA) $E_g$ & N/A & N/A & worse & worse & better\\
     \hline
   \end{tabular*}
\caption{The response of various approximations to increases in specific circuit parameters. $^*$: For $k$ identical qubits, the mutual inductance is physically bounded as $M_j \lesssim \frac{1}{k} \sqrt{L_j L_c}$, so $\alpha_j = M_j / L_j \leq \frac{1}{k} \l(E_{\tilde L_c} / E_{L_j} \r)^{-1/2}$. Physically, increasing $\l(E_{\tilde L_c} / E_{L_j} \r)$ (by decreasing the coupler length scale) should correspond to a proportional decrease in $M_j$. }
\label{appTrends}
\end{figure}
 
\subsection{ Comparison of linear and nonlinear theories}

We now consider the parameter regimes that distinguish the different theories modeling $E_g$. These regimes can be explained by the limitations of each theory's approximation. For example, while it is numerically exact, the LN theory correctly describes the effective potential to only second order in $\alpha_j$. Hence we expect it to be inaccurate where the order $O(\alpha^3)$ terms of $E_g(\varphi_x)$ are relevant. On the other hand, the NA theory incorporates the effect of $\alpha$ to all orders, but uses the harmonic approximation to describe the zero-point energy component of $E_g$. In the limit $\beta_c \rightarrow 1$ this approximation breaks down\footnote{Indeed, the harmonic approximation to the coupler zero-point energy is $E_{\tilde L_c} \zeta_c \sqrt{ 1 - \beta_c \cos(\varphi_{c}^{(*)})}$, where $\varphi_{c}^{(*)} = \varphi_x + \beta_c \sin_{\beta_c}(\varphi_x)$ is the classical minimum point determined by the total external bias $\varphi_{x}$. The limit $\varphi_x \rightarrow 0$, $\beta_c \rightarrow 1$ causes the harmonic zero-point energy to vanish.}, although the zero-point energy is a relatively small contribution to $E_g$ (for small impedance $\zeta_c$). The LA theory suffers from both limitations and should only be accurate in the limit where both previous theories agree; thus we will not focus on this theory in our comparisons. Qualitatively, the breakdown of each approximation occurs in the limit of large nonlinearity $\beta_c$, coupling $\alpha_j$, and near the maximal coupling bias $\varphi_{cx} = 0$. When all of these conditions hold, both the LN and NA theories are insufficient to describe the interaction. We shall also find intermediate regimes where one of these theories is more accurate than the other. One regime where the NA theory holds while the linear theories do not ($\beta_c = 0.95$, non-zero $\varphi_{cx}$) corresponds to non-negligible non-stoquastic and $k$-local interactions (discussed in the next section).

\begin{figure}[h!]
  \includegraphics[width = \textwidth,trim={2cm 0 1.7cm 0cm},clip]{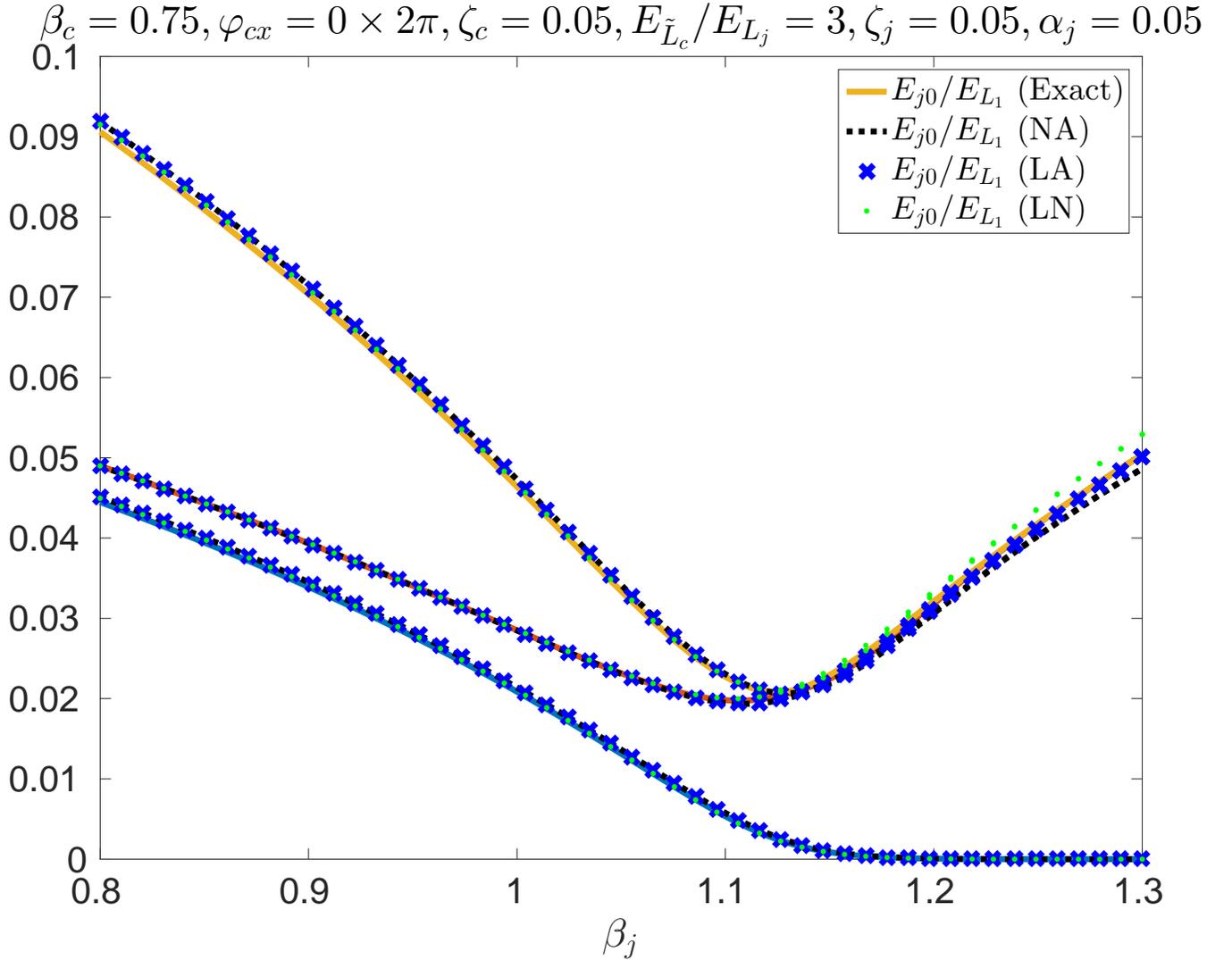}
  \caption{All Born-Oppenheimer theories accurately predict the low energy spectra in the `reference' regime. We consider a single coupler circuit interacting with two identical flux qubits for varying qubit nonlinearity $\beta_j$. (All circuits are at zero bias, $\varphi_{cx} = \varphi_{jx} = 0$.) Solid curves represent exact numerical diagonalization of the full Hamiltonian (equation~\eqref{HExact}). The black dashed, dark blue crossed, and light green dotted curves correspond to the nonlinear analytic (NA), linear analytic (LA), and linear numerical (LN) theories of the Born-Oppenheimer Approximation, respectively. (See Appendix Section~\ref{numericsMethods} for a detailed description of each calculation.) }
  \label{BOBreakdownVaryBetaJReference}
\end{figure}

\begin{figure}[h!]
  \includegraphics[width = \textwidth,trim={3cm 0 4cm 0},clip]{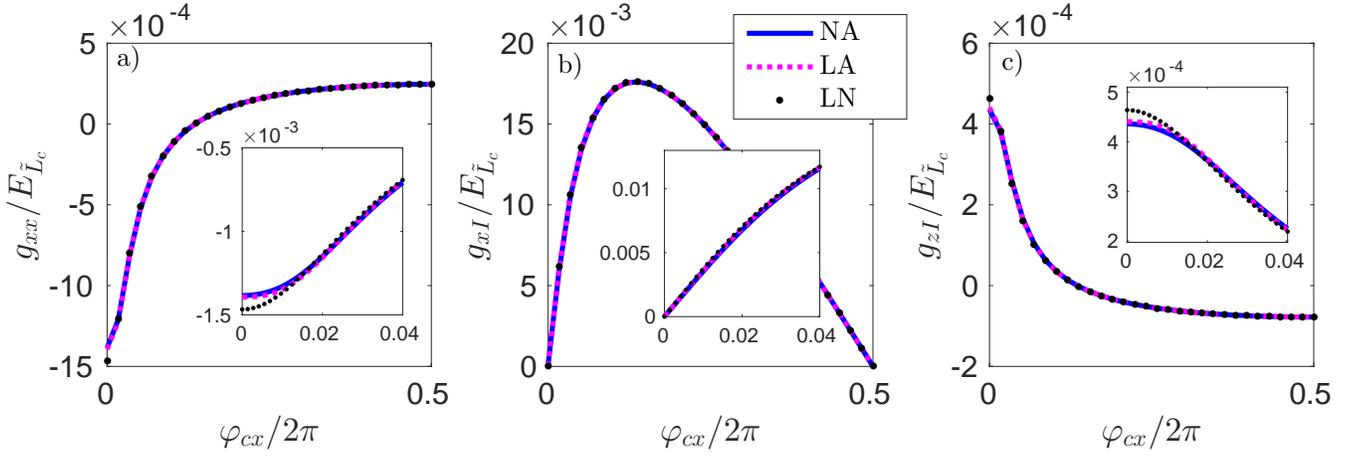}
  \caption{Excluding a small discrepancy near the maximal coupling bias $\varphi_{cx}=0$, all Born-Oppenheimer theories predict the same qubit dynamics in the reference regime. Shown are coupler-induced qubit coefficients for $\hat H_{\textrm{int}} = E_g(\hat \varphi_x)$ at the reference parameters (Fig.~\ref{BOBreakdownVaryBetaJReference}, with $\beta_j = 1.05$). The solid dark blue, dashed magenta, and dotted black curves correspond to the predictions of the nonlinear analytic (NA), linear analytic (LA), and linear numerical (LN) theories, respectively. Plots a), b), and c) correspond to the $xx$, $xI$, and $zI$ terms, respectively. All calculations were carried out in the `parity' basis (see Appendix Section~\ref{numericsMethods} for more details).}
  \label{qubitTermsAtReference}
\end{figure}

\begin{figure}[h!]
  \includegraphics[width = \textwidth,trim={3.8cm 0 4.4cm 0},clip]{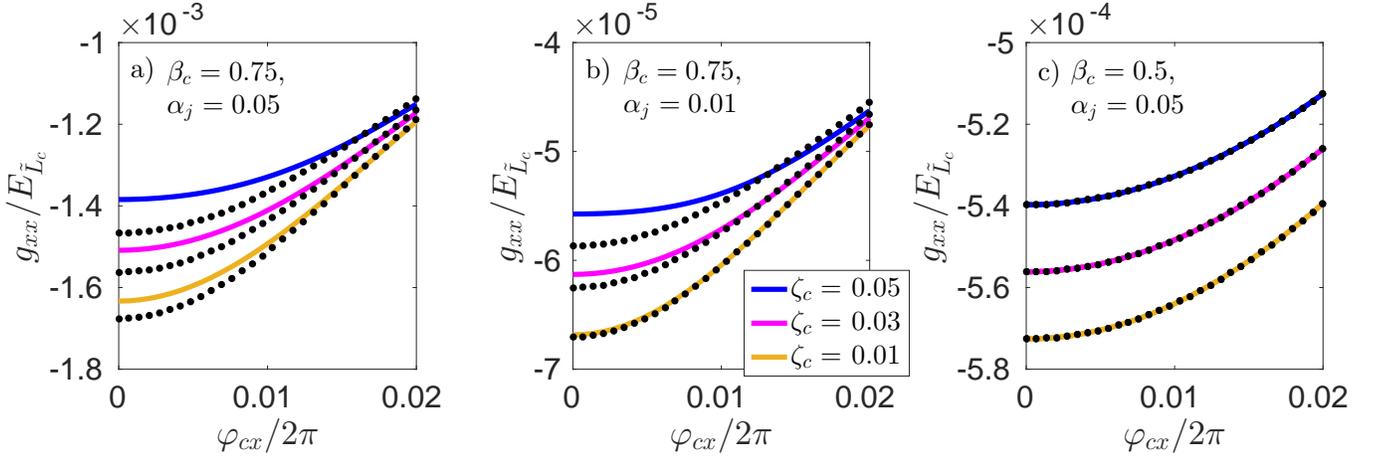}
  \caption{Discrepancy between the different Born-Oppenheimer theories near the maximal coupling bias, $\varphi_{cx} = 0$. Solid curves:  $xx$ coupling predicted by the nonlinear analytic (NA) theory, for coupler impedance $\zeta_c = 0.05$ (dark blue), $\zeta_c = 0.03$ (magenta), and $\zeta_c = 0.01$ (light orange). Overlayed dotted curves correspond to the $xx$ coupling predicted by the linear numerical (LN) theory at the same coupler parameters. The top curves in plot a) correspond to the `reference' coupler parameters described in the text ($\beta_c = 0.75,\alpha_j = 0.05, \zeta_c = 0.05$). The curves in plots b) and c) correspond to the weak coupling ($\alpha_j \rightarrow 0.01$) and low nonlinearity $\beta_c \rightarrow 0.5$ limits. In all calculations the qubit parameters were fixed at $\beta_j  =1.05, \zeta_j = 0.05, \varphi_{jx} =0$. Since the `parity' basis was used to define the Hamiltonian coefficients, the $g_{xx}$ interaction is strictly stoquastic (i.e., it is a $zz$ coupling in the computational, `persistent current' basis). All calculations were carried out as done for Fig.~\ref{qubitTermsAtReference} (see Appendix Section~\ref{numericsMethods} for more details).}
  \label{compareQubitDynamics}
\end{figure}

The qubit dynamics predicted by both LN and NA theories can be inaccurate when the coupler is tuned to maximum coupling, $\varphi_{cx} = 0$. This is true, to a small extent, even in the reference regime ($\beta_c = 0.75, \alpha_j = 0.05$, and $\zeta_c = 0.05$, Fig.~\ref{BOBreakdownVaryBetaJReference}) where all theories predict the spectrum accurately. For these coupler parameters, the qubit dynamics (i.e., the qubit Hamiltonian coefficients $g_{\bar \eta}$) predicted by each theory are close to equal at almost every coupler bias $\varphi_{cx}$ (cf. Fig.~\ref{qubitTermsAtReference}). However, there is a slight discrepancy near the maximal coupling limit $|\varphi_{cx}| \leq 0.01 \times 2 \pi$ (cf. inset of Fig.~\ref{qubitTermsAtReference}), which suggests that at least one theory is inadequate. To investigate this discrepancy, we compute the $xx$ couplings for the NA and LN theories at varying coupler impedances near $\varphi_{cx} = 0$. We first consider the classical limit of small coupler impedance, $\zeta_c \rightarrow 0$.  The zero-point energy component of $E_g$ vanishes in this limit, so that the NA prediction becomes exact. As seen in Fig.~\ref{compareQubitDynamics}(a), the NA and LN predictions still disagree in this limit. Thus the LN theory is slightly inaccurate in predicting effect on the qubit dynamics of the classical component of $E_g$. Since this contribution to $E_g$ does not change when increasing $\zeta_c$, the small error in the LN predictions persists even for $\zeta_c = 0.05$ \footnote{Note that the Born-Oppenheimer Approximation is only valid for non-zero $\zeta_c$. The predicted coupling $g_{xx}$ in the $\zeta_c \rightarrow 0$ limit therefore only illustrates the classical contribution to this coupling.}. On the other hand, we can also consider the weak coupling limit, $\alpha_j \ll 1$, where the LN theory is exact (up to order $O(\alpha^3)$). In this limit, the two theories still only agree when we also take the classical limit of small coupler impedance, $\zeta_c = 0.01$ (cf. Fig.~\ref{compareQubitDynamics}(b)).  This indicates that the NA theory also has a small but non-negligible error due to its approximation of the coupler zero-point energy (which is approximately proportional to $\zeta_c$). Thus, near the maximum coupling bias $\varphi_{cx} = 0$, both theories may be slightly inaccurate in predicting the qubit dynamics. Yet decreasing the coupler nonlinearity from $\beta_c = 0.75$ to $\beta_c = 0.5$ causes the predictions of both theories to agree, even at maximum coupling bias $\varphi_{cx} = 0$ (Fig.~\ref{compareQubitDynamics}(c)). This is not surprising, as the harmonic approximation to the zero-point energy improves as the coupler nonlinearity decreases, thereby improving the accuracy of the analytic theories\footnote{To see why this is the case, we consider the coupler Hamiltonian linearized about its classical minimum point, equation~\eqref{linHam}. At bias $\varphi_{x} = 0$, the next leading order correction is quartic, with effective potential $ \frac{(1 - \beta_c)}{2} (\hat \varphi_c - \varphi_c^*)^2 + \frac{\beta_c}{24}(\hat \varphi_c - \varphi_c^*)^4 +O(\alpha^6)$. The higher order corrections are therefore small for $\beta_c = 0.5$.} (cf. Fig.~\ref{ZPEComparison}). Similarly, the derivatives of the LA theory (equations~\eqref{U12p}, \eqref{U12pp}) suggest that the higher order corrections in $\alpha$ become less important for smaller $\beta_c$. While both theories agree in this limit, we also see in Fig.~\ref{compareQubitDynamics}(c) that the coupler zero-point energy still has a significant effect on the observed coupling. It is therefore important to account for non-zero coupler impedance, especially for high precision modeling and calibration of inductively coupled circuits.

\begin{figure}[h!]
  \includegraphics[width = \textwidth,trim={2cm 0 1.7cm 0cm},clip]{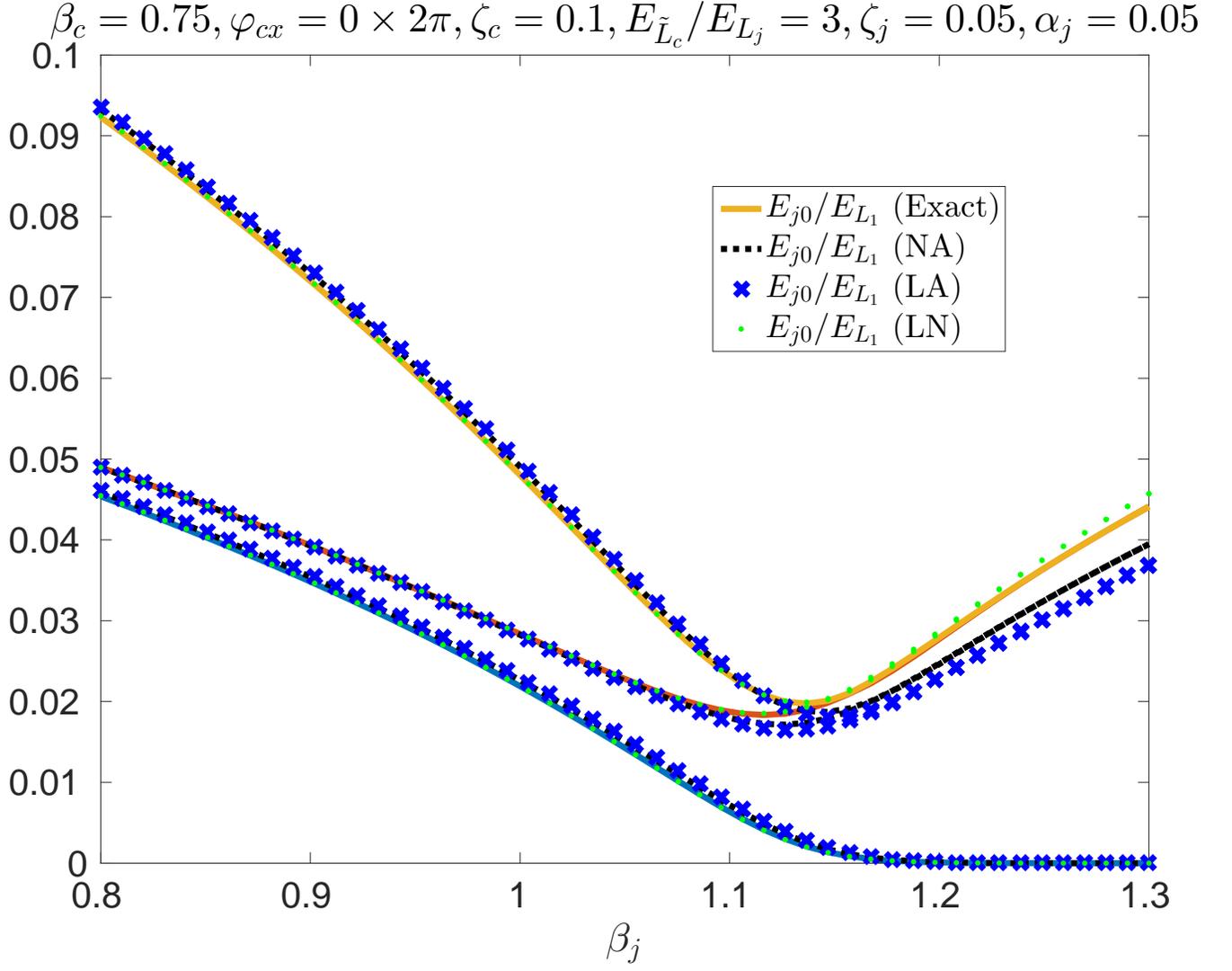}  
\caption{Increasing coupler impedance decreases the accuracy of the analytic (NA and LA) theories, while leaving the numerical theory unchanged. We consider the low energy spectrum of two coupled flux qubits, but double the coupler impedance relative to the reference regime (Fig.~\ref{BOBreakdownVaryBetaJReference}). Solid curves represent exact numerical diagonalization of the full Hamiltonian (equation~\eqref{HExact}). The black dashed, dark blue crossed, and light green dotted curves correspond to the nonlinear analytic (NA), linear analytic (LA), and linear numerical (LN) theories of the Born-Oppenheimer Approximation, respectively. (See Appendix Section~\ref{numericsMethods} for a detailed description of each calculation.)}
\label{BOBreakdownVaryBetaJBigZetaC}
\end{figure}

\begin{figure}[h!]
  \includegraphics[width = \textwidth,trim={3cm 0 4cm 0},clip]{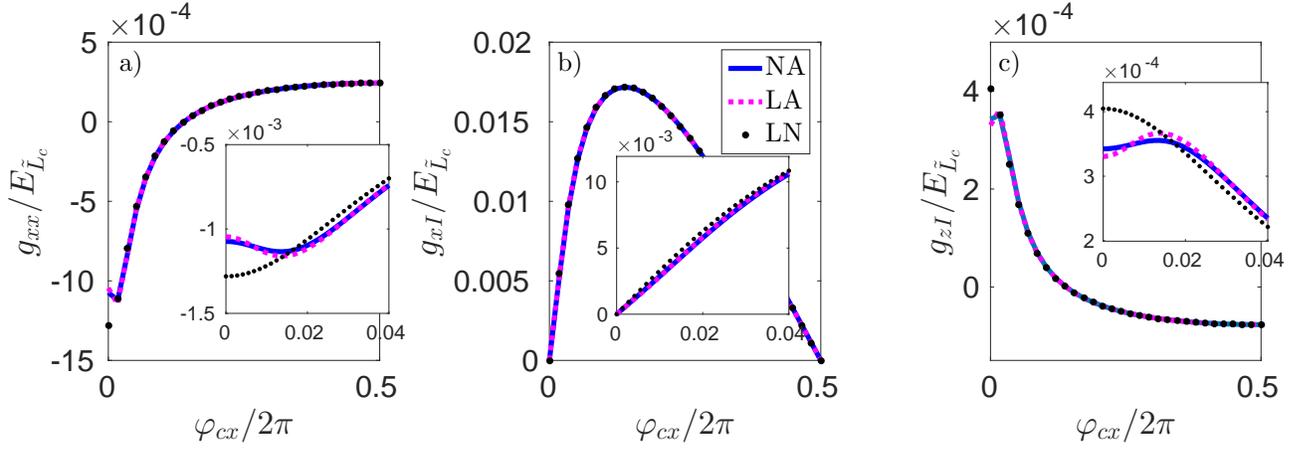}
  \caption{Increasing coupler impedance $\zeta_c$ increases discrepancy between the analytic and numerical theories (relative to the reference regime,~\Cref{qubitTermsAtReference}). For $xx$ and $z I $ terms (plots a,c), a discrepancy between analytic (NA and LA, solid dark blue and dashed magenta) and numerical (NL, dotted black) theories exists near maximum coupling, $\varphi_{cx} = 0$. The theories match closely for the local $xI$ term (plot b). Calculations were carried out for qubit parameters $ \zeta_j = \alpha_j = 0.05, \beta_j = 1.05, \varphi_{jx} = 0$ and  coupler parameters $\beta_c = 0.75, \zeta_c =0.1$ (twice the impedance of the reference regime).  All calculations were carried out in the `parity' basis (see Appendix Section~\ref{numericsMethods} for more details).}
  \label{qubitTermsLargeZetaC}
\end{figure}

The regime of high coupler impedance draws a sharper contrast between the NA and LN theories. In Fig.~\ref{BOBreakdownVaryBetaJBigZetaC} we compute the energy spectrum of the coupled qubits but increase the impedance $\zeta_c$ from $0.05$ to $0.1$. This is expected to improve the accuracy of the Born-Oppenheimer Approximation since the coupler gap is approximately doubled. At the same time, it should worsen the NA (and LA) theory because the harmonic approximation to the zero-point energy (the quantum contribution to $E_g$) becomes more significant (cf. the inset of Fig.~\ref{EgPlots}). Since the LN theory represents the zero-point energy numerically exactly (at least to second order in $\alpha$), it is insensitive to this change. We note that this discrepancy only exists near $\varphi_{cx} = 0$, since away from this point the NA theory's harmonic approximation improves (cf. Fig.~\ref{ZPEComparison}).  Indeed, for $\varphi_{cx} \gtrsim 0.05 \times 2 \pi$ we find that the predicted qubit dynamics (coefficients $g_{\bar \eta}$) of each theory all agree, as seen in Fig.~\ref{qubitTermsLargeZetaC}.

\begin{figure}[h!]
   \includegraphics[width = \textwidth,trim={2cm 0 1.4cm 0cm},clip]{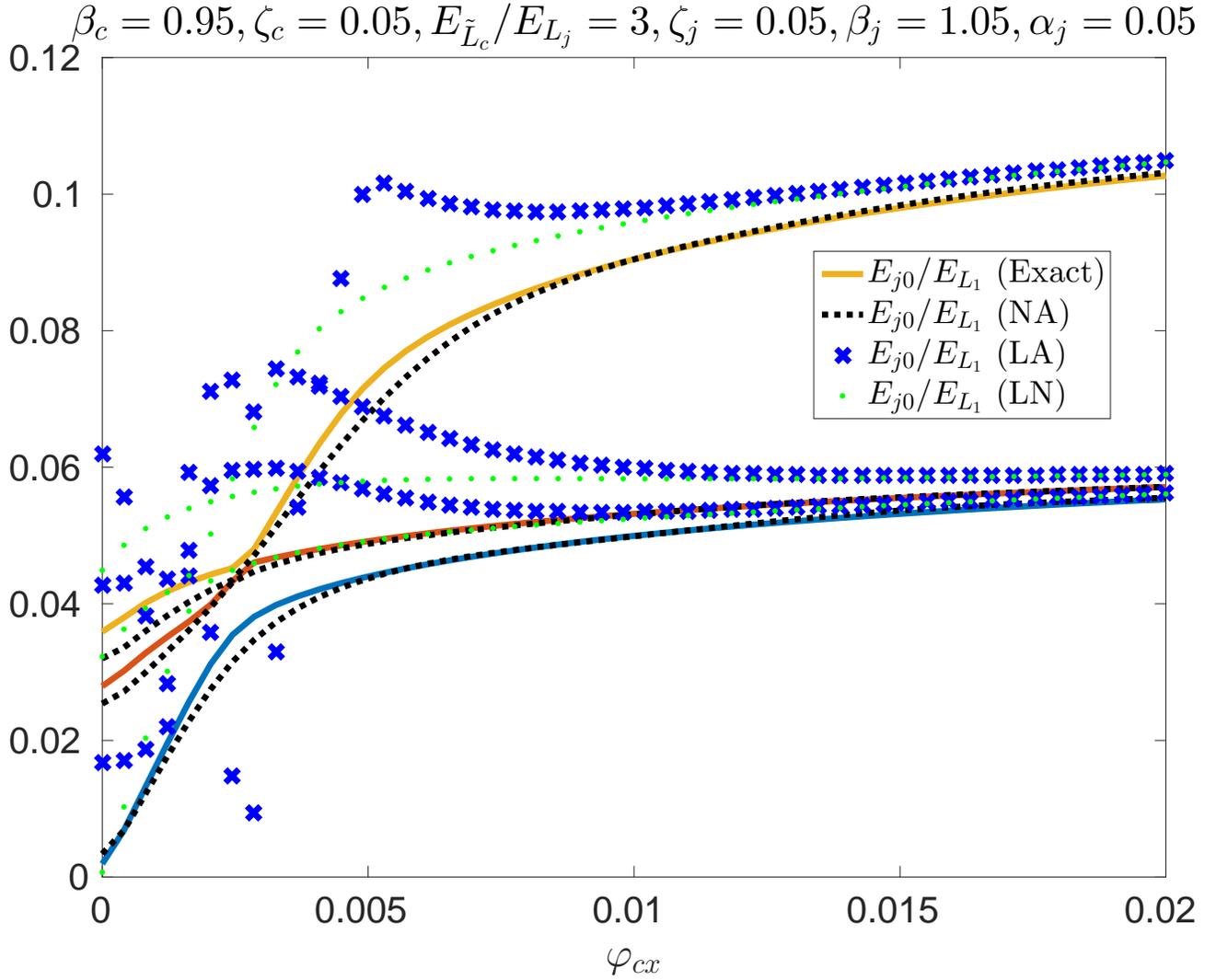} 
   \caption{Born-Oppenheimer theories fail to predict the low energy spectrum for high coupler nonlinearity (near $\varphi_{cx} = 0$). We consider a single coupler circuit interacting with two identical flux qubits for varying coupler bias, $\varphi_{cx} \ll 1$. Circuit parameters are identical to the reference regime (Fig.~\ref{BOBreakdownVaryBetaJReference}), except qubit nonlinearity is fixed at $\beta_c = 1.05$ and coupler nonlinearity $\beta_c$ is increased from $0.75$ to $0.95$. Solid curves represent exact numerical diagonalization of the full Hamiltonian (equation~\eqref{HExact}). The black dashed, dark blue crossed, and light green dotted curves correspond to the nonlinear analytic (NA), linear analytic (LA), and linear numerical (LN) theories of the Born-Oppenheimer Approximation, respectively. The NA theory agrees well with exact diagonalization for $\varphi_{cx}\gtrsim 0.01 \times 2 \pi$. The large oscillations observed in the LA spectrum are due to the divergences in the analytic expressions for the first and second derivatives of $E_g$ as $\beta_c \rightarrow 1$ (equations~\eqref{U12p} and~\eqref{U12pp}). Fig.~\ref{BOBreakdownVaryPhiCXLargerRange} shows the same calculation for a larger range of bias values, $\varphi_{cx} \in [0,0.2] \times 2 \pi$.  (See Appendix Section~\ref{numericsMethods} for a detailed description of each calculation.)}
   \label{BOBreakdownVaryPhiCX}
 \end{figure}

\begin{figure}[h!]
  \includegraphics[width = \textwidth,trim={4cm 0 4cm 0},clip]{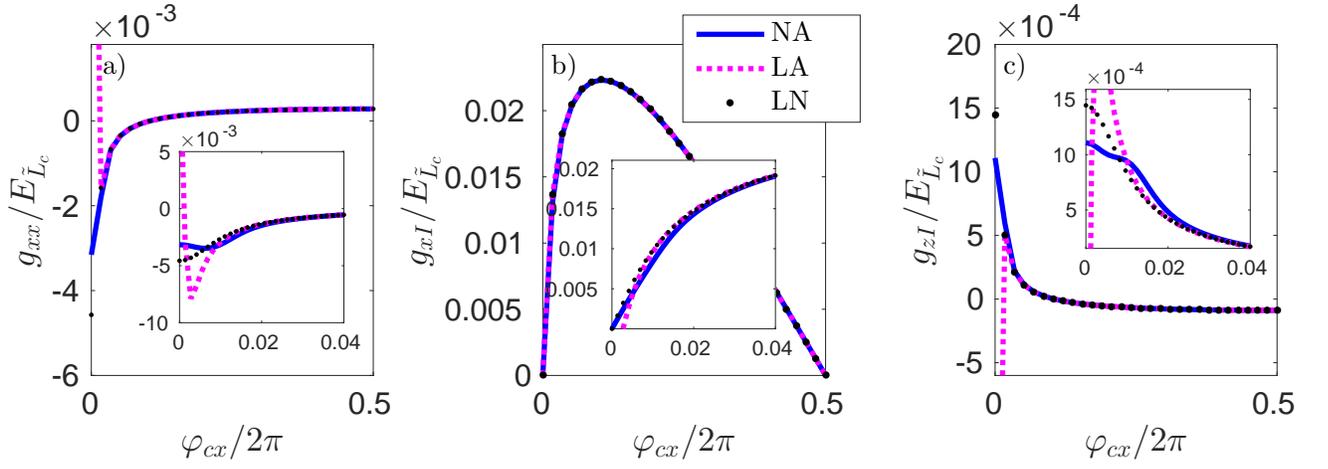}
  \caption{Increasing coupler nonlinearity $\beta_c$ increases discrepancy between the analytic and numerical theories (relative to the reference regime,~\Cref{qubitTermsAtReference}). Plots a), b), and c) correspond to the $xx$, $xI$, and $zI$ terms, respectively, with coupler nonlinearity increased from $\beta_c = 0.75$ to $\beta_c = 0.95$ relative to the reference regime. The solid dark blue, dashed magenta, and dotted black curves correspond to the predictions of the nonlinear analytic (NA), linear analytic (LA), and linear numerical (LN) theories, respectively. For $\varphi_{cx} \lesssim 0.01 \times 2 \pi$ none of the theories are expected to be accurate (Fig.~\ref{BOBreakdownVaryPhiCX}). The LA and LN theories agree for $\varphi_{cx} \gtrsim 0.01 \times 2 \pi$, indicating that the harmonic approximation to the zero-point energy converges (Fig.~\ref{ZPEComparison}). Thus the NL theory (making only the harmonic approximation) is expected to be accurate for $\varphi_{cx} \gtrsim 0.01 \times 2 \pi$. The discrepancy between the NA and LN theories for $\varphi_{cx} \approx 0.01 \times 2 \pi$ indicates that higher order terms neglected by the LN theory are significant. The divergence of the LA calculation is due to the divergences in the analytic expressions for the first and second derivatives of $E_g$ as $\beta_c \rightarrow 1$ (equations~\eqref{U12p} and~\eqref{U12pp}). All calculations were carried out in the `parity' basis. To account for higher coupler nonlinearity, the sums used in the NA calculated (Eqn. ~\eqref{gEta}) were truncated at $|\nu| \leq 200$ (see Appendix Section~\ref{numericsMethods} for more details).}
  \label{qubitTermsLargeBetaC}
\end{figure}

The regime of large coupler nonlinearity allows us to draw another contrast between the two theories. As noted previously, at the maximum coupling point $\varphi_{cx} = 0$ neither theory represents the spectrum accurately (cf. Fig.~\ref{BOBreakdownVaryBetaJBigBetaC}) when we increase $\beta_c$ from $0.75$ to $0.95$. Yet when we bias the coupler away from this point, we find that spectrum predicted by the nonlinear (NA) theory agrees with exact diagonalization past the bias point $\varphi_{cx} \gtrsim 0.01 \times 2 \pi$ (cf. Fig.~\ref{BOBreakdownVaryPhiCX}). This is explained by noting that $\varphi_{cx} = 0.01 \times 2 \pi$ is approximately point where the harmonic approximation to the coupler zero-point energy becomes accurate (up to an additive constant, as seen in Fig.~\ref{ZPEComparison}). Indeed, this also explains why, for  $\varphi_{cx} \gtrsim 0.01 \times 2 \pi$, both analytic and numerical {\it linear} theories (LA and LN) predict approximately the same spectrum in cf. Fig.~\ref{BOBreakdownVaryPhiCX}. Importantly, there is an intermediate regime ($0.01 \times 2 \pi \lesssim \varphi_{cx} \lesssim0.02 \times 2 \pi$) where the NA theory correctly predicts the spectrum while both LN and LA theories do not\footnote{For sufficiently large biases all theories correctly predict the circuit spectrum and qubit dynamics. This can be seen in Figures~\ref{BOBreakdownVaryPhiCXLargerRange} and~\ref{qubitTermsLargeBetaC}).}. This stresses the importance of including higher order terms when describing the coupler-mediated interaction, as there is also a discrepancy in the predicted qubit dynamics (cf. Fig.~\ref{qubitTermsLargeBetaC}) in this regime. Interestingly, this regime is also where we observe non-negligible non-stoquastic interactions between the qubits. We also note that, although we do not expect them to accurately predict the observed coupling $g_{xx}$ at $\varphi_{cx} \approx 0$, both NA and LN ~Fig.~\ref{qubitTermsLargeBetaC} {\it do not diverge} in the high nonlinearity limit. This is in contrast to the linear, analytic (LA) theory, which predicts an arbitrarily large value as $\beta_c \rightarrow 1$, even coming from the classical contribution to $E_g$ (equations~\eqref{U12pp} and~\eqref{gxxlin}).

\begin{figure}[h!]
  \includegraphics[width = \textwidth,trim={3cm 0 4cm 0},clip]{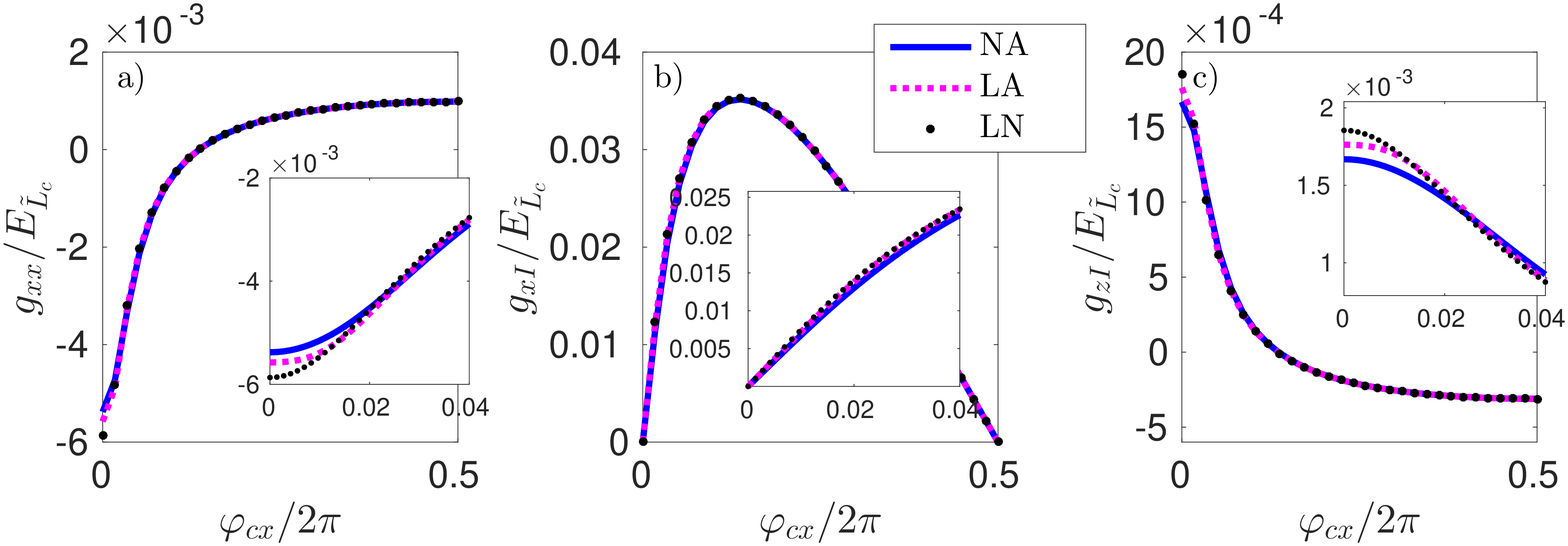}
  \caption{Coupler-induced qubit coefficients for $\hat H_{\textrm{int}} = E_g(\hat \varphi_x)$ at strong coupling $\alpha_j$. Shown are coupler-induced qubit coefficients for $\hat H_{\textrm{int}} = E_g(\hat \varphi_x)$ in the strong coupling limit (Fig.~\ref{BOBreakdownVaryBetaJReference}, with $\beta_j = 1.05$ and $\alpha_j$ increased from $0.05$ to $0.1$). The solid dark blue, dashed magenta, and dotted black curves correspond to the predictions of the nonlinear analytic (NA), linear analytic (LA), and linear numerical (LN) theories, respectively. Plots a), b), and c) correspond to the $xx$, $xI$, and $zI$ terms, respectively.  All calculations were carried out in the `parity' basis (see Appendix Section~\ref{numericsMethods} for more details).}
  \label{qubitTermsAtLargeAlphaJ}
\end{figure}

The strong coupling ($\alpha_j$) limit shows the same contrast between the NA and LN theories as the large nonlinearity limit. Again, while we find that at maximum coupling bias ($\varphi_{cx} = 0$) and $\alpha_j = 0.1$ neither theory is adequate (Fig.~\ref{BOBreakdownVaryBetaJBigAlpha}), the NA theory accurately predicts the low energy spectrum even for small, non-zero bias $\varphi_{cx}$ (Fig.~\ref{BOBreakdownVaryPhiCXBigAlpha}). There is also a similar contrast in the predicted qubit dynamics, as seen in Fig.~\ref{qubitTermsAtLargeAlphaJ}.

\subsection{$3$-body and non-stoquastic interactions}
\label{kLocalNonStoquastic}
We have also calculated the strength of some $3$-local and non-stoquastic interactions predicted by our nonlinear theory. Such interactions are absent in linear theories: The quadratic representation of $E_g$ precludes any $k$-local qubit couplings with $k>2$. Similarly, in the `parity' qubit basis an interaction of the form $\hat \varphi_1 \otimes \hat \varphi_2$ can only produce $xx$ couplings due to symmetry considerations\footnote{Equivalently, in the standard (persistent current) basis, we would only observe $zz$-type couplings.}. In order to ensure the validity of our results, we assume coupler and qubit parameter regimes for which the nonlinear, analytic Hamiltonian~\eqref{Hint} correctly reproduces the 2-qubit spectrum. We note that there are other proposals in the literature for exotic couplings involving superconducting qubits\cite{Sameti2017,Chancellor2017,Vinci2017}. Although the physical mechanisms driving these exotic couplings differ from those observed in our work, a key similarity is the need for non-linearity in the coupler device. Indeed, the interactions predicted by our analytic theory vanish in the limit of zero coupler nonlinearity, $\beta_c \rightarrow 0$.

\begin{figure}[h!]
\includegraphics[width = \textwidth,trim={0 0 .5cm 0},clip]{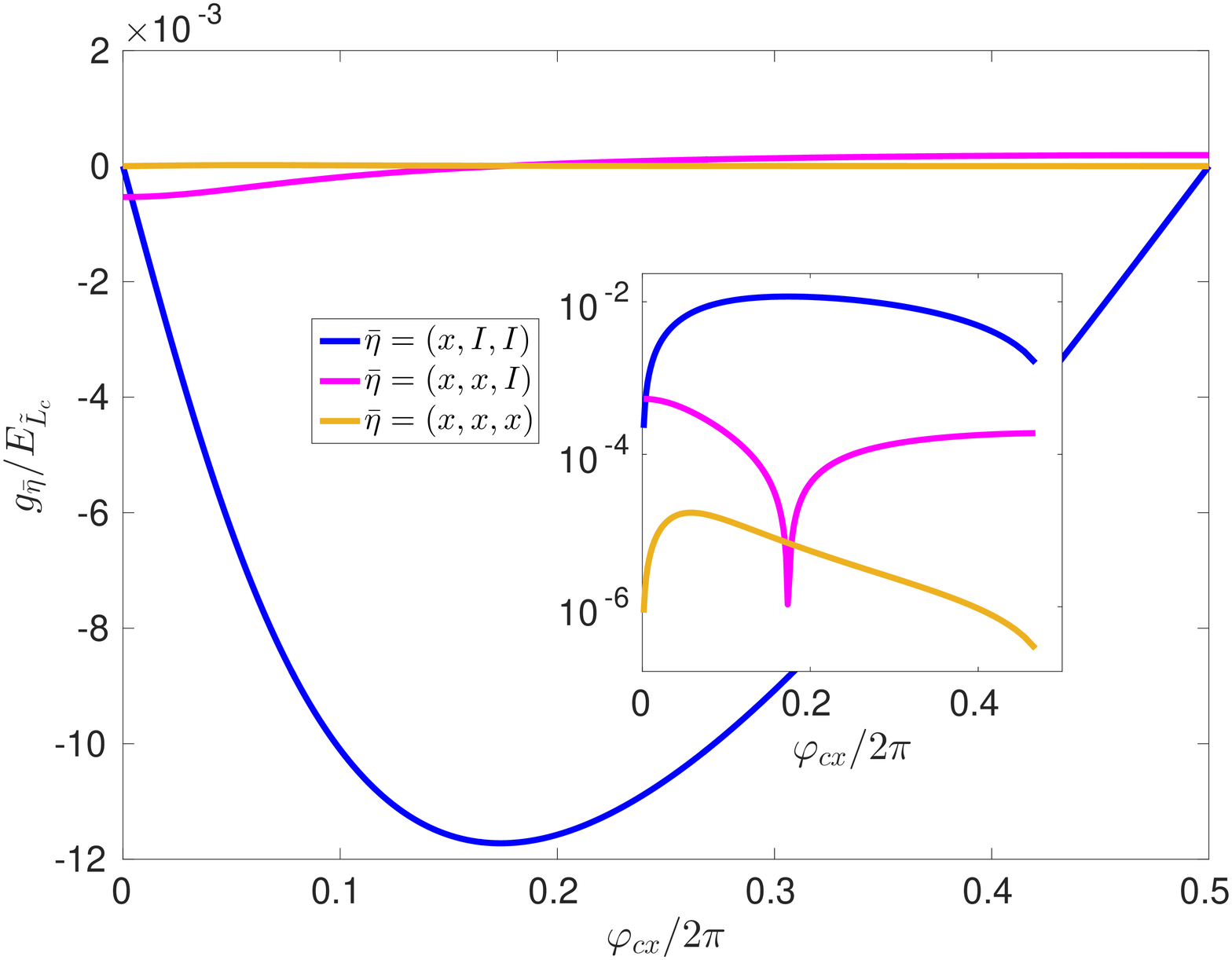}
\caption{Coupler-mediated 3-local interactions are small for typical parameter regimes. Comparison of k-qubit $x$-type couplings for three interacting qubits (in the parity basis): The value of $g_{\bar \eta}$ was computed for $\bar \eta = (x,I,I)$ (dark blue), $(x,x,I)$ (magenta), and $(x,x,x)$ (light orange) using the nonlinear, analytic theory (Section~\ref{reductionQubit}). (Inset is a semi-logarithmic plot of $|g_{\bar \eta}|/E_{\tilde L_c}$.) The qubit and coupler parameters were $\beta_j = 1.05$, $\zeta_j = 0.05$, and $\varphi_{jx} = 0$ and  $\beta_c = 0.5$ and $\zeta_c = 0.05$, respectively. All calculations were carried out in the `parity' basis (see Appendix Section~\ref{numericsMethods} for more details).}
\label{kLocalCouplings}
\end{figure}

In Fig.~\ref{kLocalCouplings} we consider a system of three flux qubits interacting with a single coupler circuit and compare the 3-qubit coupling $\sigma_x\otimes\sigma_x\otimes\sigma_x$ to analogous $1$-local and $2-$local terms. Since we have not verified that the exact spectrum of the three qubit system matches the one predicted by our approximations, we have chosen a more conservative value for the coupler nonlinearity ($\beta_c = 0.5$) relative to the reference regime discussed in the previous section ($\beta_c = 0.75$)\footnote{At the maximal coupling point $\varphi_{cx} = 0$ and impedance $\zeta_c = 0.05$, this change increases the ground state energy gap of $\hat H_c$ from $5.32 \times 10^{-2} E_{\tilde L_c}$ to $7.19 \times 10^{-2} E_{\tilde L_c}$. }. We find that the maximum 3-body coupling ($\sim 1.71 \times 10^{-5} E_{\tilde L_c}$) is more than an order of magnitude smaller than maximum 2-body coupling ($\sim 5.35 \times 10^{-4} E_{\tilde L_c}$). For qubit energy scale $E_{L_j} = 200$ GHz and given $E_{\tilde L_c} / E_{L_j} = 3$, these correspond to maximum couplings of $g_{xxx} \sim 10.3$ MHz and $g_{xxI} = 321$ MHz, compared to the bare (coupler-free) qubit splitting of $884$ MHz. We note that the computed 3-local interaction can be increased significantly by modifying the circuit parameters\footnote{For example, increasing $\beta_c$ from $0.5$ to $0.75$ increases the maximum 3-local coupling approximately five-fold, to $g_{xxx} \sim 8.63 \times 10^{-5} E_{\tilde L_c} = 51.8$ MHz. This occurs at bias $\varphi_{cx} \sim 0.0272 \times 2\pi$, where the approximation to the zero-point energy is expected to hold well (cf.~Fig.\ref{ZPEComparison}).}, although one must be careful that the approximations we have discussed are still valid.

\begin{figure}[h!]
\includegraphics[width = \textwidth,trim={.5cm 0 .5cm 0},clip]{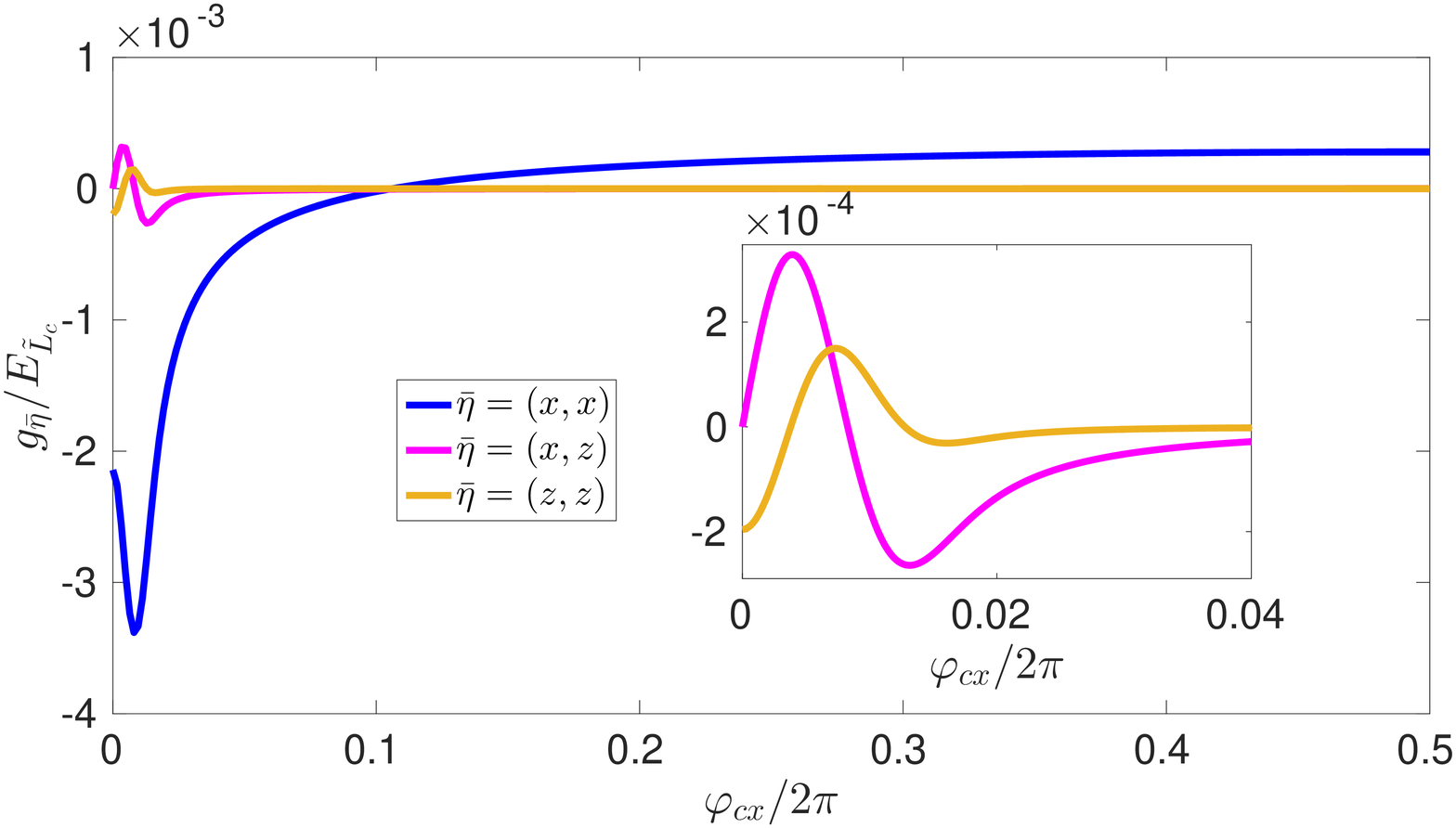}
\caption{The nonlinear theory predicts small but non-negligible non-stoquastic couplings. Main figure: Comparison of 2-qubit couplings depending on coupling type. (Inset is the same plot for the reduced bias range $\varphi_{cx} \in [0,0.04] \times 2 \pi$, focused on only the $xz$ and $zz$ couplings.) The value of $g_{\bar \eta}$ was computed for $\bar \eta = (x,x), (x,z),$ and $(z,z)$. The physical and numerical parameters used in this calculation were identical to those in Fig.~\ref{kLocalCouplings}, except that we assume a coupler $\beta_c =0.95$. Note that the interaction Hamiltonian of the nonlinear, analytic (NA) theory closely predicts the 2-qubit spectrum only for $\varphi_{cx} \gtrsim 0.01 \times 2 \pi$, cf. Fig.~\ref{BOBreakdownVaryPhiCX}. All calculations were carried out in the `parity' basis, so that the non-stoquastic interactions correspond to $(x,z)$ and $(z,z)$. To account for higher coupler nonlinearity, the sums used in the NA calculated (Eqn. ~\eqref{gEta}) were truncated at $|\nu| \leq 200$ (see Appendix Section~\ref{numericsMethods} for more details).}
\label{exoticCouplings}
\end{figure}

The nonlinear theory predicts small but non-negligible non-stoquastic couplings. These couplings are of the form $zz$ or $xz$ in our chosen `parity' basis. Like the typical (stoquastic) $xx$ couplings, we find that these terms increase with coupler nonlinearity $\beta_c$\footnote{This can be explained from the generic coupling formula~\eqref{gEta}: the local $z$ Pauli coefficients $c_z(\nu \alpha_j)$ (equation~\eqref{cDef}) vanish at $\nu = 0$ and peak in magnitude for finite values of $\nu$. The Fourier coefficients $B_\nu$ defining the interaction decay exponentially with $\nu$ but also tend to increase with increasing $\beta_c$. The coupling itself is a sum of products of these terms, so increasing the nonlinearity tends to increase the magnitude of $g_{zz}$.}.  Even so, for even large coupler nonlinearity $\beta_c = 0.95$, the non-stoquastic terms tend to be small compared to the $xx$ couplings, as seen in Fig.~\ref{exoticCouplings}. As noted previously, for such large $\beta_c$ the nonlinear, analytic theory is only accurate away from $\varphi_{cx} = 0$. Yet this region is specifically where non-stoquastic interactions are non-negligible (see inset). These interactions are of order $1-2 \times 10^{-4} E_{\tilde L_c}$, even for $\varphi_{cx}\gtrsim 0.01 \times 2 \pi$ where the nonlinear theory correctly predicts the qubit spectrum (Fig.~\ref{BOBreakdownVaryPhiCX}). For the given circuit parameters and typical $E_{L_j} = 200$ GHz, this corresponds to $xz$ and $zz$ interactions on the order of $100$ MHz.

\section{Conclusions}
We have presented a non-perturbative analysis of a generic inductive coupler circuit within the Born-Oppenheimer Approximation. This provides an explicit and efficiently computable Fourier series for any term in the effective qubit-qubit interaction Hamiltonian. We also account for finite coupler impedance (associated with the coupler's zero-point energy), which gives small but non-negligible quantum corrections to the predicted qubit Hamiltonian. Our results apply whenever the Born-Oppenheimer Approximation and harmonic approximation to the coupler ground state energy are valid (otherwise, there will be deviations as outlined in the numerical study). Importantly, the regime of large coupler nonlinearity and strong coupling $M_j/L_j$ where our results correctly predict the low energy spectrum while deviating significantly from standard linear theories. This regime corresponds to large observed qubit-qubit couplings, as well as small but non-negligible non-stoquastic interactions.  Our analysis is also able to accommodate $k$-body interactions with $k>2$. Although for the considered circuit parameters both $k$-body and non-stoquastic interactions are weak, our theory provides a means to optimize these interactions without resorting to perturbative constructions. As another avenue of investigation, in Appendix Section~\ref{generalization} we show how our theory can be generalized to more complex circuit configurations. We expect that our work will be of use in more accurately modeling existing superconducting qubit devices.

\begin{acknowledgements}
We thank Vadim N. Smelyanskiy and Mostafa Khezri for insightful discussions and helpful comments regarding the text.
\end{acknowledgements}

\bibliography{bibli}
\bibliographystyle{h-physrev}

\section{Appendix}

\begin{figure}[h!]
  \includegraphics[width = .9\textwidth,trim={0cm 0 0cm 0cm},clip]{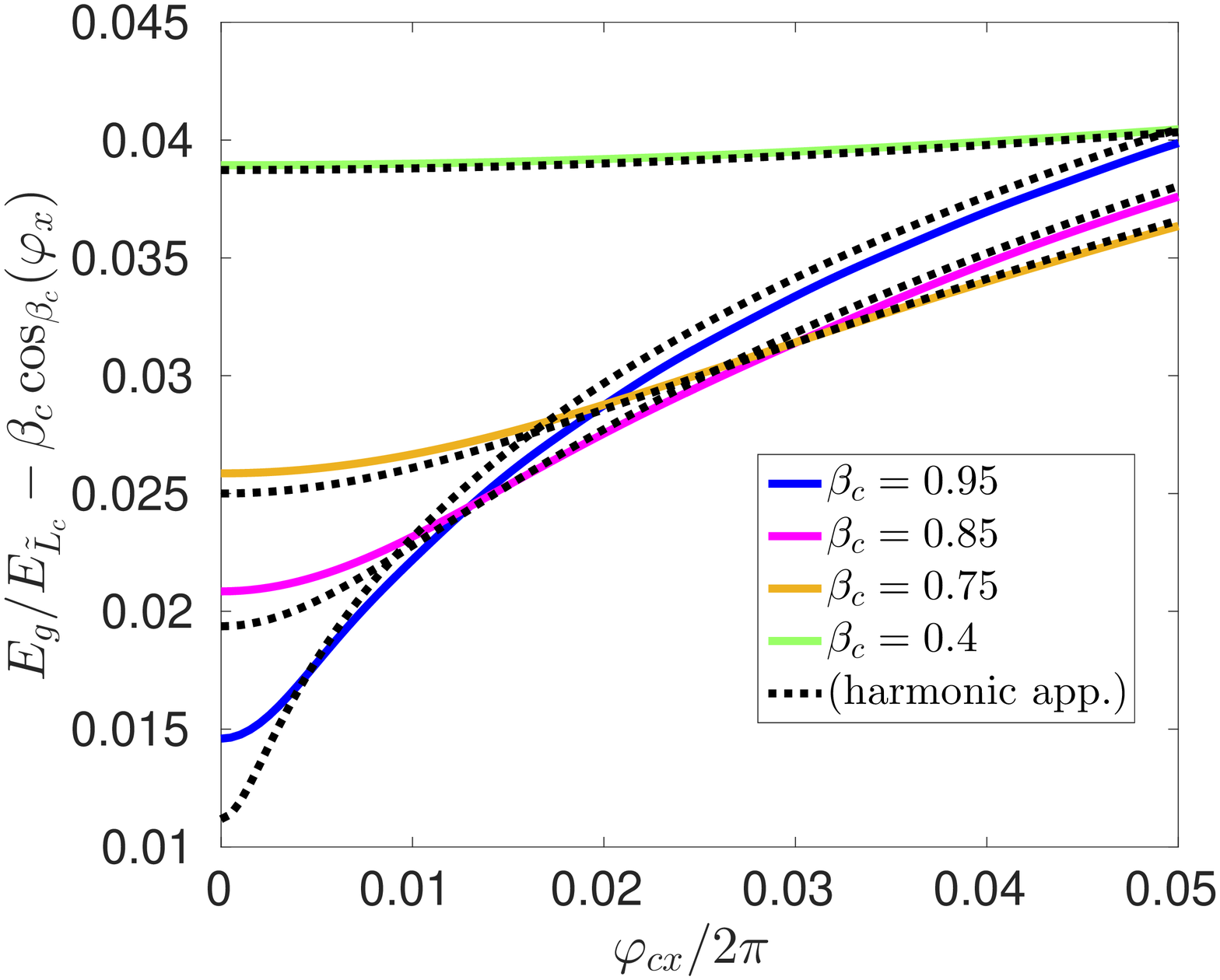}
  \caption{The harmonic approximation to the coupler zero-point energy converges for small but non-zero biases ($\varphi_{cx} \gtrsim 0.01 \times 2 \pi$). Comparison of coupler zero-point energies for different coupler nonlinearities $\beta_c$ and fixed impedance $\zeta_c = 0.05$. Solid curves correspond to the numerically exact zero-point energy. This is computed as the difference between the numerically exact ground state energy $E_g(\varphi_x)/ E_{\tilde L_c}$ and the classical potential minimum $U_{\textrm{min}}(\varphi_x) = \beta_c \cos_{\beta_c}(\varphi_x)$. Dashed curves correspond to the harmonic approximation to the zero-point energy, $\zeta_c \sqrt{1 - \beta_c \cos(\varphi_c^{(*)})}$, discussed in Section~\ref{ZPEDerivation}. From the top, each pair of solid and dashed curves corresponds to coupler nonlinearities $\beta_c = 0.4$ (very light green), 0.75 (light orange), 0.85 (magenta), and $0.95$ (dark blue), respectively. The exact calculation was carried out using 50 harmonic oscillator basis states, as discussed in Section~\ref{numericsMethods}.}
\label{ZPEComparison}
\end{figure}


\begin{figure}[h!]
   \includegraphics[width = .9\textwidth,trim={2cm 0 1.4cm 0cm},clip]{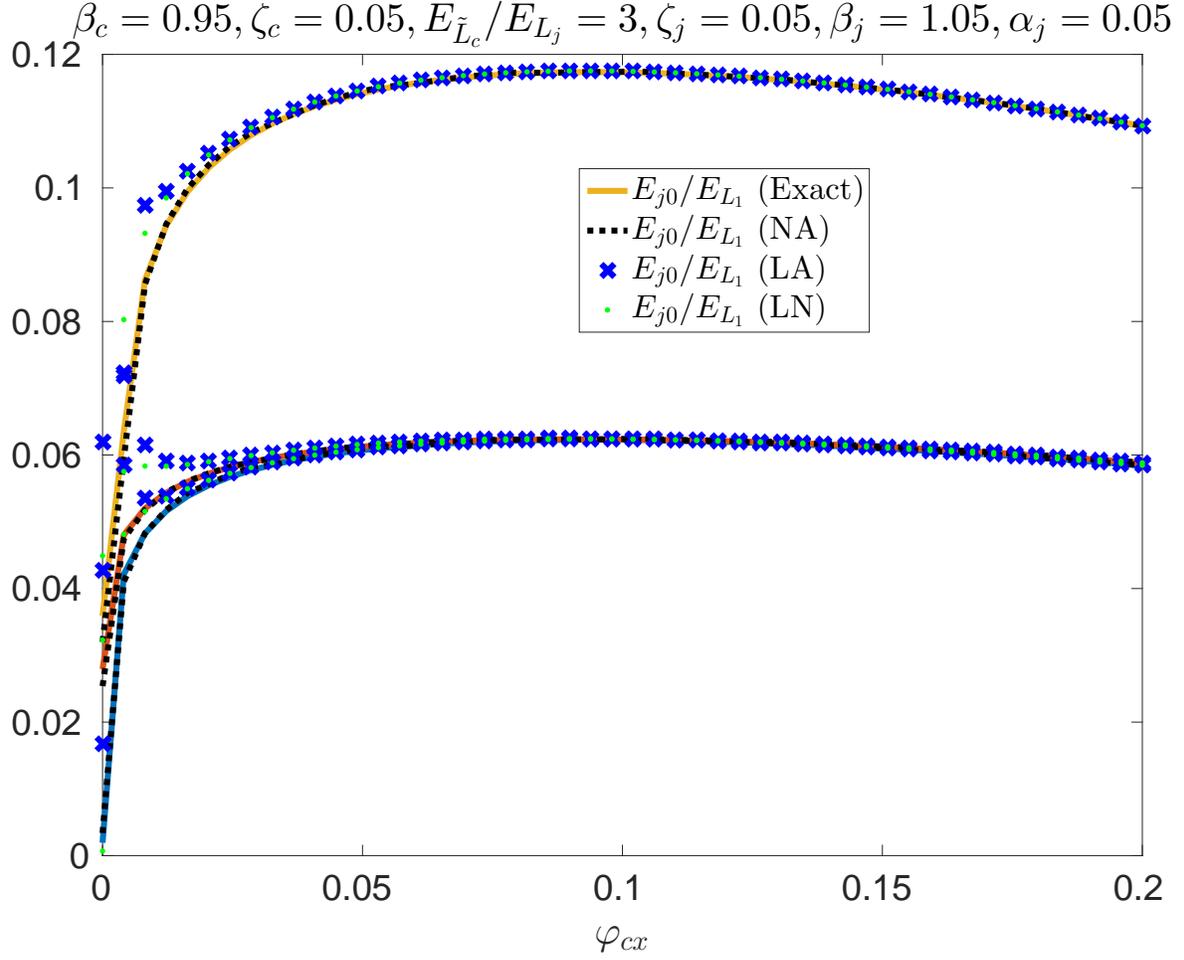} 
\caption{Even for high coupler nonlinearity, all theories predict the correct low-energy spectrum at sufficiently large coupler bias. Circuit parameters are identical to the reference regime (Fig.~\ref{BOBreakdownVaryBetaJReference}), except qubit nonlinearity is fixed at $\beta_c = 1.05$ and coupler nonlinearity $\beta_c$ is increased from $0.75$ to $0.95$. Solid curves represent exact numerical diagonalization of the full Hamiltonian (equation~\eqref{HExact}). The black dashed, dark blue crossed, and light green dotted curves correspond to the nonlinear analytic (NA), linear analytic (LA), and linear numerical (LN) theories of the Born-Oppenheimer Approximation, respectively. The NA theory agrees well with exact diagonalization for $\varphi_{cx}\gtrsim 0.01 \times 2 \pi$. The large oscillations observed in the LA spectrum are due to the divergences in the analytic expressions for the first and second derivatives of $E_g$ as $\beta_c \rightarrow 1$ (equations~\eqref{U12p} and~\eqref{U12pp}). Fig.~\ref{BOBreakdownVaryPhiCX} shows the same calculation for bias values focused near $\varphi_{cx} = 0$.  (See Appendix Section~\ref{numericsMethods} for a detailed description of each calculation.)}
\label{BOBreakdownVaryPhiCXLargerRange}
\end{figure}

\begin{figure}[h!]
  \includegraphics[width = .9\textwidth,trim={2cm 0 1.7cm 0cm},clip]{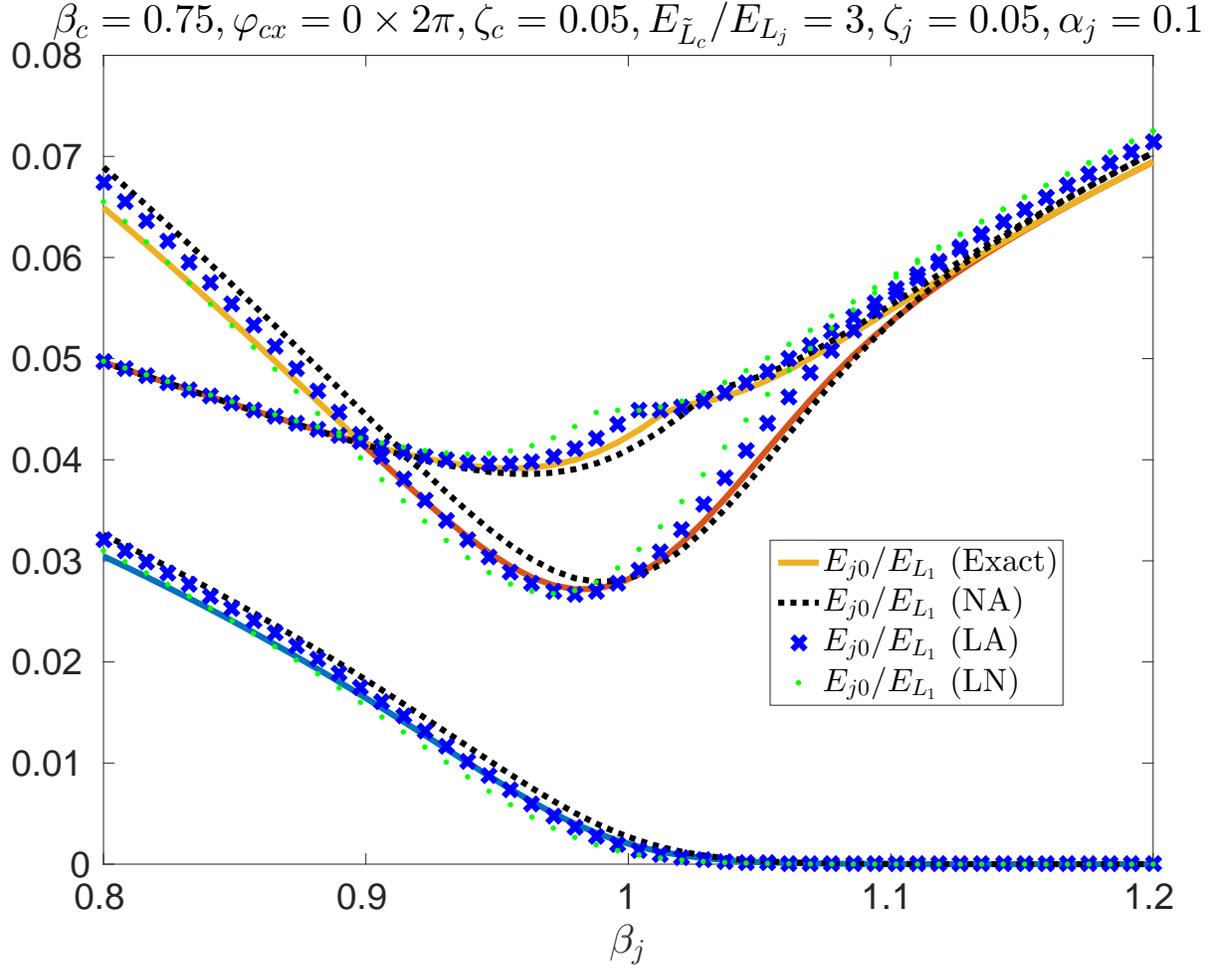}  
  \caption{Born-Oppenheimer theories fail to predict the low energy spectrum for strong coupling and at maximum bias. We consider a single coupler circuit interacting with two identical flux qubits for varying qubit nonlinearity, $\beta_j$. Circuit parameters are identical to the reference regime (Fig.~\ref{BOBreakdownVaryBetaJReference}), except the coupling strength $\alpha_j = M_j/L$ is increased from $0.05$ to $0.1$. Solid curves represent exact numerical diagonalization of the full Hamiltonian (equation~\eqref{HExact}). The black dashed, dark blue crossed, and light green dotted curves correspond to the nonlinear analytic (NA), linear analytic (LA), and linear numerical (LN) theories of the Born-Oppenheimer Approximation, respectively. Fig.~\ref{BOBreakdownVaryPhiCXBigAlpha} considers the same parameter regime, but for varying coupler bias, $\varphi_{cx}$.  (See Appendix Section~\ref{numericsMethods} for a detailed description of each calculation.)}
\label{BOBreakdownVaryBetaJBigAlpha}
\end{figure}

\begin{figure}[h!]
  \includegraphics[width = .9\textwidth,trim={2cm 0 1.4cm 0cm},clip]{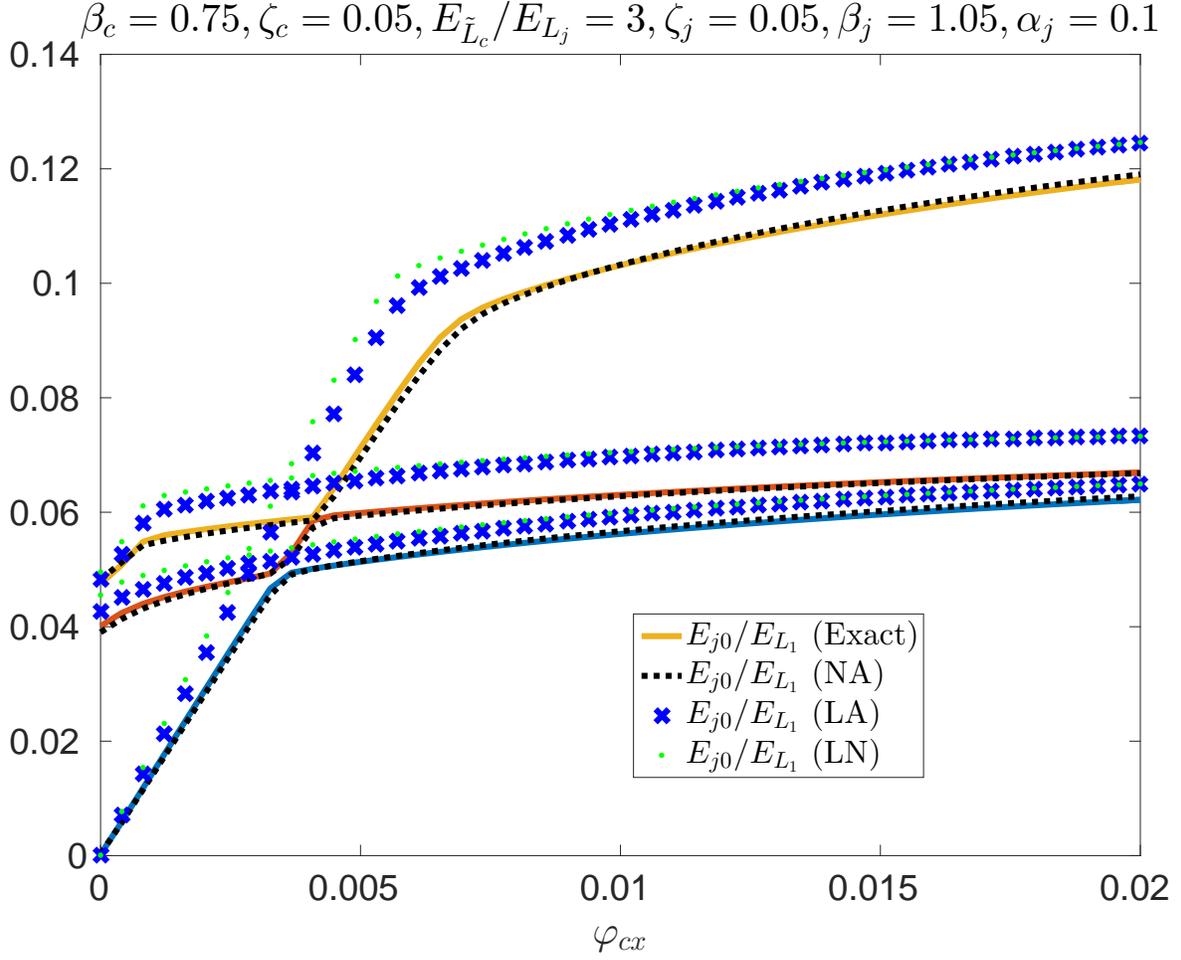} 
  \caption{Away from maximum bias, $\varphi_{cx} = 0$, the NA theory accurately predicts the low energy spectrum even for strong coupling, while the linear theories fail. (Although not shown, for sufficiently large bias, the LA and LN theories do converge to the exact spectrum.) We consider a single coupler circuit interacting with two identical flux qubits for varying coupler bias, $\varphi_{cx}$. Circuit parameters are identical to the reference regime (Fig.~\ref{BOBreakdownVaryBetaJReference}), except the coupling strength $\alpha_j = M_j/L$ is increased from $0.05$ to $0.1$ and the qubit nonlinearity is fixed at $\beta_j = 1.05$. Solid curves represent exact numerical diagonalization of the full Hamiltonian (equation~\eqref{HExact}). The black dashed, dark blue crossed, and light green dotted curves correspond to the nonlinear analytic (NA), linear analytic (LA), and linear numerical (LN) theories of the Born-Oppenheimer Approximation, respectively. Fig.~\ref{BOBreakdownVaryBetaJBigAlpha} considers the same parameter regime, but fixed at maximum coupling $\varphi_{cx} = 0$ and for varying qubit nonlinearity, $\beta_j$.  (See Appendix Section~\ref{numericsMethods} for a detailed description of each calculation.)}
\label{BOBreakdownVaryPhiCXBigAlpha}
\end{figure}

\begin{figure}[h!]
  \includegraphics[width = .9\textwidth,trim={2cm 0 1.7cm 0cm},clip]{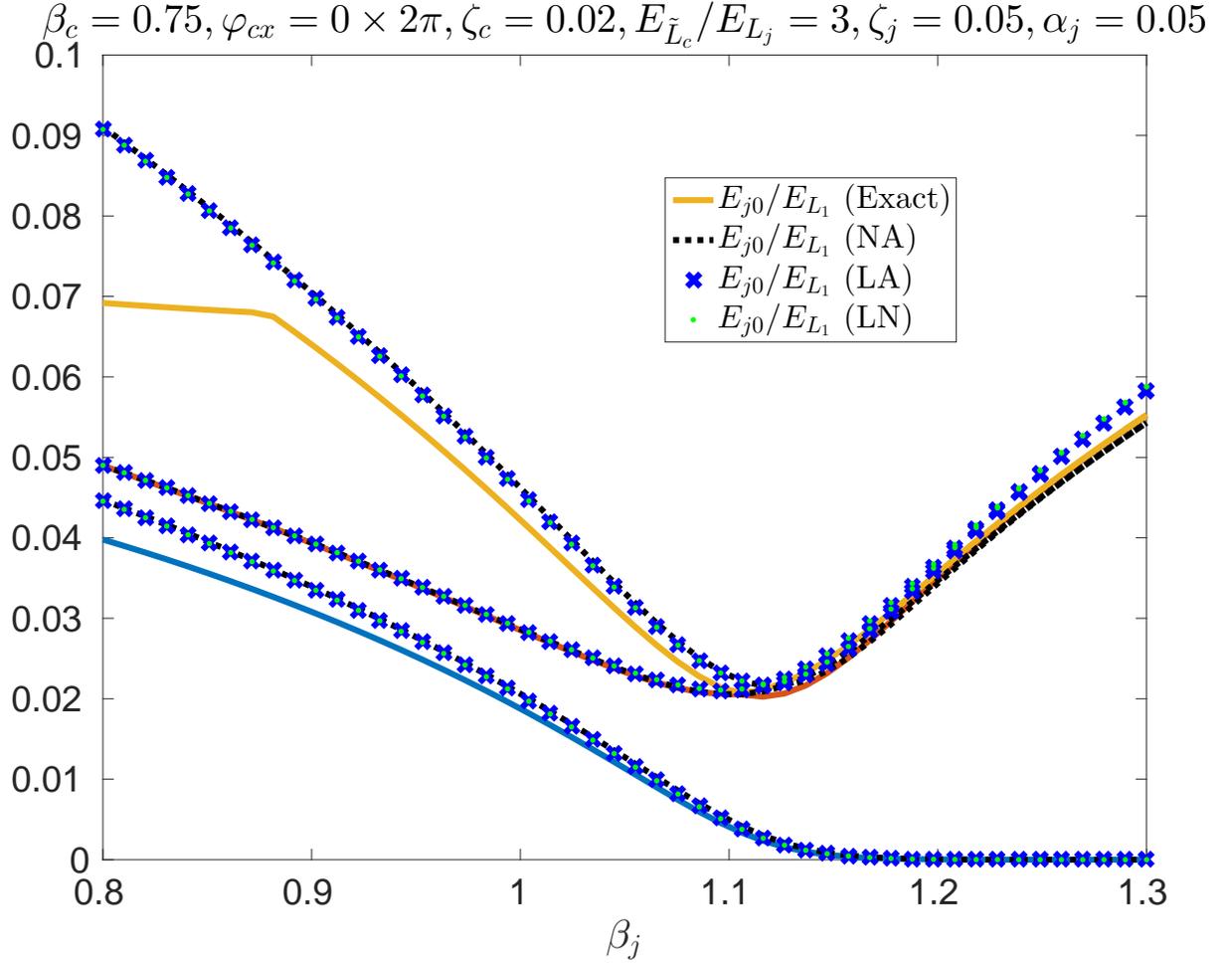}  
\caption{Born-Oppenheimer theories break down in the limit of small coupler impedance. We consider a single coupler circuit interacting with two identical flux qubits for varying qubit nonlinearity, $\beta_j$. Circuit parameters are identical to the reference regime (Fig.~\ref{BOBreakdownVaryBetaJReference}), except the coupler impedance $\zeta_c = \frac{2 \pi e}{\Phi_0} \sqrt{\tilde L_c / C }$ is decreased from $0.05$ to $0.02$. Solid curves represent exact numerical diagonalization of the full Hamiltonian (equation~\eqref{HExact}). The black dashed, dark blue crossed, and light green dotted curves correspond to the nonlinear analytic (NA), linear analytic (LA), and linear numerical (LN) theories of the Born-Oppenheimer Approximation, respectively. (See Appendix Section~\ref{numericsMethods} for a detailed description of each calculation.)}
\label{BOBreakdownVaryBetaJSmallZetaC}
\end{figure}

\begin{figure}[h!]
  \includegraphics[width = .9\textwidth,trim={2cm 0 1.4cm 0cm},clip]{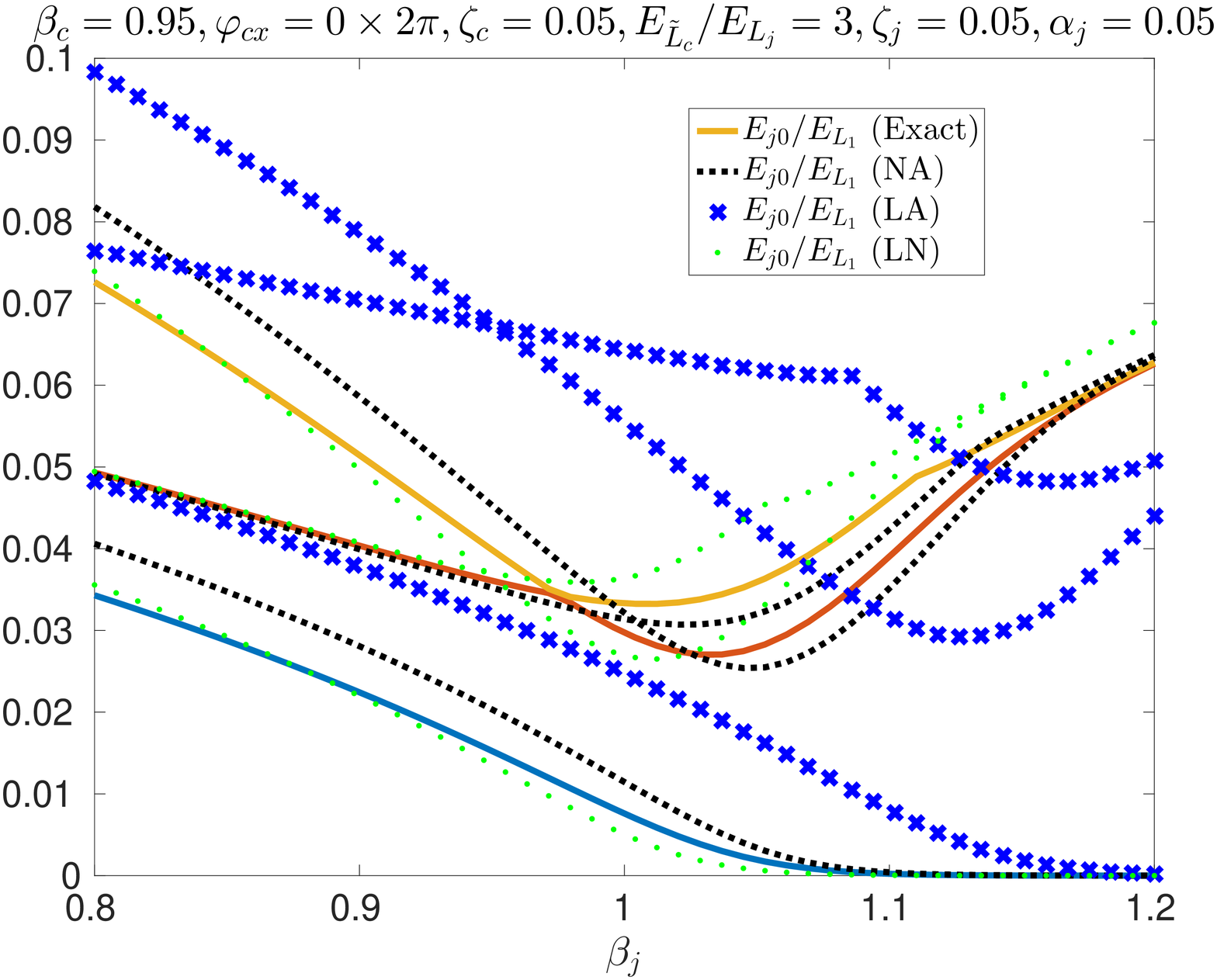}
\caption{Born-Oppenheimer theories break down in the limit of large coupler nonlinearity and at maximum bias, $\varphi_{cx} = 0$. A single coupler circuit interacting with two identical flux qubits for varying qubit nonlinearity, $\beta_j$. Circuit parameters are identical to the reference regime (Fig.~\ref{BOBreakdownVaryBetaJReference}), except the coupler nonlinearity $\beta_c$ is increased from $0.75$ to $0.95$.  Solid curves represent exact numerical diagonalization of the full Hamiltonian (equation~\eqref{HExact}). The black dashed, dark blue crossed, and light green dotted curves correspond to the nonlinear analytic (NA), linear analytic (LA), and linear numerical (LN) theories of the Born-Oppenheimer Approximation, respectively. (See Appendix Section~\ref{numericsMethods} for a detailed description of each calculation.)}
\label{BOBreakdownVaryBetaJBigBetaC}
\end{figure}


\subsection{Numerical methods}
\label{numericsMethods}
We briefly describe the numerical methods used to create the Figures~\ref{BOBreakdownVaryBetaJReference}-\ref{BOBreakdownVaryBetaJBigBetaC}. In all calculations involving matrix diagonalization, the circuit Hamiltonians are represented in a basis of harmonic oscillator eigenstates~\cite{Smith2016}. This basis is specified by the normal modes of the linear part of the Hamiltonian (i.e., the part independent of the Josephson junctions). The Hamiltonian can then be decomposed into a linear part (a sum of number operators) and a sinusoidal part (deriving either directly from a Josephson Junction or from the nonlinear theory in the main text). In general the Hamiltonian takes the form
\begin{equation}
  \label{HNM}
  \hat H = \sum_n \hbar \omega_n \l(\hat a_n^\dagger \hat a_n + 1/2 \r) + \sum_m C_m \exp\l(i  \sum_n r_{m,n} (\hat a_n + \hat a_n^\dagger)  \r)\,,
\end{equation}
where the coefficients $\omega_n, C_m$, and $r_n$ are circuit dependent. The linear part of the Hamiltonian has a diagonal representation in the harmonic oscillator basis, while the matrix elements of the exponential operators can be computed using the identity~\cite{Gradshteyn2014}
\begin{align}
  \label{expPhi}
  \begin{split}
    \bra{j}e^{i r (\hat a + \hat a^\dagger)} \ket{k} & =   \frac{e^{-\frac{r^2}{2}} }{\sqrt{ j! k!}}\sum_{l \leq 0}  l! \binom{j}{l}\binom{k}{l}  ( i  r)^{j+k-2 l}\\
    & = (i)^{3j+k} \sqrt{\frac{j!}{k!}} e^{-\frac{r^2}{2}}  r^{k-j}   L_{j}^{(k-j)}(r^2)\,.
  \end{split}
\end{align}
Here $L_{j}^{(k-j)}(r^2)$ refers to the generalized Laguerre polynomial.

{\it Spectrum calculations:} In~\Cref{BOBreakdownVaryBetaJReference,BOBreakdownVaryBetaJBigZetaC,BOBreakdownVaryPhiCX,BOBreakdownVaryPhiCXLargerRange,BOBreakdownVaryBetaJBigAlpha,BOBreakdownVaryBetaJSmallZetaC,BOBreakdownVaryBetaJBigBetaC} we compute the spectrum of two flux qubit circuits interacting with a coupler circuit. For the exact calculation, each circuit is treated as an independent degree freedom, so that the exact Hamiltonian (equation~\eqref{HExact}) is expressed as a sum of three modes in the form of equation~\eqref{HNM}. In all figures we truncate the harmonic oscillator basis at $40\times 40\times 18$ states, with the last mode corresponding to the highest frequency mode (associated primarily with coupler motion). Similarly, the spectrum calculations involving the Born-Oppenheimer Approximation truncate the reduced Hamiltonian $\hat H_1 + \hat H_2 + \hat H_{\textrm{int}}$ to $40\times 40$ basis states. For the nonlinear (NA) approximation to $E_g(\hat \varphi_x)$, we truncated the Fourier series~\eqref{Hint} at $|\nu|\leq 100$, with the inner series describing the zero-point energy (equation~\eqref{interactionSeries}) truncated at $|\mu|\leq 40$.

{\it Qubit dynamics:} In~\Cref{qubitTermsAtReference,compareQubitDynamics,qubitTermsLargeZetaC,qubitTermsLargeBetaC,kLocalCouplings,exoticCouplings} we compute the coupler's contribution to the flux qubits' Hamiltonian. This is done by projecting $\hat H_{\textrm{int}}$ into the `qubit subspace' spanned by the two lowest energy states of each independent flux qubit. Our calculations are carried out in the `parity' basis, which (for unbiased flux qubits, $\varphi_{jx} = 0$) corresponds to the (symmetric and anti-symmetric) ground and first excited state of each local qubit Hamiltonian $\hat H_j$. As is done for the other spectrum calculations, the eigenstates are computed by representing each flux qubit's Hamiltonian in the harmonic oscillator basis (truncated at $50$ basis states). The NA calculations were based on equation~\eqref{gEta}, with the sums truncated at $|\nu| \leq 60$ (unless otherwise noted) and the inner sum defining coefficients $B_\nu$ truncated at $|\mu|\leq 40$ (equation~\eqref{interactionSeries}) . The (linear) LA and LN calculations were based on equation~\eqref{gEtaLin}, using approximate analytic and exact numerical derivatives \eqref{U12p} and \eqref{U12pp}, \eqref{EigenDeriv} and \eqref{EigenDeriv2}, respectively. The details of the projections themselves are discussed in detail Sections~\ref{reductionQubit} (for the NA theory) and~\ref{EgLinearized} (for the LA and LN theories).

\subsection{Inversion of Josephson Junction relation}
\label{expMuSeries}
In this section we solve for the function $f(x) = e^{i \mu x}$ (for any integer $\mu$) as a Fourier series in $\varphi$ under the constraint
\begin{equation}
  \label{implicitEq}
   x- \varphi - \beta \sin(x) = 0\,,
\end{equation}
where $\beta$ is a scalar satisfying $|\beta|<1$. The resulting Fourier series corresponds to equation~\eqref{fourierCoefs} in the main text,
\begin{equation}
  \label{transInvertApp}
  e^{ i \mu x} =  \sum_{\nu} e^{i \nu \varphi} A_{\nu}^{(\mu)}\,,
\end{equation}
where
\begin{equation}
  \label{fourierCoefsApp}
  A_{\nu}^{(\mu)} = \l\{\begin{array}{cc}
  \delta_{\mu,0} - \frac{\beta_c}{2}(\delta_{\mu,1} + \delta_{\mu,-1})& \nu = 0\\
  \frac{\mu J_{\nu-\mu}(\beta_c \nu)}{\nu} & \nu \neq 0
  \end{array}\r.\,,
\end{equation}
The function $e^{i \mu x}$ is used to derive the Fourier Series for the coupler ground state energy, equation~\eqref{EgFinal}. 

To prove this result, observe that if $ x $  is a unique solution\footnote{The solution is unique if and only if the potential $\frac{(x-\varphi)^2}{2} + \beta \cos(x)$ has a unique extremum (for all $\varphi$). This holds if and only if it is a convex function of $x$. Taking the second derivative, we see that this holds exactly when $\Gamma(x) = 1 - \beta\cos(x) \geq 0$ for all $x$, which is equivalent to $|\beta|\leq 1$.  } to \eqref{implicitEq} then the Dirac delta function at this point satisfies
$$\delta( z -  x) = \delta(z - \varphi -  \beta \sin(z)) \Gamma(z)\,,$$
where
$$\Gamma( z) = \l|\partial_{z} \l( z -  \varphi - \beta \sin(z) \r)\r| = 1 - \beta \cos(z)\,. $$
The start of the calculation is similar to the derivation of the Lagrange Reversion Theorem~\cite{Whittaker1996}. We leave it as an exercise to the reader to justify the rearrangements of sums and integrals. 

\begin{align}
  \label{transInvertDerivation}
  \begin{split}
    f( x) = & \intd z f( z)  \delta( z - x) \\
    = & \intd z f( z)  \delta( z -  \varphi -  \beta \sin(z)) \Gamma( z)\\
    = & \intd z f( z) \Gamma( z)\intd k \frac{1}{2 \pi}  e^{i  k  ( z -  \varphi -  \beta \sin(z))} \\
    = & \intd z f( z) \Gamma( z) \intd k \frac{1}{2 \pi} \sum_{n}\frac{\l(-i k \beta \sin(z) \r)^{n}}{n!}  e^{i  k  ( z -  \varphi )} \\
    = & \intd z f( z) \Gamma( z) \intd k \frac{1}{2 \pi} \sum_{n}\frac{\l(\partial_{\varphi} \beta \sin(z)  \r)^{n}}{n!}  e^{i  k  ( z -  \varphi ) } \\
    = & \sum_{n} \intd z f( z) \Gamma( z) \frac{\l(\partial_{\varphi} \beta \sin( z) \r)^{n}}{n!} \intd k \frac{1}{2 \pi}   e^{i  k  ( z  - \varphi )} \\
    = & \sum_{n} (\partial_{\varphi})^{n} \intd z f( z) \Gamma( z) \frac{\l( \beta \sin( z) \r)^{n}}{n!} \intd k \frac{1}{2 \pi}   e^{i  k  ( z -  \varphi) } \\
    = & \sum_{n} (\partial_{\varphi})^{n} \intd z f( z) \Gamma( z) \frac{\l( \beta \sin( z) \r)^{n}}{n!}  \delta( z -  \varphi) \\
    = & \sum_{n} (\partial_{\varphi})^{n} f( \varphi) \Gamma( \varphi) \frac{\l( \beta \sin(\varphi) \r)^{n}}{n!}   \\
    = & \sum_{n} (\partial_{\varphi})^{n}  \sum_{ \nu} e^{i  \nu   \varphi} \int_{-\pi}^\pi \mbox{d}\tau \frac{e^{-i  \nu   \tau}}{2 \pi}f( \tau) \Gamma( \tau) \frac{\l( \beta \sin(\tau) \r)^{n}}{n!}    \\
    = & \sum_{n}  \sum_{ \nu} (i \nu)^{n}  e^{i  \nu   \varphi} \int_{-\pi}^\pi \mbox{d}\tau \frac{e^{-i  \nu   \tau}}{2 \pi}f( \tau) \Gamma( \tau) \frac{\l( \beta \sin(\tau) \r)^{n}}{n!}    \\
    = &  \sum_{ \nu}   e^{i  \nu   \varphi} \int_{-\pi}^\pi \mbox{d}\tau \frac{e^{-i  \nu   \tau}}{2 \pi}f( \tau) \Gamma( \tau) \sum_{n} \frac{\l(i \nu \beta \sin(\tau) \r)^{n}}{n!}   \\
     = &  \sum_{ \nu}   e^{i  \nu   \varphi} \int_{-\pi}^\pi \mbox{d}\tau \frac{e^{-i  \nu   \tau}}{2 \pi}f( \tau) \Gamma( \tau) e^{i  \nu   \beta \sin(\tau)} \\
     = &  \sum_{ \nu}   e^{i  \nu   \varphi} \l[e^{i  \nu   \beta \sin(\tau)} \Gamma( \tau)f( \tau) \r]_{ \nu} \,.
  \end{split}
\end{align}
In the last line, we have introduced the notation $[h(\tau)]_\nu = \int_{-\pi}^\pi \mbox{d}\tau \frac{e^{-i  \nu   \tau}}{2 \pi} h(\tau)$ to represent the Fourier coefficient of $h(\tau)$ corresponding to $e^{i\nu \tau}$. We note that the definition above is actually agnostic to the definition of the function $f(x)$ (except the assumption that it is periodic and smooth). 

To complete the derivation, we make the substitutions $\Gamma(\tau) = 1 - \beta \cos(\tau)$ and $f(\tau) = e^{i \mu \tau}$,
$$e^{i \mu x} = \sum_{\nu} e^{i \nu \varphi}\l[e^{i \nu \beta \sin(\tau)}(1 - \beta \cos(\tau))e^{i \mu \tau} \r]_\nu\,.$$
The product $(1 - \beta \cos(\tau)) e^{i \mu \tau}$ has Fourier coefficients
\begin{equation}
  \label{fGammau}
  [(1 - \beta \cos(\tau)) e^{i \mu \tau}]_\gamma =  \delta_{\gamma,\mu} - \frac{\beta}{2}\l(\delta_{\gamma,\mu+1} + \delta_{\gamma,\mu-1} \r)\,.
\end{equation}
Likewise, the Jacobi-Anger identity\cite[Eqn. 9.4.41]{Abramowitz1964} gives us the Fourier coefficients of $e^{i \nu \beta \sin(\tau)}$,
\begin{equation}
  \label{expBetaSinSigma}
  [e^{i \nu \beta \sin(\tau)}]_\sigma =  J_{\sigma} (\beta \nu) \,,
\end{equation}
where $J_{\sigma}(x)$ is the Bessel function of the first kind. Combining these statements, we compute
\begin{align}
  \begin{split}
    e^{i \mu x} &=  \sum_{\nu} e^{i \nu \varphi} \sum_\sigma   [e^{i \nu \beta \sin(\tau)}]_\sigma\l[(1 - \beta \cos(\tau)) e^{i \mu \tau}\r]_{\nu - \sigma}\\
    & = \sum_{\nu} e^{i \nu \varphi} \l( J_{\nu - \mu}(\beta \nu) - \frac{\beta}{2}\l(J_{\nu - \mu-1}(\beta \nu) + J_{\nu - \mu+1}(\beta \nu) \r) \r)\\
    & = \delta_{\mu,0} - \frac{\beta}{2}\l(\delta_{\mu,1} + \delta_{\mu,-1}  \r) + \sum_{\nu\neq 0} e^{i \nu \varphi} \l( J_{\nu - \mu}(\beta \nu) - \frac{\beta}{2}\l( \frac{2(\nu -\mu)}{\beta \nu}J_{\nu - \mu}(\beta \nu) \r) \r)\\
    & = \delta_{\mu,0} - \frac{\beta}{2}\l(\delta_{\mu,1} + \delta_{\mu,-1}  \r) + \sum_{\nu\neq 0} e^{i \nu \varphi}\, \frac{\mu J_{\nu - \mu}(\beta \nu)}{\nu}\,,
  \end{split}
\end{align}
where in the first line we expressed $\l[e^{i \nu \beta \sin(\tau)}(1 - \beta \cos(\tau))e^{i \mu \tau} \r]_\nu$ as a convolution. In the second line we used equation~\eqref{fGammau}, and in the third we separated the sum between $\nu = 0$ and $\nu \neq 0$ and used the identities
\begin{align*}
  J_{\sigma}(0) & = \delta_{\sigma,0}\\
  J_{\sigma-1}(x) + J_{\sigma+1}(x) & = \frac{2 \sigma}{x} J_{\sigma}(x) \quad (x \neq 0)\,.
\end{align*}
This completes the derivation of equation\eqref{fourierCoefsApp} (equation~\eqref{fourierCoefs} in the main text). In Appendix Section~\ref{generalization} we discuss the generalization of these results to circuits with more than one degree of freedom.

\subsection{Derivation of the $\cos_{\beta}$ function}
\label{cosBetaIdentity}

In this section we prove the equality of each line in equation~\eqref{cosBetaSeries}. Rewritten here, these equations define the $\cos_\beta(\varphi_x)$ function,
\begin{align}
  \label{cosBetaSeriesAppendix}
  \begin{split}
    \cos_{\beta}(\varphi_x) & \equiv  1 - \int_{0}^{\varphi_x} \sin_{\beta}(\theta) \mbox{d}\, \theta \\
    & = \frac{\beta}{2}\l( \sin_{\beta}(\varphi_x)\r)^2 + \cos(\varphi + \beta \sin_{\beta}(\varphi_x))\\
    & = 1 + \sum_{\nu >0}  \frac{2 J_{\nu}({\beta} \nu)}{  {\beta} \nu^2} \l(\cos(\nu \varphi_x) - 1\r) \\
    & =  -\frac{\beta}{4} + \sum_{\nu \neq 0}  \frac{J_{\nu}({\beta} \nu)}{  {\beta} \nu^2} e^{i \nu \varphi_x} \,.
  \end{split}
\end{align}
The equality of the first and second lines follows from the fact that both have value 1 at $\varphi_x = 0$ (since ${\sin_{\beta}(0) = 0}$) and both have the same derivative (cf. equation~\eqref{sinBetaDeriv}). The equality of the first and third lines follows from direct integration of $\sin_{\beta}(\theta)$ (cf. equation~\eqref{sinBeta}).

Finally we show that $\cos_{\beta}(\varphi_x)$ equals the last line of equation~\eqref{cosBetaSeriesAppendix}. Noting that $J_{-\nu}(- \beta \nu) = J_{\nu}(\beta \nu)$, we see that the third and fourth lines of~\eqref{cosBetaSeriesAppendix} correspond to the same Fourier cosine series for all coefficients with $\nu \neq 0$. It remains to show that the constant ($\nu = 0$) coefficients also agree. We directly compute this coefficient for the first three lines by considering the integral,
\begin{align}
  \begin{split}
    \frac{1}{2 \pi} \int_0^{2 \pi} \cos_{\beta}(\varphi) \mbox{d} \varphi & = \frac{1}{2 \pi}  \int_0^{2 \pi} \partial_{\varphi} \l( \varphi \cos_{\beta}(\varphi) \r) + \varphi \sin_{\beta}(\varphi)\, \mbox{d} \varphi \\
    & = \frac{1}{2 \pi} \l(\l. \varphi \cos_{\beta}(\varphi) \r]^{\varphi = 2 \pi}_{\varphi = 0}+ \int_0^{2 \pi} \varphi \sin_{\beta}(\varphi) \mbox{d} \varphi\r) \\
      & = 1 +  \frac{1}{2 \pi} \int_0^{2 \pi} (u - \beta \sin(u)) \sin(u)\l(1 - \beta \cos(u) \r) \mbox{d} u\\
      & = -\frac{\beta}{4}\,.
  \end{split}
\end{align}
In the first line we integrated by parts and used $\partial_\varphi \cos_{\beta}(\varphi) = - \sin_{\beta}(\varphi)$ (first line of~\eqref{cosBetaSeriesAppendix}), while in the third line we used $\cos_{\beta}(2 \pi) = 1$ (second line of~\eqref{cosBetaSeriesAppendix} and note $\sin_{\beta}(2 \pi) = 0$ using~\eqref{sinBeta}) and the change of variables,
\begin{align*}
  \varphi &= u - \beta \sin(u)\\
  \sin_{\beta}(\varphi) & = \sin(u)  \\
  \mbox{d} \varphi &= (1 - \beta \cos(u)) \mbox{d} u\,.
\end{align*}
Thus the $\nu = 0$ Fourier coefficient of the first three lines of~\eqref{cosBetaSeriesAppendix} agrees with the final line, which was all that was left to show.

\subsection{Derivation of zero-point energy Fourier series}
\label{sqrtCosFS}
In this section we derive the series of identities defining the approximate coupler zero-point energy (equation~\eqref{UZPE}),
\begin{align}
  \label{UZPEAppendix}
  \begin{split}
    U_{\textrm{ZPE}} & = \zeta_c \sqrt{1 -\beta_c\cos(\varphi_c^{(*)})} \\
    & = \zeta_c\l( G_{0}(\beta_c)- \beta_c G_{1}(\beta_c)+ \sum_{\nu\neq 0}e^{i \nu \varphi_x} \l(\frac{1 }{\nu} \sum_\mu \mu\, G_{\mu}(\beta_c) J_{\nu-\mu}(\beta_c \nu)  \r) \r)\,,
  \end{split}
\end{align}
where
\begin{equation}
  G_\mu(\beta)  = \sum_{l\geq 0} \binom{1/2}{\mu+2l}\binom{\mu+2l}{l}\l(-\frac{\beta}{2} \r)^{\mu+2l}\,.
\end{equation}
and $\varphi_x = \varphi_{cx} - \sum_j \alpha_j \varphi_j$ is a qubit-dependent flux parameter. We begin by deriving the Fourier series of the function $\sqrt{1 - \beta_c \cos(\theta)}$. This follows directly from the generalized binomial theorem\cite[Eqn. 3.6.9]{Abramowitz1964},
\begin{align}
  \label{sqrtSeries}
  \begin{split}
    \sqrt{1 - \frac{\beta_c}{2} (z + z^{-1})} & = \sum_{k\geq 0} \binom{1/2}{k}\l(-\frac{\beta_c}{2} (z + z^{-1}) \r)^k\\
    & = \sum_{k\geq 0} \binom{1/2}{k} \l( -\frac{\beta_c}{2}\r)^k \sum_{l\geq 0} \binom{k}{l} z^{k - 2 l }\\
    & = \sum_{\mu} z^{\mu } \sum_{l \geq 0} \binom{1/2}{\mu + 2 l}  \binom{ \mu + 2 l}{l}\l(- \frac{\beta_c}{2}\r)^{\mu + 2 l} \\
    & \equiv \sum_{\mu} z^{\mu } G_{\mu}(\beta_c)
  \end{split}
\end{align}
In the second line we used the binomial theorem again, while in the third line we changed to index $\mu = k - 2 l$ (which goes over both positive and negative integers). In the final line we have equated the sum over $l$ with the coefficient $G_{\mu}(\beta_c)$. Algebraic manipulations of this sum allow it to be rewritten in terms of the confluent hypergeometric function\cite[Ch. 15]{Abramowitz1964},
\begin{equation}
  G_{\mu}(\beta_c) = \l(-\frac{\beta_c}{2}\r)^\mu \binom{1/2}{\mu} {_2F_1}\l(\frac{\mu}{2}-\frac{1}{4},\frac{\mu}{2}+\frac{1}{4};1+\mu; \beta_c^2 \r)\,.
\end{equation}
(This assumes $\mu\geq 0$, though we note that $G_{\mu}(\beta_c) = G_{-\mu}(\beta_c)$; cf. the left hand side of equation~\eqref{sqrtSeries}.) The coefficients $G_{\mu}(\beta_c)$ can also be expressed in terms of the Legendre functions\cite[Ch. 8]{Abramowitz1964},
\begin{equation}
  G_{\mu}(\beta_c) = (i\, \mbox{sgn}(\beta))^\mu \frac{\Gamma(3/2)}{\Gamma(3/2-\mu)} (1 - \beta_c^2)^{1/4}P_{1/2}^{-\mu}\l(\frac{1}{\sqrt{1-\beta_c^2}}\r)\,.
\end{equation}
(This expression is valid for any integer $\mu$.)

To complete the derivation of equation~\eqref{UZPEAppendix}, we substitute $z = e^{i \varphi_c^{(*)}}$ into~\eqref{sqrtSeries} and use $\cos(\varphi_c^{(*)}) = \frac{1}{2} (e^{i \varphi_c^{(*)}} + e^{- i \varphi_c^{(*)}})$, giving
\begin{align}
  \begin{split}
    \sqrt{1 - \beta_c \cos(\varphi_c^{(*)})} & = \sum_{\mu} G_{\mu}(\beta_c) e^{i \mu \varphi_c^{(*)}}\\
    & = \sum_{\mu} G_{\mu}(\beta_c) \sum_\nu e^{i \nu \varphi_x} A_{\nu}^{(\mu)}\\
    & = \sum_{\nu} e^{i \nu \varphi_x} \sum_\mu G_{\mu}(\beta_c)  A_{\nu}^{(\mu)}\,,
  \end{split}
\end{align}
where in the second line we invoked identity~\eqref{transInvert} (derived in Appendix Section~\ref{expMuSeries}) and in the third line we rearranged the order of summation. Identity~\eqref{transInvert} expresses $e^{i \mu \varphi_c^{(*)}}$ as a Fourier series in $\varphi_x$ given the implicit relationship ${\varphi_c^{(*)} = \varphi_x + \beta_c \sin(\varphi_c^{(*)})}$. Equation~\eqref{UZPEAppendix} follows from equation~\eqref{fourierCoefs} for the Fourier coefficients $A_{\nu}^{(\mu)}$ and the fact that $G_{\mu}(\beta_c) = G_{-\mu}(\beta_c)$.

\subsection{Classical analysis of coupler circuit}
\label{classical}
In this section we carry out a classical analysis of the qubit-coupler dynamics. We show that, in the classical limit of large coupler plasma frequency, the reduced qubit interaction Hamiltonian corresponds exactly to the minimum of the coupler potential $E_{\tilde L_c } U_{\textrm{min}}(\varphi_x) = E_{\tilde L_c } \beta_c \cos_{\beta_c}(\varphi_x)$. To begin, we rewrite the first of the classical current equations~\eqref{currentEqs2} in terms of the dimensionless parameters in equation~\eqref{paramDefs}
\begin{equation}
  \label{currentEqAppendix}
  \tilde L_c C\, \ddot \varphi_c  - \beta_c \sin(\varphi_c) + \varphi_c - \varphi_x = 0\,,
\end{equation}
where
\begin{equation}
  \label{vfx}
  \varphi_x = \varphi_{cx} - \sum_j \alpha_j \varphi_j\,.
\end{equation}
Analogously to the Born-Oppenheimer Approximation in the quantum treatment, we assume that the qubit-dependent flux variables are slow compared to the coupler plasma frequency $ 1/\sqrt{\tilde L_c C}$. This allows us to approximately solve equation~\eqref{currentEqAppendix} by dropping the term proportional to $\tilde L_c C$. The coupler flux variable $\varphi_c$ is then no longer an independent variable, since it can be written as an explicit function of $\varphi_x$,
\begin{equation}
   \varphi_c =  \varphi_x + \beta_c \sin(\varphi_c) = \varphi_x + \beta_c \sin_{\beta_c}(\varphi_x)\,.
\end{equation}
This is the same inversion we carried out when solving for the minimum of the coupler-potential, $U'(\varphi_c^{(*)}) = 0$. Noting that
$$\Phi_c - \Phi_{cx} + \sum_j \alpha_j \Phi_j = \frac{\Phi_0}{2 \pi} \l(\varphi_c - \varphi_x \r) =  \frac{\Phi_0}{2 \pi} \beta_c \sin_{\beta_c} (\varphi_x),$$
we substitute directly into the second current equation~\eqref{currentEqs2}, giving
\begin{equation}
  \frac{\Phi_{j}}{L_j} + \alpha_j \beta_c \frac{1}{\tilde L_c}\frac{\Phi_0}{2 \pi}  \sin_{\beta_c} (\varphi_x)  - I^{*}_{j}  = 0\,.
\end{equation}
These reduced system of equations are independent of the coupler flux variable $\varphi_c$. Since they are the Euler-Lagrange equations for the qubit flux variables, the nonlinear term corresponds exactly to an interaction potential
\begin{equation}
  \pdp {U_{\textrm{int}}}{\Phi_j} =  \alpha_j \beta_c \frac{1}{\tilde L_c}\frac{\Phi_0}{2 \pi}  \sin_{\beta_c} (\varphi_x)\,.
\end{equation}
Using equation~\eqref{vfx}, $\Phi_j = \frac{\Phi_0}{2 \pi}\varphi_j$, and the relationship $\partial_{\varphi_x} \cos_{\beta_c} (\varphi_x) = - \sin_{\beta_c} (\varphi_x)$, we can immediately solve for $U_{\textrm{int}}$ as
\begin{equation}
  U_{\textrm{int}} = \frac{(\Phi_0/2 \pi)^2}{\tilde L_c} \beta_c \cos_{\beta_c}(\varphi_x) =  E_{\tilde L_c} \beta_c \cos_{\beta_c}(\varphi_x)\,.
\end{equation}
Hence the classical interaction potential mediated by the coupler circuit corresponds exactly to the minimum value of the coupler's potential energy, $U_{\textrm{min}}(\varphi_x)$.

\subsection{Truncation error in equation~\eqref{gEta}}
\label{truncationError}
In this section we bound the error of truncating the sum in equation~\eqref{gEta},
\begin{equation}
  \label{gEtaAppendix}
  g_{\bar \eta}/ E_{\tilde L_c} = \sum_{\nu } B_\nu e^{i \nu \varphi_{cx}} \prod_{j = 1}^k c_{\eta_j}^{(j)}(\nu \alpha_j) \,.
\end{equation}
 Noting that $\cos_{\beta_c}(0) = 1$ and $U_{\textrm{ZPE}}(\varphi_x = 0) = \zeta_c\sqrt{1 - \beta_c}$, we compare the two expressions in \eqref{EgFinal} at $\varphi_x = 0$,
\begin{equation}
 \beta_c + \zeta_c \sqrt{1 - \beta_c} = B_0 + 2\sum_{\nu>0} B_{\nu}\,,
\end{equation}
where we have used the fact that $B_\nu = B_{-\nu}$. Collecting terms dependent and independent of $\zeta_c$, we obtain the identities
\begin{align}
  \begin{split}
   2\sum_{\nu>0} B_\nu^{(0)} & = \beta_c + \frac{1}{4}\beta_c^2 \\
   2\sum_{\nu>0} B_\nu^{(1)} & =  \sqrt{1 - \beta_c} - G_0(\beta_c) + \beta_c G_1(\beta_c)
  \end{split}
\end{align}
where (using equation~\eqref{interactionSeries} for $\nu \neq 0$)
\begin{align}
  \begin{split}
    B_{\nu} & = B_\nu^{(0)} + \zeta_c B_{\nu}^{(1)}\\
    B_\nu^{(0)} & = \frac{ J_{\nu}(\beta_c \nu)}{\nu^2}\\
    B_\nu^{(1)} & =\frac{1}{\nu}\sum_{\mu} \mu G_{\mu}(\beta_c) J_{\nu-\mu}(\beta_c \nu)\,.
  \end{split}
\end{align}
Using the fact that $B_{\nu}^{(1)}\leq 0 \leq B_{\nu}^{(0)}$ for all $\nu \neq 0$, this allows us to define the truncation error bound
\begin{equation}
  R_{\nu_{\textrm{max}}} \geq |g_{\bar \eta} - g_{\bar \eta}^{(\nu_{\textrm{max}})} |/E_{\tilde L_c} \,,
\end{equation}
where $g_{\bar \eta}^{(\nu_{\textrm{max}})}$ is obtained by summing the series \eqref{gEtaAppendix} only up to $|\nu|\leq \nu_{\max}$. The bound can be computed numerically as\footnote{We assume that the convolution defining $B_{\nu}^{(1)}$ is carried out to arbitrary precision. This is a good approximation as $\mu G_{\mu}(\beta_c) $ decays exponentially in $\mu$. For example, at $\beta_c= 0.95$ we have that $\mu |G_{\mu}(\beta_c) |< 10^{-16}$ for all $|\mu|\geq 101$. }
\begin{equation}
  R_{\nu_{\max}} =  R_{\nu_{\max}}^{(0)} + \zeta_c  R_{\nu_{\max}}^{(1)} \,,
\end{equation}
where (using the fact that the product  $|e^{i \nu \varphi_{cx}} \prod_{j = 1}^k c_{\eta_j}^{(j)}(\nu \alpha_j)|<1$)
\begin{align}
  \begin{split}
    R_{\nu_{\max}}^{(0)} & = \l|\beta_c + \frac{1}{4}\beta_c^2-2\sum_{\nu=1}^{\nu_{\textrm{max}}} B_{\nu}^{(0)} \r|\\
    R_{\nu_{\max}}^{(1)} & = \l|\sqrt{1 - \beta_c} - G_0(\beta_c) + \beta_c G_1(\beta_c) -2\sum_{\nu=1}^{\nu_{\textrm{max}}} B_{\nu}^{(1)} \r|\,.
  \end{split}
\end{align}
We remark that the error bound grows quickly as $\beta_c\rightarrow 1$. For example, to achieve an error in $g_{\bar \eta}$ of at most $10^{-3}\times E_{\tilde L_c}$ for $\beta_c = 3/4$ and $\zeta_c = 1/4$, we are required to truncate at $\nu_{\max}\geq 18$, while the same bound for $\beta_c = 0.95$ requires $\nu_{\textrm{max}}\geq 187$.

\subsection{Validity of Born-Oppenheimer Approximation: Diagonal Correction}
\label{BOValidity}
In this section we discuss the approximations leading to the general coupler-mediated interaction Hamiltonian, equation~\eqref{Hint}. We begin by discussing the Born-Oppenheimer Approximation used to eliminate the coupler degree of freedom. As in the study of molecular collisions, we assume that the (fast) coupler is always in its ground state. That is, we make the following ansatz for the full wave-function in the flux operator basis \cite{Pierce1962,Tully1976},
\begin{equation}
  \label{ansatz}
  \Psi(\varphi_c,\bar \varphi_q,t) = \psi_{g}(\varphi_c; \bar \varphi_q) \, \chi(\bar \varphi_q,t)\,.
\end{equation}
Here $\bar \varphi_q = (\varphi_1,\varphi_2,\,...\,,\varphi_k)$ denotes the $k$ qubit flux variables, while $\psi_{g}(\varphi_c; \bar \varphi_q) $ is the ground state of the coupler Hamiltonian $\hat H_c$ (equation~\eqref{Hc}). Since $\hat H_c$ is parameterized by the qubit-dependent flux variable $\varphi_x$, we likewise treat $\psi_g$ as a parameterized function of $\bar \varphi_q$. The effective qubit Hamiltonian is obtained by considering the Schr\"{o}dinger equation for the ansatz wave-function,
\begin{align}
  \label{BOSEqn0}
  \begin{split}
    i \hbar \, \psi_{g}( \varphi_c; \bar \varphi_q ) \,  \partial_t    \chi(\bar \varphi_q, t)  & =  \l(\sum_j  H_j +  H_c \r) \psi_{g}(\varphi_c; \bar \varphi_q) \, \chi(\bar \varphi_q , t)  \\
    & = \l(\sum_j E_{L_j}\l(- 2 \zeta_j^2 \partial_{\varphi_j}^2  + U_j\r)  +  E_g \r)\psi_{g}(\varphi_c; \bar \varphi_q) \, \chi(\bar \varphi_q, t)  \\
    & = -\sum_j E_{L_j} 2 \zeta_j^2 \Big( \l(\partial_{\varphi_j}^2  \psi_{g}(\varphi_c; \bar \varphi_q)\r) \, \chi(\bar \varphi_q, t)   \\
    & \quad \quad + 2 \l(\partial_{\varphi_j}  \psi_{g}(\varphi_c; \bar \varphi_q) \r)\, \l(\partial_{\varphi_j} \chi(\bar \varphi_q) \r) +   \psi_{g}(\varphi_c; \bar \varphi_q) \, \partial_{\varphi_j}^2 \chi(\bar \varphi_q, t)\Big)\\
    & \quad + \l(\sum_j E_{L_j} U_j +  E_g \r) \psi_{g}(\varphi_c; \bar \varphi_q) \,\chi(\bar \varphi_q, t)\,.  \\
  \end{split}
\end{align}
Here we have assumed that the individual qubit Hamiltonians are of the generic form $E_{L_j} \l( 4 \zeta_j^2 \frac{\hat q_j^2}{2} + U_j(\hat \varphi_j) \r)$ (charge plus flux potential term), with a linear impedance $\zeta_j = \frac{2 \pi e}{\Phi_0} \sqrt{\frac{L_j}{C_j}}$, and we have used $H_c \psi_{g}(\varphi_c; \bar \varphi_q) = E_g \psi_{g}(\varphi_c; \bar \varphi_q)$. 

The Born-Oppenheimer ansatz~\eqref{ansatz} allows us to consider the reduced dynamics of the qubit systems alone. To do so, we multiply both sides of equation~\eqref{BOSEqn0} by $\psi_{g}(\varphi_c; \bar \varphi_q)^*$ and integrate over the variable $\varphi_c$. Carrying out this integration leaves a reduced Schr\"{o}dinger equation involving only the qubit wave-function $\chi(\bar \varphi_q)$,
\begin{equation}
  \label{BOSEqn}
  i \hbar \partial_t \chi(\bar \varphi_q , t) = \l(\sum_j E_{L_j}\l(- 2 \zeta_j^2 \partial_{\varphi_j}^2 +  U_j(\varphi_j) \r) +  E_g(\bar \varphi_q) + K(\bar \varphi_q) \r) \chi(\bar \varphi_q , t)\,,
\end{equation}
where we treat the coupler ground state energy $E_g$ as an explicit function of the qubit variables $\bar \varphi_q$ and introduce the Born-Oppenheimer Diagonal Correction~\cite{Tully1976,Valeev2003},
\begin{align}
  \label{BODCK}
  \begin{split}
    K(\bar \varphi_q) & = \int \mbox{d} \varphi_c \, \psi_{g}(\varphi_c; \bar \varphi_q)^* \l(- \sum_j E_{ L_j}2 \zeta_j^2 \partial_{\varphi_j}^2  \psi_{g}(\varphi_c; \bar \varphi_q)\r)\\
    & = -\l(2 \sum_j E_{ L_j} \zeta_j^2 \alpha_j^2 \r) \int \mbox{d} \varphi_c \, \psi_{g}(\varphi_c; \bar \varphi_q)^* \partial_{\varphi_x}^2 \psi_{g}(\varphi_c; \bar \varphi_q) \\
    & = \l(2 \sum_j E_{ L_j} \zeta_j^2 \alpha_j^2 \r) \bra{\partial_{\varphi_x} \psi_{g}} \partial_{\varphi_x} \psi_{g} \rangle\,.
  \end{split}
\end{align}
(This originates from the first term on the third line of~\eqref{BOSEqn0}.) In the derivation of equations~\eqref{BOSEqn} and~\eqref{BODCK} we use the fact that $\psi_g(\varphi_c; \bar \varphi_q)$ is real valued\footnote{The Hamiltonian $H_c$ is real valued in the flux operator basis, hence its eigenstates can be expressed as real functions of $\varphi_{c}$ up to a global phase.}. This fact allows us to drop  in equation~\eqref{BOSEqn} the integrals of $\psi_g(\varphi_c; \bar \varphi_q) \partial_{\varphi_j} \psi_g(\varphi_c; \bar \varphi_q)$ (which vanishes since $\psi_g(\varphi_c; \bar \varphi_q)$ has unit norm), and similarly allows us to equate $  \bra{ \psi_{g}} \partial_{\varphi_x}^2 \psi_{g} \rangle= -\bra{\partial_{\varphi_x} \psi_{g}} \partial_{\varphi_x} \psi_{g} \rangle$.

\begin{figure}[h!]
\includegraphics[width = \textwidth,trim={0 0 .5cm 0},clip]{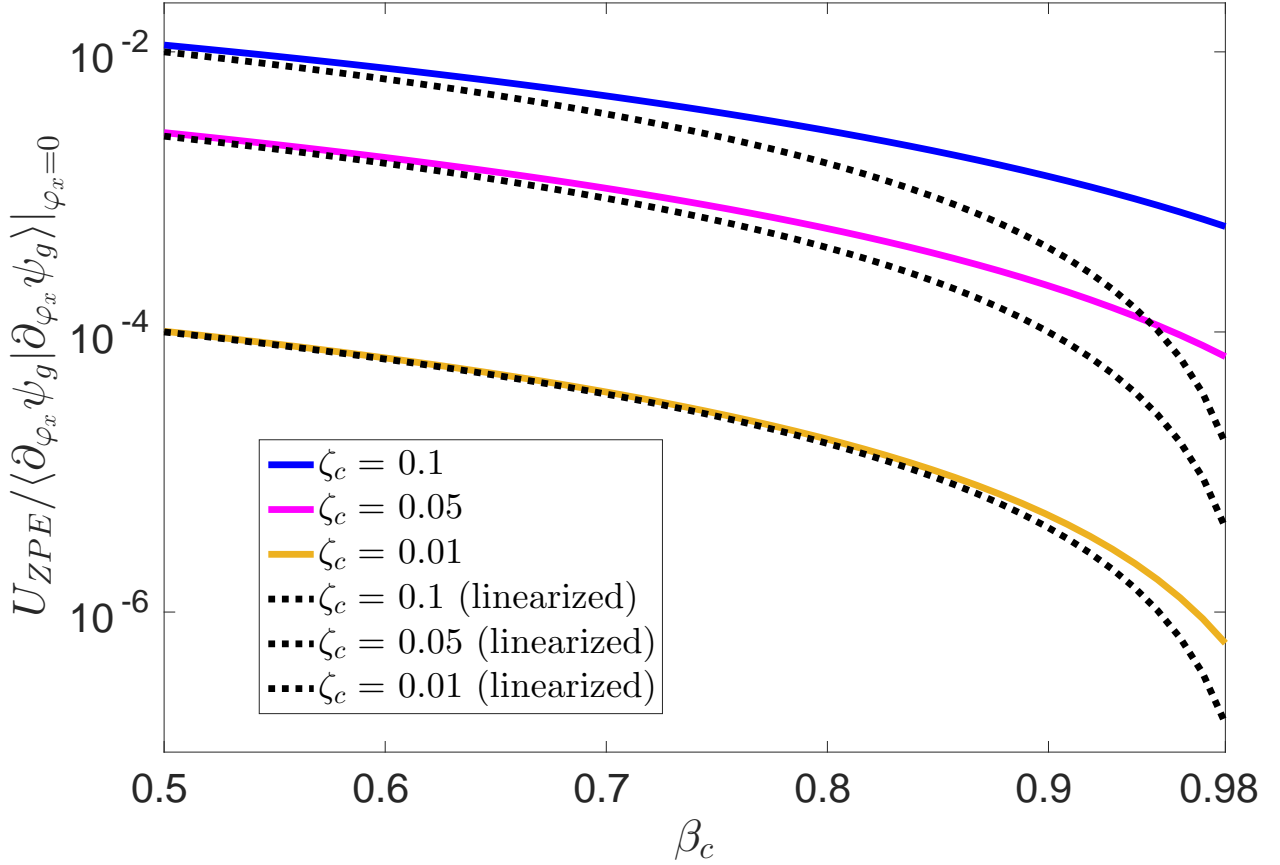}
\caption{Ratio of the zero-point energy $U_{\textrm{ZPE}}$ to the integral in the Born-Oppenheimer Diagonal Correction compared to the linearized Hamiltonian approximation, $4\zeta_c^2 (1 - \beta_c )^{2}$. These calculations we carried out at flux bias $\varphi_{cx} = 0$ (which minimizes the ratio). Solid curves (starting from the top) correspond to the numerically exact ratio at coupler impedances $\zeta_c = 0.1$ (dark blue), 0.05 (magenta), and $0.01$ (light orange), respectively. The Hamiltonian $\hat H_c$ (equation~\eqref{Hc}) was diagonalized in the harmonic oscillator basis truncated at $70$ basis states, and the vector $\partial_{\varphi_x} \ket{\psi_g}$ was then computed using equation~\eqref{psigPrime}. The value of $U_{\textrm{ZPE}}$ was computed by subtracting the classical energy contribution $\beta_c \cos_{\beta_c}(0) = \beta_c$ from the ground state energy $E_g/E_{\tilde L_c}$. Overlayed dashed curves correspond to the linear approximation, $4 \zeta_c^2 (1 - \beta_c)^2$, equation~\eqref{BODCsmall}.}
\label{BODC}
\end{figure}

In the main text we neglect the diagonal correction $K(\bar \varphi_q)$ since it is typically negligible. In order to bound its size, we approximate the integral factor $\int \mbox{d} \varphi_c \,  \l|\partial_{\varphi_x} \psi_{g}(\varphi_c; \bar \varphi_q) \r|^2 = \bra{\partial_{\varphi_x}\psi_g} \partial_{\varphi_x}\psi_g \rangle$ by linearizing the coupler Hamiltonian $\hat H_c$. Noting that $(E_g - \hat H_c) \ket{\psi_g} = 0$ for all $\varphi_x$, we take the derivative to show that
\begin{align}
  \label{psigPrime}
  \begin{split}
     \ket{\partial_{\varphi_x}\psi_g} & = - (E_g - \hat H_c)^{-1} \partial_{\varphi_x}\l((E_g - \hat H_c) \r) \ket{\psi_g}\\
    & = -\frac{E_{\tilde L_c}}{E_g - \hat H_c} \hat \varphi_c \ket{\psi_g}\,.
  \end{split}
\end{align}
(Note that $(E_g - \hat H_c)^{-1}$ represents the Moore-Penrose pseudo-inverse, which vanishes on the state $\ket{\psi_g}$.) As we did for the analysis of the zero-point energy, we now approximate $\hat H_c$ as an harmonic oscillator with characteristic frequency $E_{\tilde L_c} \sqrt{4 \zeta_c^2 U''(\varphi_c^{(*)})} = 2 E_{\tilde L_c} \zeta_c \sqrt{1 - \beta_c \cos(\varphi_c^{*})}$ (see equation~\eqref{linHam}). Using $\hat \varphi_c = \sqrt{\frac{\zeta_c}{\sqrt{1 - \beta_c \cos(\varphi_c^{*})}}}\l( \hat a + \hat a^\dagger\r)$, we obtain
\begin{equation}
  \label{psigPrimeApprox}
   \ket{\partial_{\varphi_x} \psi_g} \simeq \frac{1}{2 \sqrt{\zeta_c}  (1 - \beta_c \cos(\varphi_c^{*}))^{3/4}}  \ket{1}\,,
\end{equation}
where $\ket{1}$ is the first harmonic oscillator excited state. In fact this approximation diverges as $\beta_c \cos(\varphi_c^{*}) \rightarrow 1$, which suggests that we can only use it as an approximate upper bound for the norm of $\partial_{\varphi_x} \ket{\psi_g}$. Substituting equation~\eqref{psigPrimeApprox} into~\eqref{BODCK}, we obtain
\begin{equation}
  \label{Kbound}
  K(\bar \varphi_c)/E_{\tilde L_c} \lesssim  2 \sum_j \frac{E_{ L_j}  \zeta_j^2 \alpha_j^2}{E_{\tilde L_c}}  \frac{1}{4 \zeta_c (1 - \beta_c \cos(\varphi_c^{*}))^{3/2}}  \,.
\end{equation}
Comparing $K/E_{\tilde L_c}$ to the coupler's zero-point energy ($U_{\textrm{ZPE}}$, equation~\eqref{ZPE}) at their perspective maxima and minima ($\varphi_{x} = 0$), we see that it is valid to neglect the diagonal correction in the limit
\begin{equation}
  \label{BODCsmall}
  2 \sum_j \frac{E_{ L_j}  \zeta_j^2 \alpha_j^2}{E_{\tilde L_c}}  \ll  \frac{ U_{\textrm{ZPE}}(0)}{\bra{\partial_{\varphi_x}\psi_g} \partial_{\varphi_x}\psi_g \rangle} \simeq 4 \zeta_c^2 (1 - \beta_c )^2\,.
\end{equation}
We stress that the value $\zeta_c^2 (1 - \beta_c )^2$ on the right hand side of equation~\eqref{BODCsmall} is only a good approximation when $\beta_c \cos(\varphi_c^{*})$ is not too close to $1$ (see Fig.~\ref{BODC}). If we assume that the qubit and coupler impedances are comparable, identical qubits, and that $k$ and $\beta_c$ are not too large, then equation~\eqref{BODCsmall} simplifies to 
\begin{equation}
  E_{L_j} \alpha_j^2  \ll  E_{\tilde L_c} (1 - \beta_c )^2\,.
\end{equation}
This bound is achievable even for relatively large nonlinearity $\beta_c$ and coupling $\alpha_j$ as long as we are in the fast coupler limit, $E_{\tilde L_c} \ll E_{L_j}$.

\subsection{Non-adiabatic corrections to Born-Oppenheimer}
\label{BONonadiabatic}

We now discuss the leading non-adiabatic corrections to the Born-Oppenheimer Approximation. These corrections stem from an exact representation qubit-coupler wave-function\cite{Pelzer1932,Born1952},
\begin{equation}
  \label{PsiFull}
  \tilde \Psi(\varphi_c,\bar \varphi_q, t) = \sum_m  \psi_{m}(\varphi_c; \bar \varphi_q) \, \chi_m(\bar \varphi_q, t) \,.
\end{equation}
Here the wave-functions $\psi_{m}$ denote the (normalized) eigenstates of $\hat H_c$ parameterized by the qubit flux variables $\bar \varphi_q$ through the coupler bias, $\varphi_x$. (Our original ansatz truncated this sum at the ground state.) Repeating the same analysis as in equation~\eqref{BOSEqn0}, then multiplying by $\psi_{m} (\varphi_c; \bar \varphi_q)$ and integrating, we obtain a set of coupled equations for the functions $\chi_m$,
\begin{equation}
  \label{BOSqnFull}
  i\hbar \partial_t \chi_{m} = \l( \sum_j H_j + E_m + K_{m,m}\r) \chi_m + \sum_{m' \neq m} \l(T_{m,m'} + K_{m,m'} \r) \chi_{m'} \,. 
\end{equation}
Here $E_m$ is the energy of $\psi_{m}$ (as an eigenstate of $\hat H_c$, parameterized by $\varphi_x$) while the coupling terms $T_{m,m'} + K_{m,m'}$ are defined by
\begin{align}
  \begin{split}
    K_{m,m'} = \l(2 \sum_j E_{ L_j} \zeta_j^2 \alpha_j^2 \r) \bra{ \partial_{\varphi_{x}} \psi_{m} } \partial_{\varphi_{x}} \psi_{m'} \rangle  \,.
  \end{split}
\end{align}
and
\begin{align}
  \label{Tmm'}
  \begin{split}
    T_{m,m'} = i \sum_{j}2 E_{L_j} \zeta_j^2 \alpha_j \l[\bra{  \psi_{m} } \partial_{\varphi_{x}} \psi_{m'} \rangle,q_j  \r ]_+\,.
  \end{split}
\end{align}
These terms originate in the integrals of the third and fourth lines of~\eqref{BOSEqn0} (generalized to wave-function~\eqref{PsiFull}). Notice that $K_{g,g}$ corresponds to the diagonal correction discussed previously, while $T_{m,m} = 0$ for all $m$ since $\bra{  \psi_{m} } \partial_{\varphi_{x}} \psi_{m} \rangle = \partial_{\varphi_x} \l(\bra{  \psi_{m} }  \psi_{m} \rangle/2\r) = 0$. Also we have expressed $T_{m,m'}$ as an anti-commutator involving the charge operators ${q_j = -i \partial_{\varphi_j} = -i \alpha_j \partial_{\varphi_x}}$.

From the Schr\"{o}dinger equation~\eqref{BOSqnFull} we can interpret the qubit wave-functions $\chi_{m}(\bar \varphi_q)$ as residing in different subspaces associated with each eigenstate of $\hat H_c$.  The original Born-Oppenheimer Approximation is equivalent to neglecting the coupling terms $T_{m,m'} + K_{m,m'}$ (which cause transitions between these subspaces) and assuming that the qubits start in the ground state subspace $m = g$. Thus, in order for the Born-Oppenheimer Approximation to be valid the effect of these couplings must be small. To see when this is the case, we first observe that $|\bra{ \partial_{\varphi_{x}} \psi_{m} } \partial_{\varphi_{x}} \psi_{m'} \rangle|^2 \leq  \bra{ \partial_{\varphi_x} \psi_{m} } \partial_{\varphi_{x}} \psi_{m} \rangle \bra{ \partial_{\varphi_x} \psi_{m'} } \partial_{\varphi_{x}} \psi_{m'} \rangle $ by the Cauchy-Schwarz inequality. Hence for $m = g$ we expect the coupling corrections $K_{g,m'}$ to be comparable to the diagonal correction $K$. Thus assuming a non-negligible gap $E_1 - E_g$ on the order of the coupler's zero point energy, we may ignore $K_{g,m}$ whenever it is valid to ignore $K$ (condition~\eqref{BODCsmall}). The other non-adiabatic coupling terms ($T_{m,m'}$) may have a non-negligible effect on the qubit dynamics, although a detailed study of these corrections is beyond the scope of this work.

\subsection{Generalization to more complicated circuits}
\label{generalization}
The techniques used in this paper can also be used to study more complicated circuit configurations. Specifically, the derivation of equation~\eqref{transInvertDerivation} can be immediately generalized to multivariate functions under the more general constraint,
\begin{equation}
  \label{transGeneral}
  \bar x - \bar \varphi - \bar F(\bar x) = 0\,.
\end{equation}
In this case we assume that $\bar F(\bar x)$ is a smooth, periodic function of all variables $x_i$, and that its Jacobian matrix $(D \bar F)_{i j} = \partial_{x_j} F_i(\bar x)$, has bounded norm $||D F||<1$ for all $\bar x$\footnote{This ensures that for every value of $\bar \varphi$, the solution $\bar x$ to \eqref{transGeneral} is unique.}. The generalized version of equation~\eqref{transInvertDerivation} is then
\begin{equation}
  \label{generalResult}
  f(\bar x) =\sum_{\bar \nu} e^{i \bar \nu \cdot \bar \varphi} [e^{i \bar \nu \cdot \bar F(\bar \tau)} \Gamma(\bar \tau) f(\bar \tau)]_{\bar \nu}\,,
\end{equation}
with $\Gamma(\bar \tau) = \mbox{det}\l(I - D F(\bar \tau) \r)$. In this case
$$[h(\bar \tau)]_{\bar \nu} = \int_{-\pi}^\pi \mbox{d}^n \tau \frac{e^{-i  \bar \nu \cdot \bar  \tau}}{(2 \pi)^n} h(\bar \tau)$$
denotes the Fourier coefficient of the multi-variate function $h(\bar \tau)$ corresponding to the index vector $\bar \nu$.\footnote{A further generalization can be made in the case where $\bar F(\bar x)$ is not periodic. This corresponds to replacing the Fourier series~\eqref{generalResult} with a Fourier transform.}

\begin{figure}
\includegraphics[width = \textwidth]{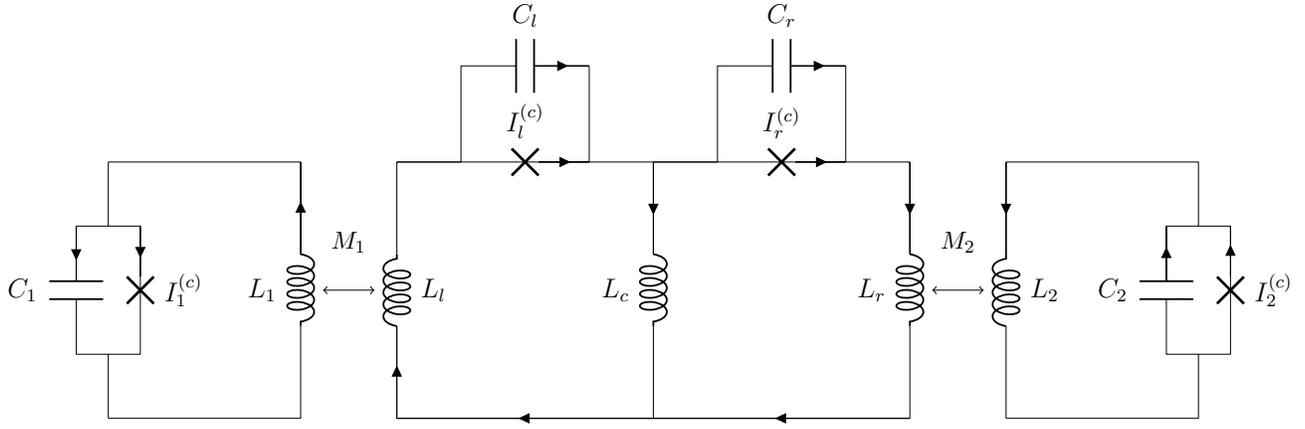}
\caption{A more complicated coupler implementation involving two distinct junctions.}
\label{diagram2JJcoupler}
\end{figure}

As an example, we may apply our general result~\eqref{generalResult} to the two-junction coupler circuit seen in Fig~\ref{diagram2JJcoupler}, which has two independent, interacting degrees of freedom, $\bar \varphi_c = (\varphi_l,\varphi_r)$. As we did in the main text, to study this circuit we would compute the flux configuration $\bar \varphi_c ^{(*)} = (\varphi_l^{(*)},\varphi_r^{(*)})$ corresponding to the minimum of its potential. Although we do not work it out here, one can show that the gradient equations $\nabla_{\bar \varphi_c} U(\bar \varphi_c; \bar \varphi_x) = 0$ corresponding to this minimum are of the form
\begin{equation}
  \bar x - \bar \varphi_x - \mathcal{B} \sin(\bar x) = 0\,.
\end{equation}
In this case $(\sin(\bar x))_j = \sin(x_j)$ and the vector $\bar \varphi_x$ corresponds to the external flux biases associated with each coupler loop. Similarly, $\mathcal{B}$ (analogous to $\beta$) is a matrix relating the coupler's critical currents and linear inductances. Generalizing our analysis for finding the coupler potential minimum (i.e., the classical part of the ground state energy) corresponds to setting $F(\bar x) = \mathcal{B} \sin(\bar x)$ and $f(\bar x) = U(\bar x; \bar \varphi_x)$  in equation~\eqref{generalResult}. The coupler zero-point energies may be approximated similarly to what is done in Section~\ref{ZPEDerivation}, though this is more challenging as now there more than one effective normal mode frequencies.

\end{document}